\documentclass{natureprintstyle}
\usepackage{epsfig,caption}
\usepackage{color}
\usepackage{bm}
\usepackage{longtable}
\usepackage{amssymb}
\usepackage{rotating}

\usepackage{graphicx, bm, amssymb, xcolor}
\usepackage{verbatim}
\usepackage{hyperref}
\usepackage{hypcap}
\usepackage[fleqn]{amsmath}
\usepackage{xspace}
\usepackage{lineno}

\newcommand{\apj}{Astrophys. J.}

\newcommand{\pasp}{Publ. Astron. Soc. Pac.}
\newcommand{\apjs}{Astrophys. J. Supp.}
\newcommand{\araa}{Annu. Rev. Astron. Astrophys.}
\newcommand{\mnras}{Mon. Not. R. Astron. Soc.}
\newcommand{\apjl}{Astrophys. J. Let.}
\newcommand{\aap}{Astron. Astrophys.}
\newcommand{\aj}{Astron. J.}
\newcommand{\nat}{Nature}

\newcommand{\msun}{\mbox{M$_\odot$}}

\newcommand{\myr}{\mbox{${\rm Myr}$}}
\newcommand{\gyr}{\mbox{${\rm Gyr}$}}
\newcommand{\pc}{\mbox{${\rm pc}$}}

\newcommand{\kpc}{\mbox{${\rm kpc}$}}
\newcommand{\kms}{\mbox{${\rm km}~{\rm s}^{-1}$}}
\newcommand{\cmc}{\mbox{${\rm cm}^{-3}$}}
\newcommand{\dex}{\mbox{${\rm dex}$}}

\newcommand{\tsf}{t_{\rm sf}}
\newcommand{\tref}{t_{\rm H\alpha,ref}}
\newcommand{\tstar}{t_{\rm H\alpha}}
\newcommand{\tgas}{t_{\rm CO}}
\newcommand{\tover}{t_{\rm fb}}
\newcommand{\esf}{\epsilon_{\rm sf}}
\newcommand{\etafb}{\eta_{\rm fb}}
\newcommand{\vfb}{v_{\rm fb}}

\newcommand{\mh}{[{\rm M}/{\rm H}]}
\newcommand{\tdep}{t_{\rm dep}}

\newcommand{\hii}{H{\sc ii}\xspace}
\newcommand{\halpha}{H$\alpha$\xspace}
\newcommand{\co}{\mbox{CO(1-0)}\xspace}

\newcommand{\clumpfind}{{\sc Clumpfind}\xspace}
\newcommand{\be}{\begin{equation}}
\newcommand{\ee}{\end{equation}}

\title{Fast and inefficient star formation due to short-lived molecular clouds and rapid feedback}

\author{J.~M.~Diederik Kruijssen$^{1,2}$,
Andreas Schruba$^3$,
M\'{e}lanie Chevance$^1$,
Steven~N.~Longmore$^4$,
Alexander~P.~S.~Hygate$^{2,1}$,
Daniel~T.~Haydon$^1$,
Anna F.~McLeod$^{5,6}$,
Julianne J.~Dalcanton$^7$,
Linda J.~Tacconi$^3$
\& Ewine F.~van Dishoeck$^{8,3}$}

\usepackage{graphicx}
\makeatletter
\let\saved@includegraphics\includegraphics
\AtBeginDocument{\let\includegraphics\saved@includegraphics}
\renewenvironment*{figure}{\@float{figure}}{\end@float}
\makeatother

\begin{document}

\maketitle

\let\thefootnote\relax\footnote{

\begin{affiliations}
\item Astronomisches Rechen-Institut, Zentrum f\"{u}r Astronomie der Universit\"{a}t Heidelberg, M\"{o}nchhofstra\ss e 12-14, 69120 Heidelberg, Germany

\item Max-Planck Institut f\"{u}r Astronomie, K\"{o}nigstuhl 17, 69117 Heidelberg, Germany

\item Max-Planck Institut f\"{u}r Extraterrestrische Physik, Giessenbachstra\ss e 1, 85748 Garching, Germany

\item Astrophysics Research Institute, Liverpool John Moores University, IC2, Liverpool Science Park, 146 Brownlow Hill, Liverpool L3 5RF, United Kingdom

\item Department of Astronomy, University of California Berkeley, Berkeley, CA 94720, USA

\item Department of Physics \& Astronomy, Texas Tech Univ., PO Box 41051, Lubbock, TX 79409, USA

\item Department of Astronomy, University of Washington, Box 351580, Seattle, WA 98195, USA

\item Leiden Observatory, Leiden University, P.O.\ Box 9513, NL-2300 RA, Leiden, the Netherlands

\end{affiliations}

} 

\vspace{-3.5mm}
\begin{abstract}
The physics of star formation and the deposition of mass, momentum, and energy into the interstellar medium by massive stars (`feedback') are the main uncertainties in modern cosmological simulations of galaxy formation and evolution\cite{somerville15,naab17}.
These processes determine the properties of galaxies\cite{scannapieco12,hopkins13b}, but are poorly understood on the $\mathbf{\lesssim}$100~pc scale of individual giant molecular clouds (GMCs)\cite{kennicutt12,krumholz14} resolved in modern galaxy formation simulations\cite{grand17,hopkins18}.
The key question is why the timescale for depleting molecular gas through star formation in galaxies ($\tdep\approx2~\gyr$)\cite{bigiel08,calzetti12} exceeds the dynamical timescale of GMCs by two orders of magnitude\cite{zuckerman74b}. Either most of a GMC's mass is converted into stars over many dynamical times\cite{koda09}, or only a small fraction turns into stars before the GMC is dispersed on a dynamical timescale\cite{leisawitz89,elmegreen00}.
Here we report our observation that molecular gas and star formation are spatially de-correlated on GMC scales in the nearby flocculent spiral galaxy NGC300, contrary to their tight correlation on galactic scales\cite{kennicutt12}. We demonstrate that this de-correlation implies rapid evolutionary cycling between GMCs, star formation, and feedback.
We apply a novel statistical method\cite{kruijssen14,kruijssen18} to quantify the evolutionary timeline and find that star formation is regulated by efficient stellar feedback, driving GMC dispersal on short timescales ($\mathbf{\sim}$1.5~Myr) due to radiation and stellar winds, prior to supernova explosions.
This feedback limits GMC lifetimes to about one dynamical timescale ($\mathbf{\sim}$10~Myr), with integrated star formation efficiencies of only 2$\mathbf{-}$3\%.
Our findings reveal that galaxies consist of building blocks undergoing vigorous, feedback-driven lifecycles, that vary with the galactic environment and collectively define how galaxies form stars.
Systematic applications of this multi-scale analysis to large galaxy samples will provide key input for a predictive, bottom-up theory of galaxy formation and evolution.
\end{abstract}

We carry out an empirical measurement to explain why $\tdep=\tsf/\esf\approx2~\gyr$, where $\tsf$ and $\esf$ represent degenerate quantities: $\tsf$ is the timescale over which gas is turned into stars and $\esf$ represents the fraction of mass converted into stars (the `star formation efficiency' or SFE). If star formation within GMCs is slow and efficient (long $\tsf$, high $\esf$), GMCs and young stars are co-spatial for many dynamical times. If star formation is fast and inefficient (short $\tsf$, low $\esf$), they should rarely coincide. Therefore, measurements of the spatial correlation between gas and young stars discriminate between these two scenarios.

\begin{figure*}
\centerline{
\includegraphics[width=\textwidth]{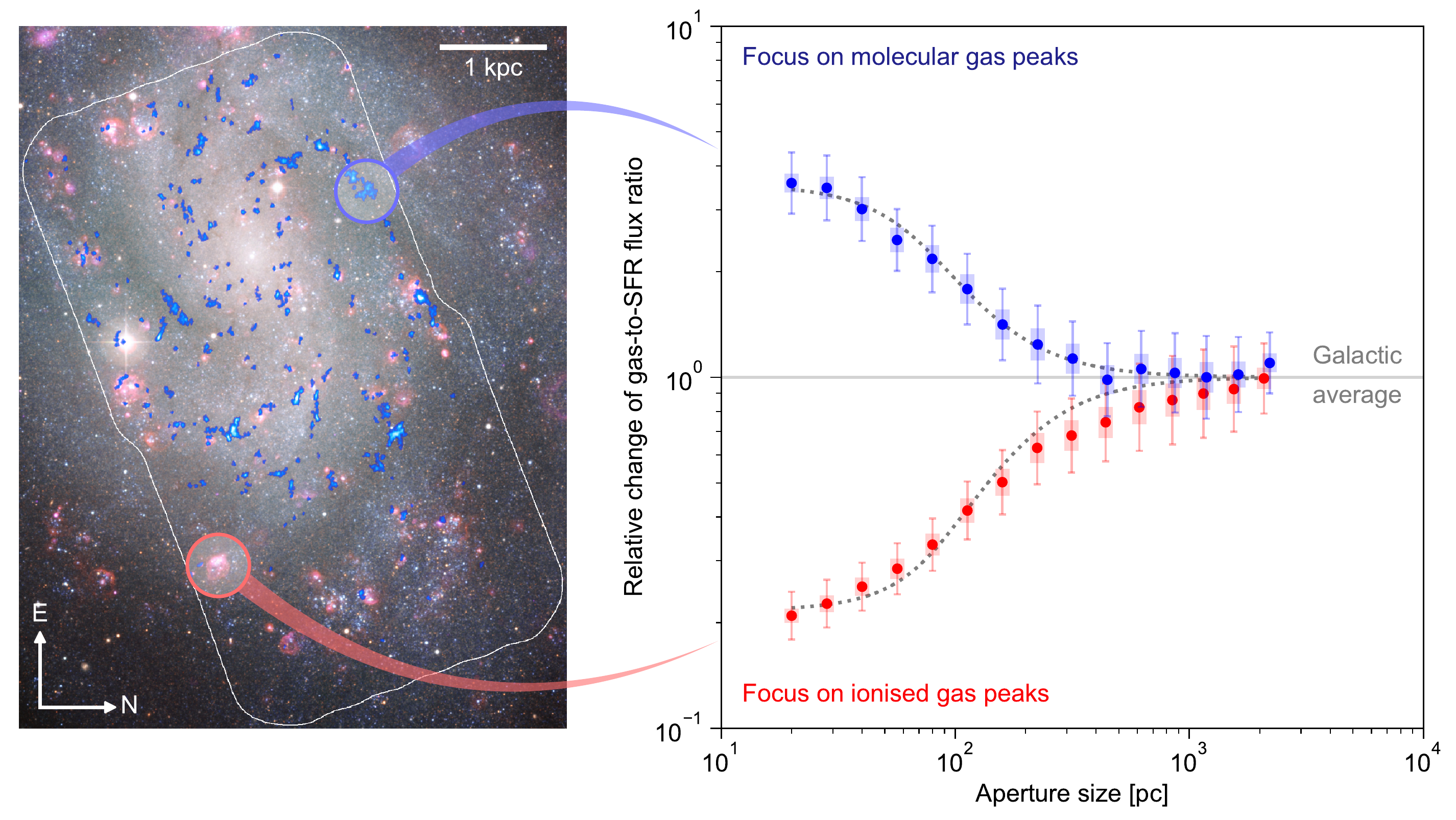}
}
\vspace{-2mm}
\caption{\label{fig:tuningfork}\textbf{De-correlation of molecular gas and young stellar emission on sub-kpc scales.} The ALMA \co map is shown in the left panel as a blue overlay on top of an optical composite image of NGC300 taken with the MPG/ESO \mbox{2.2-m} telescope (see Methods). The white contour indicates the extent of the CO map. Emission from young stars traced by \halpha is shown in pink. The scale bar shows the projected size scale, uncorrected for inclination. The change of the CO-to-\halpha flux ratio relative to the galactic average is shown as a function of spatial scale in the right panel, for apertures placed on CO emission peaks (top branch) and \halpha emission peaks (bottom branch). The error bars indicate the $1\sigma$ uncertainty on each individual data point, whereas the shaded areas indicate the effective $1\sigma$ uncertainty range that accounts for the covariance between the data points\cite{kruijssen18}. The evolutionary timeline of the GMC lifecycle is constrained by fitting a model indicated by the dotted lines. Also see Methods and Supplementary Videos~1 and~2.}
\vspace{-4mm}
\end{figure*}

We characterise the lifecycle of GMCs and star-forming regions by applying a new statistical method\cite{kruijssen18} to maps of the molecular gas and emission from young massive stars in NGC300. This method requires observational data at high sensitivity and resolution over a large field-of-view, now available with the Atacama Large Millimeter/submillimeter Array (ALMA). NGC300 is the perfect target for the first application of this method, as it is the closest ($D=2$~Mpc), face-on, star-forming disc galaxy accessible from the southern hemisphere. \autoref{fig:tuningfork}~(left) shows the molecular gas traced by our high-resolution ($2''=20~\pc$) ALMA map of \co. We combine this with a matched-resolution map of \halpha-emitting \hii regions from the MPG/ESO \mbox{2.2-m} telescope to trace recent star formation. The use of \halpha means that we define `star formation' to refer to an unembedded stellar population, with a mass of at least $200~\msun$ and a normal stellar initial mass function (see Methods).

We characterise the correlation between GMCs and star formation by placing apertures on peaks of \co or \halpha emission, and measuring how the enclosed CO-to-\halpha flux ratios are elevated or suppressed, respectively, relative to the galactic average as the aperture size is changed (\autoref{fig:tuningfork} and Supplementary Video~1)\cite{schruba10,kruijssen14}. The shorter-lived of these two tracers will be rare compared to the longer-lived, more common one. Only a small number of apertures are required to cover the complete sample of rare, short-lived emission peaks. These will encompass a relatively small part of the galaxy and contain few of the many long-lived emission peaks. This results in a CO-to-\halpha flux ratio that differs significantly from the galactic average. Conversely, covering the long-lived emission peaks requires numerous apertures that will also include many of the short-lived emission peaks, resulting in a modest deviation from the galactic CO-to-\halpha ratio.

We fit a model describing this statistical behaviour\cite{kruijssen18} (see Methods and Supplementary Video~2) to measure how long GMCs live and form stars ($\tgas$), how long feedback takes to evacuate residual gas ($\tover$, the time for which GMCs and \hii regions coexist), and the mean separation length between independent star-forming regions ($\lambda$), below which the CO-to-\halpha ratio significantly deviates from the galactic average. These evolutionary time and length scales define the fraction of gas converted into stars (the cloud-scale integrated SFE), the time-averaged mass outflow rate per unit star formation rate (SFR) (the mass loading factor $\etafb$), and the feedback outflow velocity ($\vfb$). Extensive tests\cite{kruijssen18} show that these quantities are measured with an accuracy of better than $30\%$. The quantities $\tgas$, $\tover$, $\lambda$, and $\vfb$ are independent of the adopted CO-to-H$_2$ and \halpha-to-SFR conversion factors.

\begin{figure*}
\centerline{
\includegraphics[width=\textwidth]{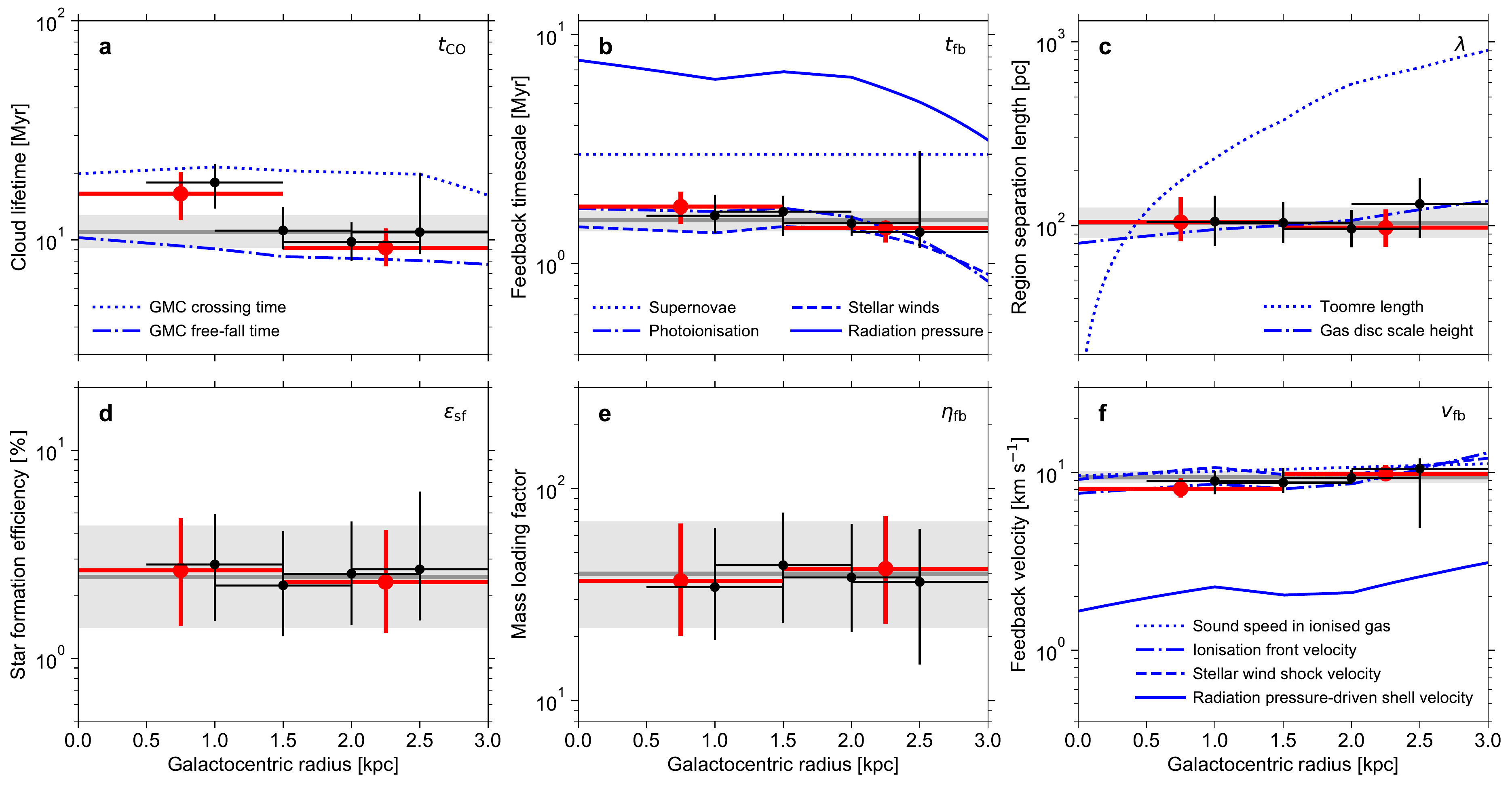}
} 
\vspace{-2mm}
\caption{\label{fig:profiles}\textbf{Radial profiles of constrained quantities in comparison to theoretical predictions.} The top row shows the three free parameters of the model fitted to the data in \autoref{fig:tuningfork}, i.e.\ the GMC lifetime ({\bf a}), the feedback timescale ({\bf b}), and the region separation length ({\bf c}). The bottom row shows quantities derived from the free parameters (see the Methods section), i.e.\ the integrated SFE ({\bf d}), the mass loading factor ({\bf e}), and the feedback outflow velocity ({\bf f}). The horizontal solid lines indicate the galactic average values, with the grey-shaded area indicating the $1\sigma$ uncertainties. The data points indicate the measurements in bins of galactic radius, with vertical error bars indicating the $1\sigma$ uncertainties and horizontal error bars spanning the radial bin. Red and black symbols represent non-overlapping and overlapping radial bins, respectively. Blue lines mark theoretical predictions for various physical mechanisms or reference values (see the legends), showing that the GMC lifecycle in NGC300 takes place on a dynamical time and is regulated by early (pre-supernova) feedback.}
\vspace{-4mm}
\end{figure*}

\autoref{fig:tuningfork}~(left) demonstrates that \co and \halpha emission are rarely co-spatial; they do not trace each other on the cloud scale. We quantify this in \autoref{fig:tuningfork}~(right), which shows the CO-to-\halpha flux ratio as a function of the aperture size for apertures placed on either \co or \halpha emission peaks (cf.~Supplementary Video~1). We find a symmetric de-correlation between \co and \halpha emission on small (${<}150~\pc$) scales, implying that molecular gas and star formation are distinct, subsequent phases of the GMC and \hii region lifecycle with similar durations. The aperture size at which the two branches diverge shows that NGC300 consists of independent, cloud-scale building blocks separated by $100{-}150~\pc$ (cf.~\autoref{fig:profiles}) that exist in a state of vigorous evolutionary cycling. This empirical result implies that star formation is fast and inefficient; the long gas depletion times observed on galactic scales are not due to slow star formation on the cloud scale.

To infer which physics drive the observed evolutionary cycling, we fit the statistical model described in the Methods\cite{kruijssen18} to the data and summarise the constrained quantities in \autoref{tab:fit}. We analyse both the entire field of view and several bins in galactic radius, visualised in \autoref{fig:profiles}. We measure GMC lifetimes spanning $\tgas=9{-}18~\myr$, with a galactic average of $\tgas=10.8^{+2.1}_{-1.7}~\myr$. These lifetimes fall between the GMC crossing and gravitational free-fall times (see Methods) at all galactic radii. GMC lifetimes are the longest in the central region ($R<1.5~\kpc$), most likely caused by higher disc stability (see~\ref{fig:galaxy}) and the correspondingly increased influence of galactic shear in supporting the GMCs against collapse\cite{dobbs13,jeffreson18}. The GMCs live and form stars for a dynamical timescale, after which they are dispersed relative to the new-born stellar population. This implies that the global evolution of GMCs is not significantly slowed down by support from magnetic fields and that GMCs do not live long enough to be affected by spiral arm crossings\cite{koda09} or cloud-cloud collisions\cite{tan00} (see Methods).\looseness=-1

GMC dispersal is likely feedback-driven in NGC300. By construction, our analysis measures the cumulative GMC lifetime up to (and including) the onset of massive star formation traced by \halpha emission, integrating over multiple cycles if these are unassociated with \halpha. The GMCs in NGC300 have high virial ratios (see Methods) and have a cumulative lifetime of about one crossing time. This short timescale does not permit multiple GMC lifecycles during which GMCs disperse dynamically (which each would take a turbulent crossing time) prior to massive star formation. It is therefore most plausible that the GMC lifetime is limited by feedback from massive stars. GMCs and \hii regions coexist on average for $\tover=1.5^{+0.2}_{-0.2}~\myr$, independently of galactic radius, implying that GMCs are dispersed before the first supernovae explode ($\sim3~\myr$, see Methods). Instead, early feedback by photoionisation, stellar winds, or radiation pressure is needed, none of which are currently included in large-scale cosmological simulations of galaxy formation\cite{schaye15,pillepich18}. The quantitative comparison to predicted GMC dispersal timescales (\autoref{fig:profiles}) shows that GMCs in NGC300 are primarily dispersed by photoionisation and stellar winds. The importance of stellar feedback in regulating the GMC lifecycle is further supported by the measured region separation lengths of \mbox{$\lambda=100{-}150~\pc$}. These closely match the disc scale height across the range of galactic radii, which is expected if the interstellar medium is structured by feedback-driven bubbles that depressurise when they break out of the disc\cite{mckee77,hopkins12}.

The short coexistence timescale of 1.5~Myr refers to that of unembedded, massive stars; low-mass stars or embedded massive stars could form and coexist with the parent GMC for longer. While we cannot trace low-mass stars in NGC300, we search for embedded massive star-forming regions using a {\it Spitzer} $24~\mu$m map, which does not suffer from extinction. We find only $4{-}7$ regions bright at $24~\mu$m that are not associated with \halpha emission, out of a total number of 224 identified \halpha peaks (or $\sim3$\%). This small number of embedded massive star-forming regions in NGC300 can be explained by the low GMC surface densities ($\sim20~\msun~\pc^{-2}$, implying an \halpha extinction of only $\sim0.5$~mag), and the observed spatial offset of $24~\mu$m emission and GMCs (consistent with observations of the Milky Way\cite{vutisalchavakul14}). This means that our extinction-corrected \halpha map recovers most of the SFR, and rules out any significant impact of embedded massive stars on the conclusions of this work (also see Methods).

The GMCs in NGC300 achieve average integrated SFEs of only $\esf=2.5^{+1.9}_{-1.1}\%$. The SFE is constant with radius, indicating that local variations in $\tdep=\tgas/\esf$ reflect changes of the GMC lifetime. Feedback disperses the GMCs whenever the observed SFE reaches a few percent. Due to the low SFE, the mass loading factor on the cloud scale is high, with $\etafb=40^{+31}_{-18}$. On galactic scales, $\etafb$ is about an order of magnitude lower\cite{bolatto13,muratov15}, indicating that most of the expelled material will not escape NGC300, but will cool to form a new generation of GMCs. This is confirmed by the measured feedback velocities of $\vfb=8.1{-}10.5~\kms$, which match the predicted photoionisation and stellar wind bubble velocities (including their increase with galactic radius) and fall well below the local escape velocities\cite{westmeier11} of $50{-}120~\kms$. The feedback velocity is derived from $\tover$ and the GMC radius and thus represents a prediction that can be verified independently with integral-field spectroscopy of the ionised gas kinematics.\looseness=-1

Numerical simulations of galaxy formation and evolution often describe cloud-scale star formation with prescriptions based on the galactic-scale relation between the (molecular) gas mass and SFR\cite{schaye15,pillepich18}. However, our results demonstrate that star-forming galaxies undergo rapid evolutionary cycling on spatial scales ${<}150~\pc$. Gas and young stars are not instantly related and the appearance of galaxies constantly changes. This fundamental behaviour of observed galaxies should be reproduced by simulations, rather than adopting the galactic-scale relationships to describe star formation on GMC scales. Our results show that GMC dynamics and pre-supernova feedback mechanisms need to be modelled to achieve this. Otherwise, galaxy simulations reproduce the macroscopic properties of galaxies for the wrong reasons. Initial steps towards including early feedback are promising in this regard\cite{agertz13,hopkins18}.

\autoref{fig:profiles} (top left) shows that the lifecycle of GMCs and \hii regions is not universal, but within NGC300 exhibits a factor-of-2 variation with galactic environment. GMC lifetimes may vary depending on the importance of galactic dynamics relative to cloud-scale dynamics\cite{jeffreson18}, whereas the impact of stellar feedback is predicted to depend on the ambient gas density (see Methods). Future GMC-scale observations of large galaxy samples will reveal dynamical cycling processes similar to those identified here and in previous Galactic observations\cite{kennicutt12}. Our statistical method will facilitate the interpretation of these observations by constraining the physics setting the GMC lifecycle in galaxies as a function of their properties.

\begin{figure}
\centerline{
\includegraphics[width=0.5\textwidth]{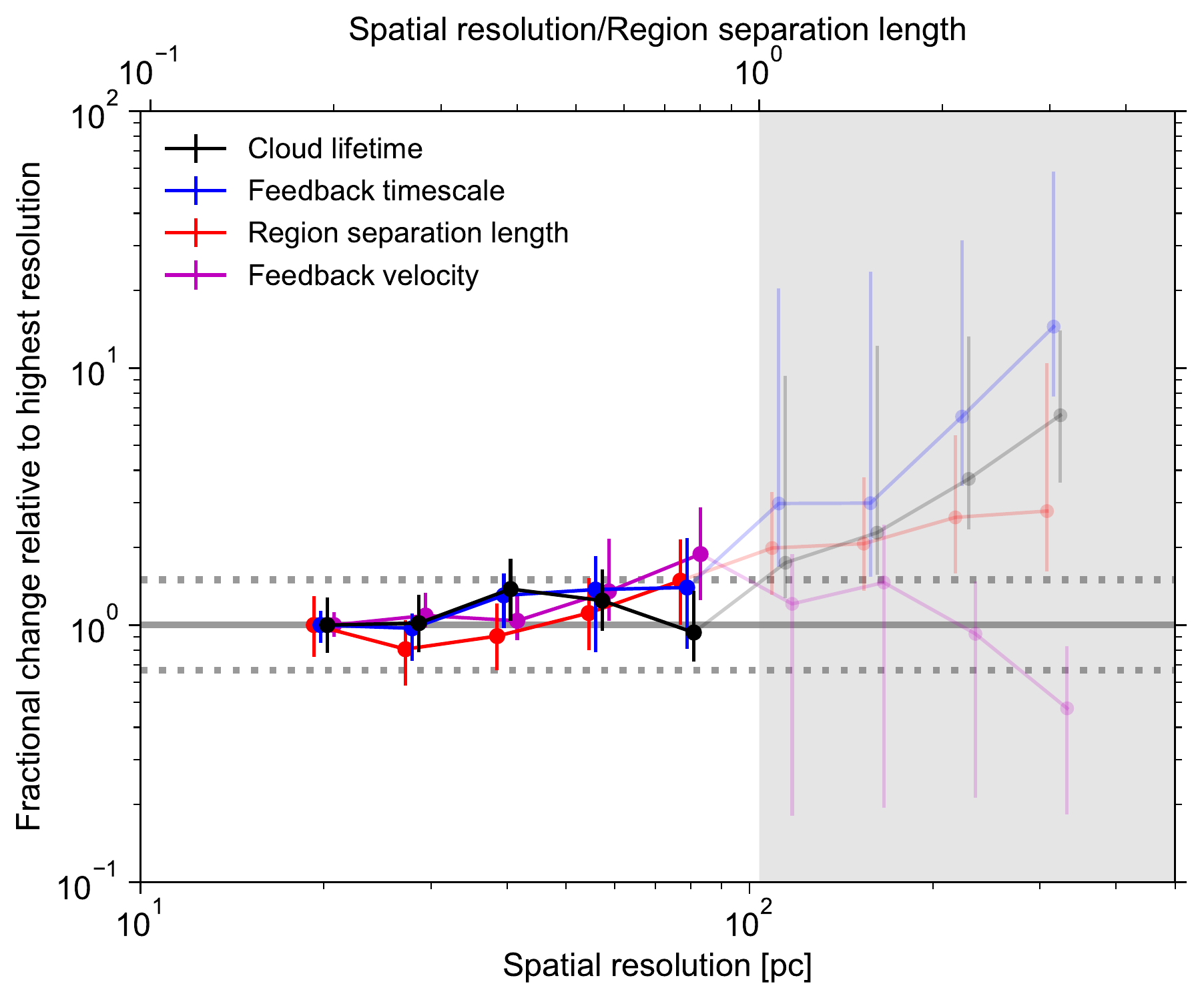}
}
\vspace{-2mm}
\caption{\label{fig:res}\textbf{Dependence of measured quantities on spatial resolution.} The lines and symbols show how four measured quantities (see the legend) change as a function of the spatial resolution to which the analysed maps have been convolved, relative to their values at native resolution. The grey-shaded area indicates spatial resolutions at which the region separation length is not resolved (see top axis). The horizontal solid line marks unity. An accurate measurement (i.e.\ within 50\%, marked by horizontal dotted lines) requires resolving the region separation length. For resolutions better than $2/3\times\lambda$, the data points are statistically consistent with unity to within the $1\sigma$ uncertainties (shown by the vertical error bars).}
\vspace{-4mm}
\end{figure}

\autoref{fig:res} shows the performance of our method\cite{kruijssen18} as a function of spatial resolution (and thus galaxy distance). Owing to its statistical nature, this method does not require resolving individual GMCs or star-forming regions, but only their separation $\lambda$, offering a major benefit relative to previous methods for characterising evolutionary cycling on sub-galactic scales. Local galaxies like NGC300 require a resolution of \mbox{$100{-}150$}~pc, which \autoref{fig:profiles} shows corresponds to the disc scale height; the cloud-scale physics of star formation and feedback can therefore be characterised for hundreds of galaxies within tens of Mpc at $1''$ resolution with current observatories. At high redshift ($z>1$), the high gas content implies elevated scale heights and collapse length scales of\cite{genzel11} ${\sim}1$~kpc (below which fragmentation can still occur), implying that the dynamical cycling scale may be accessible by modern observatories (at $0.1''$) across cosmic time. This possibility is promising in view of the environmental dependences identified here. Mapping the GMC lifecycle across the Universe will eventually enable the representation of galaxies as ensembles of vigorously evolving building blocks and motivate a bottom-up theory of how galaxies grow and form stars, from high redshift to the present day.


\begin{addendum}

\item[Supplementary Information] is linked to the online version of the paper at \url{www.nature.com/nature}.

\item [Acknowledgements] J.M.D.K.\ and M.C.\ acknowledge funding from the German Research Foundation (DFG) in the form of an Emmy Noether Research Group (grant number KR4801/1-1). J.M.D.K.\ acknowledges funding from the European Research Council (ERC) under the European Union's Horizon 2020 research and innovation programme via the ERC Starting Grant MUSTANG (grant agreement number 714907). A.P.S.H.\ and D.T.H.\ are fellows of the International Max Planck Research School for Astronomy and Cosmic Physics at the University of Heidelberg (IMPRS-HD). We thank C.~Faesi for providing his version of the MPG/ESO \mbox{2.2-m} \halpha map of NGC300. We thank B.~W.~Keller and M.~R.~Krumholz for discussions.

\item[Author Contributions] J.M.D.K.\ led the project, carried out the experiment design, developed the analysis method, and wrote the text, to which A.S., M.C., and S.N.L.\ contributed. A.S.\ performed and reduced the ALMA observations and prepared all observational data for their physical analysis. J.M.D.K.\ and M.C.\ carried out the physical analysis of the data, to which A.S., S.N.L., A.P.S.H., and D.T.H.\ contributed. All authors contributed to aspects of the analysis, the interpretation of the results, and the writing of the manuscript.

\item[Author Information]  Reprints and permissions information is available at www.nature.com/reprints. The authors declare that they have no competing financial interests. Correspondence and requests for materials should be addressed to J.M.D.K.\ (kruijssen@uni-heidelberg.de).

\end{addendum}

\clearpage

\begin{table*}
 \centering
  \begin{minipage}{118mm}
  \begin{tabular}{cccccccc}
   \hline\\[-1.5ex]
   Radial interval & $\tgas$ & $\tover$ & $\lambda$ & $\esf$ & $\etafb$ & $\vfb$ & $\chi_{\rm red}^2$  \\ 
     $[\kpc]$ & $[\myr]$ & $[\myr]$ & $[\pc]$ & $[\%]$ & $[-]$ & $[\kms]$ & $[-]$ \\[1ex] 
   \hline\\[-2ex]
   $0.0$--$3.0$ & $10.8^{+2.1}_{-1.7}$ & $1.5^{+0.2}_{-0.2}$ & $104^{+22}_{-18}$ & $2.5^{+1.9}_{-1.1}$ & $40^{+31}_{-18}$ & $9.4^{+0.8}_{-0.7}$ & $0.71$ \\[2ex]
   $0.0$--$1.5$ & $16.2^{+4.2}_{-3.9}$ & $1.8^{+0.3}_{-0.3}$ & $105^{+38}_{-22}$ & $2.6^{+2.1}_{-1.2}$ & $37^{+32}_{-17}$ & $8.1^{+1.3}_{-0.9}$ & $0.61$  \\[0.5ex] 
   $1.5$--$3.0$ & $9.2^{+2.1}_{-1.6}$ & $1.4^{+0.1}_{-0.2}$ & $98^{+25}_{-21}$ & $2.3^{+1.8}_{-1.0}$ & $42^{+33}_{-19}$ & $9.8^{+1.2}_{-0.6}$ & $0.58$  \\[2ex] 
   $0.5$--$1.5$ & $18.3^{+3.9}_{-4.4}$ & $1.6^{+0.4}_{-0.3}$ & $105^{+40}_{-28}$ & $2.8^{+2.1}_{-1.3}$ & $34^{+31}_{-15}$ & $9.0^{+1.2}_{-1.4}$ & $0.50$  \\[0.5ex] 
   $1.0$--$2.0$ & $11.0^{+3.1}_{-2.0}$ & $1.7^{+0.3}_{-0.4}$ & $104^{+30}_{-23}$ & $2.2^{+1.9}_{-1.0}$ & $44^{+33}_{-20}$ & $8.8^{+1.8}_{-1.1}$ & $0.34$  \\[0.5ex] 
   $1.5$--$2.5$ & $9.8^{+2.2}_{-1.8}$ & $1.5^{+0.1}_{-0.2}$ & $96^{+26}_{-20}$ & $2.6^{+2.0}_{-1.1}$ & $38^{+30}_{-17}$ & $9.3^{+1.0}_{-0.6}$ & $0.63$  \\[0.5ex] 
   $2.0$--$3.0$ & $10.8^{+9.4}_{-2.2}$ & $1.4^{+1.7}_{-0.2}$ & $131^{+49}_{-45}$ & $2.7^{+3.7}_{-1.2}$ & $36^{+28}_{-22}$ & $10.5^{+1.5}_{-5.6}$ & $0.75$  \\[1ex] 
   \hline
  \end{tabular} 
  \end{minipage}
  \caption{\label{tab:fit}\textbf{Constrained quantities describing cloud-scale star formation and feedback.}  For various (independent and overlapping) intervals in galactocentric radius, the table lists the GMC lifetime ($\tgas$), the feedback timescale ($\tover$), the region separation length ($\lambda$), the integrated SFE ($\esf$), the mass loading factor ($\etafb$), the feedback outflow velocity ($\vfb$), and the $\chi_{\rm red}^2$ statistic of the model fit.}
\end{table*}

\clearpage

\clearpage

\clearpage

\clearpage


\setcounter{page}{1}
\setcounter{figure}{0}
\setcounter{table}{0}
\renewcommand{\thefigure}{S\arabic{figure}}
\renewcommand{\thetable}{S\arabic{table}}


\begin{center}
{\bf \Large \uppercase{Methods} }
\end{center}

\noindent {\bf Observational data.}
We use observations from ALMA to trace molecular gas and from the MPG/ESO \mbox{2.2-m} telescope to map the H$\alpha$ line.

\vspace{1.5mm} \noindent \emph{ALMA.}
We observe the \mbox{$^{12}$CO($J=1-0$)} transition of two adjacent regions in NGC~300 using ALMA during Cycle~2 and~3. The full presentation of the data is the subject of another paper (A.S.\ \emph{et~al.}, manuscript in preparation). Observations were carried out for $10$ hours with the \mbox{12-m} main array and for $92$ hours with the \mbox{7-m} and \mbox{12-m} antennas of the ALMA Compact Array (ACA) between June~6, 2014 and November~19, 2016. The ALMA receivers were tuned to the ground-state rotational transition of carbon monoxide, CO(1-0). The \mbox{12-m} array was used in configurations C34-2 and C36-2 that provide maximum (unprojected) baselines of $0.330$~km. Two adjacent mosaic fields were observed, each with 
149 pointings of the \mbox{12-m} array and 52 pointings of the \mbox{7-m} array, covering a combined projected area of $10.5' \times 5.5' \approx 6.1~\textrm{kpc} \times 3.2~\textrm{kpc}$ at a distance of $D = 2.0$ Mpc. The observing field is centred about $2.6' = 1.5$~kpc north-west of the nucleus of NGC~300 and covers galactic radii out to $8.2' \approx 4.8$~kpc. The observations were performed with a spectral resolution of $122$~kHz per channel (0.32~km~s$^{-1}$) and a total bandwidth of 468.750~MHz per baseband. In each observing session, one of the sources J0006-0623, J0210-5101, or J2357-5311 was used as the bandpass calibrator and one of the sources J0045-3705, J0051-4226, or J0106-4034 was used as the phase and amplitude calibrator. The absolute flux scale was set with observations of Mars, Venus, Uranus, or Neptune. We adopt ALMA's fiducial absolute flux calibration uncertainty of $5\%$.

We process the data using the ALMA reduction software CASA; calibration is performed in version \mbox{4.2.2} (Cycle~2) and \mbox{4.6.0} (Cycle~3); mapping and deconvolution in version \mbox{4.7.2}. We adopt the pipeline-delivered calibrated data sets. We combine the \mbox{12-m} and \mbox{7-m} array data using an empirical noise-based weighting and deconvolve them according to a ``Briggs'' scheme with robustness parameter $r = 0.5$, as well as a multi-scale clean at 1, 2, 5, 10 times the \mbox{12-m} beam with a threshold at 3~times the root mean square (\mbox{r.m.s.}) noise level. The total power observations are calibrated and mapped using a modified version of the ALMA reduction pipeline. In particular, we perform the temperature calibration, sky subtraction and baseline fitting on an antenna basis, but adopt a single antenna efficiency (Kelvin to Jansky conversion) per observing session.
We combine the interferometric and total power data cubes in Fourier space making our observations sensitive to all spatial scales. The resulting beam size is $2.5'' \times 1.8'' \approx 20$~pc, the \mbox{r.m.s.} noise level is $6.7$ mJy per beam and per 2~km~s$^{-1}$ channel. 

We calculate molecular (H$_2$) gas masses from CO(1-0) data, adopting a metallicity dependent CO-to-H$_2$ conversion factor of\cite{bolatto13b} $\alpha_{\rm CO} = 4.35\,(Z/{\rm Z}_\odot)^{-1}$ ${\rm M}_\odot$ ${({\rm K~km~s}^{-1}~{\rm pc}^2)}^{-1}$, where $Z/{\rm Z}_\odot=10^{[{\rm M}/{\rm H}]}$ is the metallicity relative to solar. This conversion factor is about twice the Galactic value given the half-solar metallicity of NGC300 (see {\bf Structural parameters}), and matches the estimate in the similar mass and metallicity galaxies LMC\cite{wong11} and M33\cite{druard14}. It also includes a factor $1.36$ to account for heavy elements. This implies a $5\sigma$ sensitivity in molecular gas surface density of $13~{\rm M}_\odot~{\rm pc}^{-2}$ or point source sensitivity of $5600~{\rm M}_\odot$ when integrated over $10~{\rm km~s}^{-1}$. Throughout the analysis, we adopt an uncertainty on $\alpha_{\rm CO}$ of 50\%. We reiterate that the inferred values of $t_{\rm CO}$, $t_{\rm fb}$, $\lambda$, and $v_{\rm fb}$ are insensitive to the choice of $\alpha_{\rm CO}$ and the absolute flux scale.

\vspace{1.5mm} \noindent \emph{MPG/ESO \mbox{2.2-m}.}
The optical observations were obtained with the Wide Field Imager (WFI) on the MPG/ESO \mbox{2.2-m} telescope at La Silla observatory. These consist of narrowband observations of the H$\alpha$ line (with a total exposure time of $50$~min) and the nearby continuum. The raw data are available in the ESO data archive; here we use reduced data kindly provided by C.~Faesi. In summary, their data reduction\cite{faesi14} includes instrumental calibration, a combination of exposures, continuum subtraction, and Galactic extinction correction (adopting $A_R = 0.027$~mag). The flux calibration has been performed on the standard star \mbox{LTT17379} and verified with literature fluxes for several H{\sc ii} regions. The point spread function (PSF) is determined using several bright stars to be $1.35'' \approx 13$~pc. This implies an uncertainty in the absolute calibration of $4\%$.

In addition, we correct the H$\alpha$ map for the following effects. The WFI H$\alpha$ filter includes two nearby {\sc [Nii]} lines. We use optical spectroscopic observations of NGC300 to quantify the contribution from {\sc [Nii]} to the H$\alpha$ filter. The 35-pointing integral field data of NGC 300 (PI A.~McLeod, ESO program 098.B-0193(A)) were taken with the MUSE instrument mounted on the Very Large Telescope and consist of a $7' \times 5' \approx 4.1~\textrm{kpc} \times 2.9~\textrm{kpc}$ mosaic. Each pointing was observed three times using a 90$^{\circ}$ rotation dither pattern and an exposure time of 900~s. The data were reduced using the MUSE pipeline under the \textsc{Esorex} environment, using the available calibration files from the ESO archive for the relevant night. We find that {\sc [Nii]} is closely associated with H{\sc ii} regions and that their combined flux is typically $20\%$ of the H$\alpha$ flux (consistent with Ref.~\citenum{roussel05}). We remove the contribution of {\sc [Nii]} from our H$\alpha$ map scaling it by $5/6$.

We further correct for internal extinction of H$\alpha$ by $A_{\rm H\alpha}=0.5$~mag as derived from the Balmer decrement of H{\sc ii} regions\cite{roussel05}. The \mbox{r.m.s.} noise level of the corrected H$\alpha$ map is $0.74 \times 10^{-16}$ erg~s$^{-1}$ cm$^{-2}$ arcsec$^{-2}$ at a PSF size of $1.35''$. We verified the extinction correction with our MUSE data, which provides sufficient signal-to-noise in the H$\beta$ line to derive\cite{luridiana15} $A_{\rm H\alpha}$ for 46 H{\sc ii} regions, resulting in $A_{\rm H\alpha}=0.37\pm0.24$~mag (median $\pm$ standard deviation). This amount of extinction is consistent with our correction of 0.5~mag, as well as with the extinction expected at the mean surface densities of the GMCs in NGC300 ($\sim20~{\rm M}_\odot~{\rm pc}^{-2}$, corresponding to $A_{\rm H\alpha}\sim0.5$~mag at the half-solar metallicity of NGC300). It might underestimate the attenuation towards the GMC density peaks, where the surface density is $>20~{\rm M}_\odot~{\rm pc}^{-2}$. The fact that we measure an extinction of only $0.5$~mag could either be due to the well-known bias that the Balmer decrement can only be evaluated for the least extincted H{\sc ii} regions, or because young stellar populations are spatially offset from the extinction peaks of GMCs\cite{vutisalchavakul14}. In either case, the maximum H$\alpha$ extinction we measure is $\sim1$~mag.

We test whether the H$\alpha$ extinction correction obtained from the Balmer decrement and the mean GMC surface density (which are both determined at $10{-}20$~pc scales) are biased towards low values due to missing heavily attenuated H{\sc ii} regions near the GMC density peaks. As discussed in the main text, we compare the H$\alpha$ map to the \emph{Spitzer} MIPS $24~\mu$m map to identify any embedded star formation missed by the H$\alpha$ map, showing that most of the $24~\mu$m emission comes from bright H{\sc ii} regions. Only $4{-}7$ regions are bright at $24~\mu$m without associated H$\alpha$ emission, which would imply only a $\sim3\%$ increase of the total number of massive star-forming regions. Finally, the H$\alpha$ and $24~\mu$m emission are approximately linearly related. In summary, the above means that the extinction-corrected H$\alpha$ map recovers most of the SFR, and embedded star formation or extinction only affect our conclusions by less than the quoted uncertainties.

We have also verified that the H$\alpha$ emission is not significantly affected by the underlying stellar absorption. If the stellar initial mass function (IMF) is well sampled, the H$\alpha$ absorption and emission both increase linearly with the mass of a stellar population, implying that the ratio between both is controlled entirely by the stellar age. Using the Starburst99 stellar population model\cite{leitherer99}, we derive in Appendix~A of Ref.~\citenum{haydon18} that the effects of absorption become important for ages of $\sim10~{\rm Myr}$, which is at least double the timescale associated with H$\alpha$ emission (see {\bf Analysis method and input parameters}). To complement this, we looked for any significant H$\alpha$ absorption features in our MUSE data of NGC300. Irrespective of the position (H{\sc ii} region or background field), we detect no signs of H$\alpha$ absorption.

We estimate SFRs from the corrected H$\alpha$ map adopting the calibration\cite{murphy11}: $\log \textrm{SFR} [M_\odot~\mathrm{yr}^{-1}] = (\log L_\mathrm{H\alpha} - \log 41.27) [\mathrm{erg~s}^{-1}]$. This calibration assumes a fully sampled Kroupa\cite{kroupa01} IMF, Starburst99 stellar population model\cite{leitherer99}, continuous star formation over an H$\alpha$ emission timescale (about $5$~Myr; see {\bf Analysis method and input parameters}), and solar abundance. This way, we measure a total SFR of $0.092~{\rm M}_\odot~{\rm yr}^{-1}$ in NGC300, consistent with previous, multi-wavelength measurements\cite{kang16}. We adopt an uncertainty on the H$\alpha$-to-SFR conversion factor of 20\%. All of the above scaling factors do not affect the inferred values of $t_{\rm CO}$, $t_{\rm fb}$, $\lambda$, and $v_{\rm fb}$, because these are derived from the {\it relative} change of the CO-to-H$\alpha$ flux ratio as a function of spatial scale.

Our analysis further requires sensitivity to individual, young stellar populations. At the scale of individual star-forming regions ($\lambda = 104$~pc; \autoref{tab:fit}), the H$\alpha$ map has a $5\sigma$ luminosity sensitivity of $2 \times 10^{36}$ erg~s$^{-1}$. 
For a stellar population formed in an instantaneous burst $5~{\rm Myr}$ ago, which roughly matches the H{\sc ii} region lifetime $t_{\rm H\alpha}$ (see {\bf Analysis method and input parameters}), this luminosity sensitivity corresponds to the ionising radiation luminosity\cite{haydon18} of a $200~{\rm M}_\odot$ stellar population. This sensitivity is well-matched to the stellar populations formed by the smallest detected GMCs (see {\bf Consistency tests}).

\vspace{1.5mm} \noindent {\bf Structural parameters.}
We adopt the global parameters and radial profiles for NGC300 as follows. All of these quantities are used at various points of the presented analysis.

The distance of NGC300 is taken to be $D = 2.0$~Mpc, derived from Hubble Space Telescope observations of the tip of the red giant branch\cite{dalcanton09}. The center of the optical disc is at right ascension $\alpha = \textrm{00h54m53.48s}$ and declination $\delta = \textrm{-37d41m03.8s}$ (NED). This is coincident to within 1~arcsec to the centre of a disc model fit to an optical $V$-band image\cite{westmeier11} from the MPG/ESO \mbox{2.2-m} telescope. NGC300's disc has an inclination of $i = 42 \pm 0.5$ degrees and a position angle of $P.A. = 111 \pm 0.5$ degrees, derived from a 21~cm atomic hydrogen (H{\sc i}) emission map obtained with the Australia Telescope Compact Array (ATCA) telescope\cite{westmeier11}.

The H{\sc i} data have a beam size of $180'' \times 88'' \approx 1220$~pc and channel width of $8$~km~s$^{-1}$. The \mbox{r.m.s.} noise level is $3.5$~mJy per beam, corresponding to a $5\sigma$ H{\sc i} column density sensitivity of $1.0 \times 10^{19}$~cm$^{-2}$ per spectral channel or a $5\sigma$ atomic gas surface density sensitivity of $0.20$~M$_\odot$~pc$^{-2}$ over $50~{\rm km}~{\rm s}^{-1}$ (including heavy elements and deprojected to be face-on). The rotation curve is determined by fitting a tilted ring model to the velocity field of the ATCA H{\sc i} data\cite{westmeier11}. Rings are separated by $100'' \approx 970$~pc. For each ring, the galaxy centre and systemic velocity were kept fixed, while rotation velocity, position angle, and inclination were left as free parameters.

Gas surface densities are derived for atomic gas from the ATCA H{\sc i} data\cite{westmeier11} and for molecular gas from the ALMA+ACA observations. These include a factor $1.36$ to account for heavy elements and are deprojected using the adopted disc inclination. The gas velocity dispersion is assumed to be $6~{\rm km}~{\rm s}^{-1}$ as determined from H{\sc i} observations with the Karl~G.\ Jansky Very Large Array (VLA) at 80~pc scales in M33\cite{koch18}, a galaxy of comparable stellar mass, gas mass, and morphology as NGC300. This is motivated by the fact that the interstellar medium of NGC300 is dominated by atomic gas (see \ref{fig:galaxy}).

Stellar surface densities are derived from a \emph{Spitzer} IRAC $3.6~\mu$m image\cite{dale09} from which emission from hot dust and polycyclic aromatic hydrocarbons has been subtracted\cite{querejeta15}. We do not correct for contamination by post-main sequence stars, because these contribute at the $<5\%$ level when averaging over $>{\rm kpc}$ areas as is done here\cite{meidt12}. We adopt a mass-to-light ratio\cite{leroy08} of 1~MJy~sr$^{-1} = 280~{\rm M}_\odot~{\rm pc}^{-2} \times \textrm{cos}(i)$, equivalent to $\Upsilon_\star^{3.6} = 0.4~{\rm M}_\odot / {\rm L}_\odot$. This number is only used to calculate the gas disc scale height (see {\bf Theoretical predictions}). The literature contains values between\cite{leroy08,meidt14,querejeta15} $\Upsilon_\star^{3.6} = 0.4{-}0.6~{\rm M}_\odot / {\rm L}_\odot$, with NGC300 likely residing in the bottom half of the range due to its low mass. The full range of mass-to-light ratios translates to a $<10\%$ uncertainty on the inferred scale height. The stellar velocity dispersion is calculated by assuming vertical hydrostatic equilibrium, using the observed stellar surface density and assuming a stellar disc scale height $h_{\rm star}$. This scale height is inferred\cite{faesi18} using the empirical relation between scale height and maximum circular velocity for spiral galaxies\cite{vanderkruit11}. For the maximum circular velocity in NGC300 from the ATCA H{\sc i} data\cite{westmeier11} of $v_{\rm c,max}=98.8~{\rm km}~{\rm s}^{-1}$, we obtain $h_\star=305~{\rm pc}$, which we assume is constant with galactic radius\cite{vanderkruit11}.

We adopt a metallicity of $[{\rm M}/{\rm H}] = -0.23 \pm 0.06$~dex at the centre of NGC300 and a metallicity gradient of $-0.058 \pm 0.014$~dex~kpc$^{-1}$ as derived from colour-magnitude distribution (CMD) fitting of stellar populations of age $4{-}100$~Myr\cite{gogarten10}. Finally, we adopt an electron temperature (i.e.\ the temperature of ionised gas in H{\sc ii} regions) of $T_\mathrm{e}~[\mathrm{K}] = 6920 + 839 R~[\mathrm{kpc}]$ varying with galactic radius\cite{bresolin09}.

Many of the above quantities describing the structure of NGC300 are visualised as a function of galactic radius in~\ref{fig:galaxy}.

In addition, we compare the integrated properties of NGC300 (i.e.\ its SFR, stellar mass, molecular gas mass, and atomic gas mass) to those of nearby galaxies (with redshifts $0.01<z<0.05$) from the xCOLDGASS\cite{saintonge17} and xGASS\cite{catinella18} surveys in \ref{fig:galpop}. For NGC300, the stellar mass and atomic gas mass are taken from Ref.~\citenum{westmeier11}, whereas the SFR and molecular gas mass are obtained by integration over our observations out to $R=5$~kpc and $R=3$~kpc, respectively. The comparison of \ref{fig:galpop} shows that NGC300 is a `normal' galaxy for its mass. Its specific SFR (i.e.\ the SFR per unit stellar mass) resides within a factor of two of the star-forming galaxy main sequence, and its molecular and atomic gas depletion times are consistent with the $1\sigma$ scatter of the xCOLDGASS and xGASS samples. In an absolute sense, NGC300 is a low-mass galaxy relative to the bulk of the actively star-forming galaxy population, as most stars in the local Universe are born in Milky Way-mass galaxies\cite{moster13} (where the Milky Way has a stellar mass of\cite{blandhawthorn16} $5\times10^{10}~{\rm M}_\odot$). It is plausible that the molecular cloud lifecycle inferred in this work varies across the galaxy population (see {\bf Relation to other work}), underlining that our analysis should be extended to a large and representative galaxy sample.

\vspace{1.5mm} \noindent {\bf Cloud-scale quantities.}
Molecular clouds are identified and their macroscopic properties determined from the ALMA+ACA data using an updated version of the {\sc Cprops} code\cite{rosolowsky06,leroy15}. This is done as follows.

Clouds are identified as local emission peaks with signal-to-noise $(S/N) \geq 5$ in two adjacent channels with a volume defined by adjacent pixels at $S/N \geq 2$. Cloud spatial and spectral (velocity) dispersions are derived from the intensity-weighted second moments and have been corrected\cite{rosolowsky06,leroy15} to include emission at $S/N < 2$. Cloud radii are defined as $r_{\rm GMC} = 1.91 r_{\rm stdev}$, where $r_{\rm stdev}$ is the standard deviation of a two-dimensional Gaussian\cite{solomon87}. Cloud masses are derived from the CO(1-0) luminosity, corrected\cite{rosolowsky06,leroy15} to include emission at $S/N < 2$, as $M = \alpha_{\rm CO} L_{\rm CO}$ for the adopted metallicity dependent conversion factor $\alpha_{\rm CO} = 4.35$ $(Z/{\rm Z}_\odot)^{-1}$ ${\rm M}_\odot$ ${({\rm K~km~s}^{-1}~{\rm pc}^2)}^{-1}$.

Mean surface densities are defined as $\Sigma_{\rm GMC} = 0.77 M/\pi r_{\rm GMC}^2$, within the FWHM of a two-dimensional Gaussian distribution, and mean volume densities as $\rho_{\rm GMC} = 1.26 M  / (4/3 \pi r_{\rm GMC}^3)$, within the FWHM of a three-dimensional Gaussian distribution. The virial parameters of the clouds are defined as $\alpha_{\rm vir}=5\sigma_{\rm GMC}^2r_{\rm GMC}/GM_{\rm GMC}$. The cloud free-fall collapse time is $\tau_{\rm ff} {\rm~[Myr]} = \sqrt{3 \pi / 32 G \rho_{\rm GMC}} = 8.09 / \sqrt{\rho_{\rm GMC} {\rm~[M_\odot~pc^{-3}]}}$, and the turbulent crossing time is $\tau_\mathrm{cross} = 2r_{\rm GMC} / \sigma_\mathrm{GMC}$ for a one-dimensional velocity dispersion $\sigma_\mathrm{GMC}$.

For many of the above quantities describing the properties of GMCs in NGC300, we show the population average as a function of galactic radius in~\ref{fig:gmcs}.

\vspace{1.5mm} \noindent {\bf Analysis method and input parameters.}
We apply the method from Ref.~\citenum{kruijssen18}, where it is described and validated in detail. Here, we briefly summarise the main steps.

\noindent \emph{Observables.}
We use the two maps of CO(1-0) and H$\alpha$ emission in NGC300 described in {\bf Observational data}. In each of the maps, we apply {\sc Clumpfind}\cite{williams94} to identify emission peaks, on which apertures of a given size are placed to measure the CO-to-H$\alpha$ ratio on that size scale. The aperture integration is achieved by convolving the maps with a top-hat kernel representing the aperture. The CO and H$\alpha$ fluxes at the positions of the identified peaks are then added and divided to calculate the population-averaged CO-to-H$\alpha$ ratio. The above procedure is illustrated in \ref{fig:maps} and Supplementary Video~1. The relative change of the CO-to-H$\alpha$ ratio ${\cal B}$ as a function of aperture size ($l_{\rm ap}$) is thus defined as
\begin{equation}
\label{eq:bias}
{\cal B}(l_{\rm ap})=\frac{{\cal F}_{\rm H\alpha,tot}}{{\cal F}_{\rm CO,tot}}
\frac{\sum_i{\cal F}_{{\rm CO},i}(l_{\rm ap})}{\sum_i{\cal F}_{{\rm H\alpha},i}(l_{\rm ap})} ,
\end{equation}
where ${\cal F}_{\rm H\alpha,tot}$ and ${\cal F}_{\rm CO,tot}$ represent the total H$\alpha$ and CO(1-0) fluxes in the maps, and ${\cal F}_{{\rm H\alpha},i}(l_{\rm ap})$ and ${\cal F}_{{\rm CO},i}(l_{\rm ap})$ represent the flux within an aperture of diameter $l_{\rm ap}$ around an emission peak $i$. The summations in this equation iterate over the CO peaks for calculating the top branch in \autoref{fig:tuningfork} and over the H$\alpha$ peaks for calculating the bottom branch. In both cases, overlapping apertures are avoided through Monte-Carlo sampling from the emission peak sample and taking the average over $10^3$ such samples before calculating the ratio in the above equation. See Section~3.2.9 of Ref.\ \citenum{kruijssen18} for further details.

The uncertainties on the measured CO-to-H$\alpha$ ratios are calculated as described in detail in Section~3.2.10 of Ref.\ \citenum{kruijssen18}. In summary, these uncertainties account for the noise in the maps and for the fact that the luminosity functions of the molecular clouds and H{\sc ii} regions are not delta functions, but have a finite width. They also account for the covariance between CO and H$\alpha$ emission and are shown as thin error bars in \autoref{fig:tuningfork} (right). Finally, the shaded error bars in \autoref{fig:tuningfork} (right) additionally account for the covariance between the flux in apertures of different sizes that are placed on the same emission peaks.

\noindent \emph{Statistical model.}
To obtain the evolutionary timeline of the molecular clouds and H{\sc ii} regions, we fit a model to the observed CO-to-H$\alpha$ ratios as a function of the aperture size. The model predicts these ratios as a function of the underlying timescales (see Supplementary Video~2) and is given by the equations
\begin{equation}
\label{eq:biasstar}
{\cal B}_{\rm H\alpha}(l_{\rm ap})=\frac{f_{\rm CO}\left[1+\beta_{\rm CO}^{-1}\left(\frac{t_{\rm CO}}{t_{\rm fb}}-1\right)\right]^{-1}+\frac{t_{\rm H\alpha}}{\mbox{$\tau$}}\left(\frac{l_{\rm ap}}{\mbox{$\lambda$}}\right)^2}
       {f_{\rm H\alpha}+\frac{t_{\rm H\alpha}}{\mbox{$\tau$}}\left(\frac{l_{\rm ap}}{\mbox{$\lambda$}}\right)^2} ,
\end{equation}
for apertures placed on H$\alpha$ emission peaks and
\begin{equation}
\label{eq:biasgas}
{\cal B}_{\rm CO}(l_{\rm ap})=\frac{f_{\rm CO}+\frac{t_{\rm CO}}{\mbox{$\tau$}}\left(\frac{l_{\rm ap}}{\mbox{$\lambda$}}\right)^2}
       {f_{\rm H\alpha}\left[1+\beta_{\rm H\alpha}^{-1}\left(\frac{t_{\rm H\alpha}}{t_{\rm fb}}-1\right)\right]^{-1}+\frac{t_{\rm CO}}{\mbox{$\tau$}}\left(\frac{l_{\rm ap}}{\mbox{$\lambda$}}\right)^2} ,
\end{equation}
for those placed on CO emission peaks. In these equations, $t_{\rm CO}$ represents the lifetime of CO-bright regions, $t_{\rm H\alpha}$ is the lifetime of H$\alpha$-bright regions, $t_{\rm fb}$ is the duration for which these coexist, $\tau=t_{\rm CO}+t_{\rm H\alpha}-t_{\rm fb}$ is the total duration of the evolutionary timeline, $\lambda$ is the characteristic separation length between regions, $f_{\rm H\alpha}$ and $f_{\rm CO}$ represent the fraction of the central peak flux contained within an aperture of size $l_{\rm ap}$, and $\beta_{\rm H\alpha}$ and $\beta_{\rm CO}$ are the mean flux ratio between regions coexisting with the other tracer relative to those that do not. The free parameters of these expressions are $t_{\rm CO}$, $t_{\rm fb}$, and $\lambda$. The quantities $f_{\rm H\alpha}$, $f_{\rm CO}$, $\beta_{\rm H\alpha}$, and $\beta_{\rm CO}$ are functions of these three parameters (see Sections 3.2.9 and 3.2.11 of Ref.\ \citenum{kruijssen18}) and the aperture size $l_{\rm ap}$ is specified. Finally, the lifetime of H$\alpha$-bright regions is $t_{\rm H\alpha}=t_{\rm H\alpha,ref}+t_{\rm fb}$, where $t_{\rm H\alpha,ref}$ represents the known `reference timescale' that is used to turn the relative timescales constrained by fitting the above two equations to the observed CO-to-H$\alpha$ ratios into absolute timescales. This timescale has been calibrated using synthetic emission maps of H$\alpha$ emission from simulated galaxies\cite{haydon18}, accounting for its dependence on metallicity and the SFR surface density (reflecting how well the stellar IMF has been sampled\cite{dasilva12,krumholz15b}). The relation between $t_{\rm H\alpha,ref}$ and metallicity $Z$ is\cite{haydon18} 
\begin{equation}
\label{eq:tstar}
t_{\rm H\alpha,ref}=(4.32\pm0.16~{\rm Myr})\times(Z/{\rm Z}_\odot)^{-0.086\pm0.017} ,
\end{equation}
which for the observed metallicity gradient\cite{gogarten10} in NGC300 results in values between $t_{\rm H\alpha,ref}=4.53\pm0.14$ and $4.69\pm0.13~{\rm Myr}$ for galactic radii $R=0{-}3~{\rm kpc}$.

The best-fitting model is obtained by carrying out a reduced-$\chi^2$ fit to the observations, which returns a three-dimensional probability distribution function (PDF) of the free parameters. Marginalisation yields the one-dimensional PDFs, which are shown in the top row of \ref{fig:pdfs}. The uncertainties quoted in the main body of this paper refer to the 16th and 84th percentiles of the PDFs.

\noindent \emph{Influence of evolutionary timeline.}
\autoref{fig:tuningfork} shows the best-fitting model together with the observed CO-to-H$\alpha$ flux ratio as a function of aperture size for the entire field-of-view of NGC300. The obtained evolutionary timescales are (see \autoref{tab:fit} and \ref{fig:pdfs}) $t_{\rm CO}=10.8^{+2.1}_{-1.7}~{\rm Myr}$ and $t_{\rm fb}=1.5^{+0.2}_{-0.2}~{\rm Myr}$. This relatively short GMC lifetime contrasts strongly with earlier reports of lifetimes up to $50{-}100~{\rm Myr}$ in the spiral galaxy M51\cite{koda09}. We therefore repeat the right-hand panel of \autoref{fig:tuningfork} in \ref{fig:tuningfork_lines} and demonstrate how a long GMC lifetime of $t_{\rm CO}=50~{\rm Myr}$ would change the model fit. To this end, two additional models are included. In the first, the GMC lifetime is extended, but the feedback timescale during which GMCs coexist with H{\sc ii} regions is kept fixed ($t_{\rm CO}=50~{\rm Myr}$ and $t_{\rm fb}=1.5~{\rm Myr}$). In the second, the GMC lifetime is extended by increasing the time for which GMCs and H{\sc ii} regions coexist ($t_{\rm CO}=50~{\rm Myr}$ and $t_{\rm fb}=40.7~{\rm Myr}$). These alternative models represent two extremes and show that long GMC lifetimes are fundamentally incompatible with the observed CO-to-H$\alpha$ ratios in NGC300. In the first alternative model, the dissimilarity between $t_{\rm CO}$ and $t_{\rm H\alpha}$ would require a strong vertical asymmetry of the model curves, which is not observed. In the second alternative model, the long timescale over which CO and H$\alpha$ coexist would imply a negligible deviation from the galactic average CO-to-H$\alpha$ flux ratio at all size scales. This is not observed either. \ref{fig:tuningfork_lines} thus illustrates how the model fit in  \autoref{fig:tuningfork} constrains the GMC lifecycle (also see Supplementary Video~2). The difference in GMC lifetimes between NGC300 and M51 may either be physical in nature or result from differences in experiment design (see {\bf Relation to other work}).

\noindent \emph{Derived quantities.}
From the free parameters $t_{\rm CO}$, $t_{\rm fb}$, and $\lambda$, we derive three additional physical quantities, of which the PDFs are obtained through Monte-Carlo error propagation (see the bottom row of \ref{fig:pdfs}). These derived quantities are the total SFE, i.e.\ the fraction of the molecular cloud mass converted into stars
\begin{equation}
\label{eq:esf}
\epsilon_{\rm sf}=\frac{t_{\rm CO}}{t_{\rm dep}},
\end{equation}
the mass loading factor, i.e.\ the time-averaged mass outflow rate in units of the SFR
\begin{equation}
\label{eq:etafb}
\eta_{\rm fb}=\frac{1-\epsilon_{\rm sf}}{\epsilon_{\rm sf}},
\end{equation}
and the feedback outflow velocity
\begin{equation}
\label{eq:vfb}
v_{\rm fb}=\frac{r_{\rm CO}}{t_{\rm fb}}.
\end{equation}
In these equations, $t_{\rm dep}$ is the gas depletion time, defined as the ratio between the total molecular gas mass and the total SFR, and $r_{\rm CO}$ is the mean radius of the CO emission peaks, which is defined as the dispersion of a two-dimensional Gaussian emission peak\cite{kruijssen18}. We calculate the feedback velocity using the CO peak radius, which is motivated by the fact that a displacement of residual gas by this amount (either kinematically or by a phase change) is sufficient for it to be decoupled from the young stellar population in the context of our statistical method\cite{hygate18}. Due to their dependence on $t_{\rm dep}$, the quantities $\epsilon_{\rm sf}$ and $\eta_{\rm fb}$ are sensitive to absolute scaling factors of the gas mass and SFR, such as the CO-to-H$_2$ and H$\alpha$-to-SFR conversion factors.

\vspace{1.5mm} \noindent \emph{Diffuse emission filtering.}
Finally, we remove diffuse emission from the input maps on size scales $>10\lambda$ by using a high-pass Gaussian filter in Fourier space\cite{hygate18}. This is necessary, because the adopted method assumes that all emission comes from independent regions taking part in an evolutionary lifecycle. Diffuse emission from size scales much larger than the region separation length violates that assumption. After removing the diffuse emission, we also apply a noise mask with a threshold at twice the \mbox{r.m.s.} noise level. The entire analysis is then repeated on the filtered maps until convergence is reached. This typically happens after a single iteration, but four are carried out to guarantee convergence. At the end of the analysis, the CO map is unaffected and the H$\alpha$ map has had 33\% of the emission removed. The filtered map is then scaled up to add the diffuse emission to the concentrated emission regions. This assumes that the diffuse H$\alpha$ originally emerged from H{\sc ii} regions and that the amount of flux escaping an H{\sc ii} region is proportional to its observed flux. If the first of these assumptions is inaccurate, the presented values of $\epsilon_{\rm sf}$ and $\eta_{\rm fb}$ should be corrected by factors of $0.67$ and $1.5$, respectively. As before, the inferred values of $t_{\rm CO}$, $t_{\rm fb}$, $\lambda$, and $v_{\rm fb}$ are unaffected by this constant scaling. Finally, we emphasise that the conclusions of this work are not sensitive to the filtering of the diffuse H$\alpha$ emission, as its quantitative effect on the inferred quantities ranges from $7{-}37\%$. This is considerably smaller than the change needed to affect the physical interpretation of the measurements (see \autoref{fig:profiles}).

\vspace{1.5mm} \noindent \emph{Code.}
The above analysis is automated in the {\sc Heisenberg} code and is performed using the default input parameters listed in Tables~1 and~2 of Ref.~\citenum{kruijssen18}. The only exceptions are the following quantities. We adopt a distance, inclination angle, position angle, and coordinates of the galactic centre as listed in {\bf Structural Parameters} above. The aperture sizes range from a spatial resolution $l_{\rm ap}=20$ to $2560~{\rm pc}$ in steps of a factor of $\sqrt{2}$. The emission peaks on which the apertures are placed are identified over a flux range of $2.0~{\rm dex}$ below the brightest peak in each map, using flux contours at intervals of $0.5~{\rm dex}$ to separate adjacent peaks. Finally, across the range of galactic radii  $R=0{-}3~{\rm kpc}$, we perform our measurements in radial bins of width $\Delta R=1$ and $1.5~{\rm kpc}$, centred on multiples of $R=0.5$ and $0.75~{\rm kpc}$, respectively, as well as over the entire radial range (see \autoref{fig:profiles}).

All results of the above analysis are shown in \autoref{fig:profiles}. Additionally, we provide a comprehensive summary of the results in \autoref{tab:fit}.

\vspace{1.5mm} \noindent {\bf Assumptions and criteria for reliable applications.}
The analysis method has been tested systematically through thousands of applications on simulated or synthetic galaxy maps\cite{kruijssen18,haydon18,hygate18}. This has resulted in a set of quantitative requirements for reliable applications, all of which are satisfied by the observations of NGC300 used in this work.

The pair of maps used should contain tracers that are connected by a Lagrangian evolutionary process\cite{kruijssen18}, i.e.\ an emission peak in one of the maps should eventually evolve into an emission peak in the other. It is well established that star formation (traced by H$\alpha$) is strongly correlated with molecular gas (traced by CO) on galactic scales\cite{bigiel08,schruba11,leroy13}. This is widely interpreted such that molecular clouds form stars\cite{kennicutt12}. The only exception could be presented by CO or H$\alpha$ emission residing in a diffuse form\cite{pety13}. However, the molecular gas map contains negligible diffuse emission, and the little (33\%) diffuse emission present in the H$\alpha$ map is removed by our filtering method (see above). As a result, the assumption of an evolutionary connection between CO clouds and H{\sc ii} regions is reasonable.

Any systematic astrometric offsets $\Delta$ between the maps should be less than one third of the resolution or emission peak size\cite{hygate18}, i.e.\ $\Delta\leq2\sqrt{2\ln{2}}/3\times\min{\{r_{\rm H\alpha},r_{\rm CO}\}}\approx 1''$, where $r_{\rm H\alpha}$ and $r_{\rm CO}$ represent the standard deviations of Gaussian peaks of H$\alpha$ and CO emission, respectively. Ensuring a shared astrometry between H$\alpha$ and CO maps is challenging, because these do not have any point sources in common. However, we have visually checked for systematic offsets between compact H{\sc ii} regions and CO clumps on the pixel scale ($0.4''$). While small (${\sim}1''$) offsets exist in most of these cases, these are likely physical in nature, because they do not share a common direction. In addition, the H$\alpha$ astrometry has been verified against the USNO-B1.0 stellar astrometric catalogue. Again, no offsets in excess of the pixel size were found.

The sensitivity of the maps should be sufficient to obtain a near-complete census of the (luminosity-weighted) GMC and H{\sc ii} region population. As discussed in {\bf Observational data}, this is comfortably achieved by the adopted observations, which detect all GMCs down to $5600~{\rm M}_\odot$ and individual H{\sc ii} regions down to stellar masses of $200~{\rm M}_\odot$.

In order to translate the inferred relative timescales to an absolute evolutionary timeline, one of the timescales must be known, so that it can act as a reference timescale. As discussed in {\bf Analysis method and input parameters}, this is satisfied by the H$\alpha$ timescale\cite{haydon18}, which we fix using its dependence on metallicity given in equation~\ref{eq:tstar}.

The method returns reliable measurements free of major biases or large uncertainties as long as a sufficient number ($\geq35$) of emission peaks is identified in both maps\cite{kruijssen18}. We adopt a conservative limit to minimise the statistical uncertainties and omit any radial bins for which $N_{\rm peak,H\alpha}<50$ or $N_{\rm peak,CO}<50$. This occurs only for the galactic radius interval $R=\{0,1\}~{\rm kpc}$, after which we are left with $N_{\rm peak,H\alpha}\geq62$ and $N_{\rm peak,CO}\geq65$ for all fields-of-view listed in Table~\ref{tab:fit}.

The CO-to-H$\alpha$ flux ratio diagram (\autoref{fig:tuningfork}, right) should reach sufficiently large aperture sizes to converge to the galactic average. Otherwise, the obtained timescales may be biased\cite{kruijssen18}. To within $1\sigma$, the CO-to-H$\alpha$ flux ratio in units of the galactic average is consistent with unity at the maximum aperture size ($l_{\rm ap,max}=2560~{\rm pc}$) for all measurements performed in this work.

The most reliable measurements are obtained when the emission lifetimes of gas and young stars are similar\cite{kruijssen18}. Quantitatively, we require $|\log_{10}{(t_{\rm H\alpha}/t_{\rm CO})}|\leq1$. This is comfortably achieved across all of our measurements, which fall in the range $|\log_{10}{(t_{\rm H\alpha}/t_{\rm CO})}|=0.18{-}0.47$, implying that $t_{\rm H\alpha}$ and $t_{\rm CO}$ always fall within a factor of 3 of each other.

When the feedback timescale (i.e.\ the time for which emission from CO and H$\alpha$ coexists) is close to zero or to the total duration of the evolutionary timeline, the method can only provide upper or lower limits, respectively\cite{kruijssen18}. Quantitatively, we require $0.05<t_{\rm fb}/\tau<0.95$ for obtaining an absolute measurement. Across all measurements at native ($20~{\rm pc}$) resolution, we have $\min{(t_{\rm fb}/\tau)}=0.07$ and $\max{(t_{\rm fb}/\tau)}=0.11$. When including the degraded resolution experiments of \autoref{fig:res}, instead we have $\max{(t_{\rm fb}/\tau)}=0.30$. All of these deviate from zero or unity by a sufficient amount to provide a reliable measurement of the feedback timescale.

As demonstrated in \autoref{fig:res} and previously found for simulated galaxy maps\cite{kruijssen18}, the spatial resolution of the maps needs to be high enough to resolve the region separation length by a factor of $N_{\rm res}=1{-}2$, such that $\lambda\geq N_{\rm res}l_{\rm ap,min}/\cos{i}$. This is achieved across all measurements, as the smallest obtained separation length is $\min{(\lambda)}=96_{-20}^{+26}~{\rm pc}$ and $N_{\rm res}l_{\rm ap,min}/\cos{i}\approx54~{\rm pc}$ for $N_{\rm res}=2$, $l_{\rm ap,min}=20~{\rm pc}$, and $i=42^\circ$.

To avoid any spatial blending between emission peaks or the presence of considerable diffuse emission, the average region filling factors $\zeta\equiv 2r/\lambda$ of the CO and H$\alpha$ peaks should satisfy $\max{(\zeta_{\rm H\alpha}, \zeta_{\rm CO})}<0.5$. At the native ($20~{\rm pc}$) resolution of the maps, this requirement is satisfied with $\max{(\zeta_{\rm H\alpha})}=0.31_{-0.05}^{+0.07}$ and $\max{(\zeta_{\rm CO})}=0.30_{-0.04}^{+0.05}$ across both the entire field-of-view and all radial bins. Also visually, no significant blending of emission peaks is noticeable at the native resolution.

\vspace{1.5mm}  \noindent {\bf Consistency tests.}
We now discuss two choices made during the analysis and how these might influence the results. We also verify that the sensitivity limits and completeness are consistent between both maps.

As explained in {\bf Analysis method and input parameters}, we use the publicly available code {\sc Clumpfind}\cite{williams94} to identify emission peaks. In short, \clumpfind loops over a predefined set of flux levels, starting at the highest level, and identifies closed contours at each level. The pixel with the highest flux value within the closed contour is taken to represent the peak position. If this position has not previously been identified (i.e.\ it is absent at higher flux levels), it is saved as a new emission peak, provided that it is detected at $\geq5\sigma$. The choice of flux levels to use in this operation is important, and we carried out a number of tests to verify that the presented results are insensitive to this choice.

The flux levels are defined in logarithmic space by a total range ($2.0$~dex) and an interval between the levels ($0.5$~dex). Firstly, we carried out a visual inspection and found that this choice of parameters successfully identifies all obvious emission peaks (see \ref{fig:maps}). Secondly, we varied the parameters defining the flux levels, in the range $1{-}2~\dex$ for the total range and $0.25{-}1~\dex$ for the interval, to verify how this affects the quantities measured during the analysis. With the exception of cases for which a visual inspection clearly shows that obvious emission peaks are missed, these experiments give results that are consistent with the numbers in Table~\ref{tab:fit} to within the uncertainties, or ${\sim}30\%$, whichever is largest (also see Ref.~\citenum{kruijssen18}). The peak identification does not need to be complete, but should recover a representative fraction of the total population of emission peaks. This flexible requirement arises, because equation~\ref{eq:bias} defines how a relative quantity (i.e.\ the average CO-to-\halpha flux ratio) of emission peaks changes relative to the galactic average, rather than calculating how an absolute quantity (e.g.\ the total CO flux in peaks) compares to a galaxy's integrated CO luminosity.

The second important choice is that we calibrate the absolute timeline using a reference timescale given by equation~\ref{eq:tstar}, without accounting for any effects due to the possibly sparse sampling of the stellar IMF. If \hii regions in NGC300 host on average only few high-mass stars, that would effectively require a shorter reference timescale. We define a characteristic mass scale\cite{haydon18}
\be
\label{eq:mref}
M_{\rm ref}=\pi\left(\frac{\lambda}{2}\right)^2\tau\Sigma_{\rm SFR} ,
\ee
where $\Sigma_{\rm SFR}$ is the SFR surface density. This characteristic mass scale is related (but not equal) to the average stellar mass powering \hii regions. By applying stellar population synthesis models that account for IMF sampling to simulated galaxy maps\cite{haydon18}, we found that $\tref$ falls within $<20\%$ of its value for a perfectly sampled IMF if $M_{\rm ref}\geq300~\msun$. For the numbers in Table~\ref{tab:fit}, we find $M_{\rm ref}=300{-}500~\msun$, implying that the adopted \halpha reference timescales are not significantly affected by any sparse sampling of the stellar IMF.

Thirdly, the results obtained through our analysis are the most straightforward to interpret when the populations of GMCs and \hii regions have similar completeness in a Lagrangian evolutionary sense, i.e.\ the lowest-mass GMC that is typically detected would evolve into the faintest \hii region that is typically observed. The $5\sigma$ point source sensitivity limit of the CO map corresponds to $5600~\msun$ (see {\bf Observational data}), which in combination with the observed SFEs in the range $\esf=2.2{-}2.8\%$ would translate to a minimum \hii region mass of $120{-}160~\msun$. The $5\sigma$ sensitivity limit of the \halpha map used here translates to an instantaneously formed stellar mass of $200~\msun$ at an age of $5~\myr$ (see {\bf Observational data}), which provides a reasonable match to the above minimum \hii region mass.

Fourthly, we test whether the measured region separation length is consistent with the observed radii of the \hii regions and GMCs in NGC300. Our results imply feedback-regulated star formation, such that feedback-heated gas breaks out of the disc when reaching the scale height and the midplane ISM is compressed at a similar distance from the \hii region. In turn, this would set the separation length equal to the sum of the \hii region radius and the GMC radius. However, the \hii regions fade with time and GMCs are likely not traced by CO across their entire extent in the low-metallicity environment of NGC300\cite{bolatto13b}. This implies that the sum of the \hii region radius and the CO cloud radius should be smaller than the region separation length. Indeed, we find that $1.91(r_{\rm H\alpha}+r_{\rm CO})/\lambda=0.56^{+0.07}_{-0.05}<1$ (where the factor of 1.91 converts the Gaussian dispersion radii to full radii, see {\bf Cloud-scale quantities}), showing that the region separation length satisfies the lower limit imposed by the sum of the radii of GMCs and \hii regions.

Finally, we verify that the measured region separation length is consistent with the nearest neighbour distances of the emission peaks identified in the \halpha and CO maps. The separation length represents the mean geometric distance between regions in the vicinity of identified peaks, which for a local number surface density $\Sigma_{\rm reg,loc}$ is $\lambda=2/\sqrt{\pi\Sigma_{\rm reg,loc}}$. In the case of complete spatial randomness, the probability distribution function of the nearest neighbour distance $r_{\rm n}$ is given by
\be
\label{eq:nnpdf}
\frac{{\rm d}p}{{\rm d}r_{\rm n}} = 2\pi r_{\rm n}\Sigma_{\rm reg,loc}{\rm e}^{-\pi r_{\rm n}^2\Sigma_{\rm reg,loc}} ,
\ee
which integrates to an expectation value of $\langle r_{\rm n}\rangle=\sqrt{\pi}\lambda/4\approx0.443\lambda$. We measure $\lambda=104^{+22}_{-18}~\pc$ (see \autoref{tab:fit}), implying that $\langle r_{\rm n}\rangle_{\rm min}=46^{+10}_{-8}~\pc$. This is a lower limit to the typical nearest neighbour distance of the identified emission peaks, because $\lambda$ is derived from the (complete) information provided by the spatially resolved CO-to-\halpha ratio. The sample of identified peaks is incomplete by definition, resulting in a larger nearest neighbour distance than implied by $\lambda$.

An upper limit to the typical nearest neighbour distance of the identified emission peaks is obtained by calculating the mean geometric distance between peaks across the entire observed field of view. This sets an upper limit to $\lambda$ (and thus the nearest neighbour distance), because it assumes the regions are randomly distributed throughout the entire galaxy and neglects galaxy-scale spatial substructure such as (flocculent) spiral arms or any other clustering of the regions. We identify a total of 455 emission peaks across the \halpha and CO maps. A fraction $\tover/\tau\approx0.11$ of these are expected to correspond to the same regions, resulting in an effective number of $N_{\rm reg}=430$ identified regions. The inclination-corrected field of view has an area of $A=19.6~\kpc^2$, implying a mean geometric distance of $\lambda_{\rm geo}=2\sqrt{A/\pi N_{\rm reg}}=241~\pc$ and a corresponding upper limit to the nearest neighbour distance of $\langle r_{\rm n}\rangle_{\rm max}=107~\pc$. \ref{fig:neighbours} shows the probability distribution of nearest neighbour distances of the identified peaks, demonstrating that the median and mean $r_{\rm n}$ are indeed enclosed by the upper and lower limits derived above. This means that the nearest neighbour distances of the identified emission peaks are consistent with the measured region separation lengths.

\vspace{1.5mm} \noindent {\bf Theoretical predictions and interpretation.}
In \autoref{fig:profiles}, we compare the derived cloud lifetimes, feedback timescales, region separation lengths, and feedback velocities to theoretical predictions for various physical mechanisms. These predictions are based on the observed properties of NGC300 and its GMC population listed in {\bf Structural parameters} and {\bf Cloud-scale quantities}. Here we list and briefly discuss the expressions used.

\vspace{1.5mm} \noindent \emph{Cloud lifetime.}
The measured cloud lifetimes are compared to the GMC crossing time and its gravitational free-fall time. Both of these timescales are defined in {\bf Cloud-scale quantities} above. In addition to these characteristic timescales, cloud destruction can potentially be driven by galactic dynamical processes, either through dynamical dispersal or by dynamically-induced star formation and subsequent stellar feedback. We use the analytical expressions from Ref.~\citenum{jeffreson18} to derive the timescales associated with GMC destruction by epicyclic perturbations\cite{jeffreson18}, cloud-cloud collisions\cite{tan00}, mid-plane free-fall\cite{krumholz12}, and galactic shear\cite{elmegreen87}. For the range of galactic radii in NGC300 spanned by our observations, the resulting timescales are all larger than $40~\myr$. This is well in excess of the longest inferred GMC lifetime ($\tgas=18.3^{+3.9}_{-4.4}~\myr$, see \autoref{tab:fit}) and shows that the GMCs in NGC300 evolve in a way that is decoupled from the large-scale galactic dynamics. Most likely, this insensitivity to galactic dynamics results from the low (molecular) gas surface density in NGC300 and does not hold universally in other galaxies (see {\bf Relation to other work}). 

\vspace{1.5mm} \noindent \emph{Feedback timescale.}
The measured feedback timescales are compared to the supernova timescale, the photoionisation timescale, the stellar wind timescale, and the radiation pressure timescale. These are defined as follows.

The key time-limiting factor for GMC destruction by supernovae is the supernova delay time. Supernova feedback does not start acting until at least $3~\myr$ after the stars formed\cite{leitherer14}, which is well after the typical feedback timescales of $\tover=1.4{-}1.8~\myr$ measured here. This shows that supernovae are not responsible for GMC dispersal in NGC300.

The timescale for GMC dispersal by photoionisation is obtained by equating the measured GMC radii to the radius evolution of the ionisation shock\cite{spitzer78,hosokawa06} and solving for the time. This yields
\be
\label{eq:tphot}
t_{\rm phot}=\frac{4}{7}\left(\frac{3}{4}\right)^{1/2}\frac{r_{\rm S}}{c_{\rm s}}\left[\left(\frac{r_{\rm stdev}}{r_{\rm S}}\right)^{7/4}-1\right] ,
\ee
where $r_{\rm stdev}=r_{\rm GMC}/1.91$ is the adopted GMC radius, defined as the standard deviation of a Gaussian emission peak in order to match the definition of $r_{\rm CO}$ in equation~\ref{eq:vfb}. This definition is also consistent with the finding that a displacement of one standard deviation is sufficient for our method to consider two emission peaks as independent\cite{hygate18}. The Str\"{o}mgen radius $r_{\rm S}$ is given by
\be
\label{eq:rs}
r_{\rm S}=\left(\frac{3m_{\rm H}^2}{4\pi \alpha_B X_{\rm H}^2}\frac{\dot{N}_{\rm LyC}}{\rho_{\rm GMC}^2}\right)^{1/3} ,
\ee
where $\dot{N}_{\rm LyC}/{\rm s}^{-1}=10^{46.5}(M_\text{H{\sc ii}}/\msun)$ is the Lyman continuum photon emission rate\cite{leitherer14} for an \hii region powered by a stellar mass $M_\text{H{\sc ii}}$, $m_{\rm H}=1.7\times10^{-27}~{\rm kg}$ is the mass of the hydrogen atom, $\alpha_{\rm B}=2.56\times10^{-19}~{\rm m}^3~{\rm s}^{-1}\times(T/10^4~{\rm K})^{-0.83}$ is the case-B recombination rate\cite{tielens05} with $T$ the electron temperature, and $X_{\rm H}=0.7$ is the hydrogen mass fraction. The sound speed in ionised gas $c_{\rm s}$ is given by
\be
\label{eq:cs}
c_{\rm s}=\left(\frac{k_{\rm B}T}{\mu m_{\rm H}}\right)^{1/2} ,
\ee
where $k_{\rm B}=1.381\times10^{-23}~{\rm m}^2~{\rm kg}~{\rm s}^{-2}~{\rm K}^{-1}$ is the Boltzmann constant and $\mu=0.62$ is the mean molecular weight in ionised gas.

Likewise, the timescale for GMC dispersal by stellar winds follows from equating the measured GMC radii to the radius evolution of an energy-driven wind shock\cite{weaver77} and rearranging:
\be
\label{eq:twind}
t_{\rm wind}=\left(\frac{154\pi}{125}\frac{\rho_{\rm GMC}}{L_{\rm w}}\right)^{1/3}r_{\rm stdev}^{5/3} ,
\ee
where $L_{\rm w}/{\rm J}~{\rm s}^{-1}=10^{27}\times(M_\text{H{\sc ii}}/\msun)$ is the mechanical luminosity of the wind\cite{leitherer14} driven by a stellar mass $M_\text{H{\sc ii}}$. The cooling time of the hot bubble is\cite{maclow88}
\be
\label{eq:tcool}
\frac{t_{\rm cool}}{\myr}=2.9\left(\frac{Z}{{\rm 0.5Z}_\odot}\right)^{-35/22}\left(\frac{L_{\rm w}}{10^{30}~{\rm J}~{\rm s}^{-1}}\right)^{3/11}\left(\frac{\rho_{\rm GMC}}{20~\cmc}\right)^{-8/11} ,
\ee
with $Z$ the metallicity. This expression has been scaled to units appropriate for GMCs and \hii regions in NGC300 and provides a cooling time longer than the measured feedback timescale ($\tover=1.4{-}1.8~\myr$). This means that the energy-driven regime described by equation~\ref{eq:twind} is appropriate for comparing to the theoretical prediction of stellar wind-driven GMC dispersal.

The timescale for GMC dispersal by radiation pressure is obtained by solving the momentum equation for the radiation pressure force $F_{\rm rad}=(1+\tau_{\rm IR})L_{\rm bol}/c$ within a GMC of constant volume density. Here, $L_{\rm bol}/{\rm J}~{\rm s}^{-1}=10^{29.5}\times(M_\text{H{\sc ii}}/\msun)$ is the bolometric luminosity of a $\mh=-0.5$~dex stellar population\cite{leitherer14}, $c=2.998\times10^8~{\rm m}~{\rm s}^{-1}$ is the speed of light, and $\tau_{\rm IR}$ is the infrared optical depth. When solving the time evolution of a radiation pressure-driven shell, the key question is whether the GMC is optically thick ($\tau_{\rm IR}>1$) or thin ($\tau_{\rm IR}<1$) to infrared radiation. We write $\tau_{\rm IR}=\phi_{\rm tr}\kappa_{\rm R}\Sigma_{\rm GMC}$, where $\phi_{\rm tr}$ is the fraction of infrared radiation that is trapped at an optical depth $\tau_{\rm IR}=1$, $\kappa_{\rm R}=\kappa_0T_{\rm GMC}^2$ is the Rosseland mean opacity\cite{thompson05} (valid for $T_{\rm GMC}<200~{\rm K}$), with $\kappa_0=2.4\times10^{-5}~{\rm m}^2~{\rm kg}^{-1}~{\rm K}^{-2}$ a proportionality constant and $T_{\rm GMC}$ the GMC temperature, and $\Sigma_{\rm GMC}$ is the GMC surface density. Scaling to $T_{\rm GMC}=20~{\rm K}$, an infrared radiation trapping fraction of $\phi_{\rm tr}=0.2$ found in numerical simulations\cite{krumholz12b}, and the mean GMC surface density in NGC300 (see~\ref{fig:gmcs}), we obtain
\be
\label{eq:tau}
\tau_{\rm IR}=5.2\times10^{-5}\left(\frac{\phi_{\rm tr}}{0.2}\right)\left(\frac{T_{\rm GMC}}{20~{\rm K}}\right)^2\left(\frac{\Sigma_{\rm GMC}}{13~\msun~\pc^{-2}}\right) ,
\ee
which clearly shows that GMCs in NGC300 are optically thin to infrared radiation. Even if all radiation were trapped at $\tau_{\rm IR}=1$ (i.e.\ $\phi_{\rm tr}=1$) and the GMCs had temperatures of $T_{\rm GMC}=200~{\rm K}$, the optical depth would still be $\tau_{\rm IR}=0.03\ll1$. The low optical depth of GMCs in NGC300 allows us to write $F_{\rm rad}=L_{\rm bol}/c$, and the solution of the momentum equation becomes
\be
\label{eq:trad}
t_{\rm rad}=\left(\frac{2\pi c}{3}\frac{\rho_{\rm GMC}}{L_{\rm bol}}\right)^{1/2}r_{\rm stdev}^2 .
\ee

All of the timescales in equations~\ref{eq:tphot}, \ref{eq:twind}, and~\ref{eq:trad} require the mass of the stellar population powering the \hii region $M_\text{H{\sc ii}}$ to set the Lyman continuum photon emission rate, the mechanical luminosity of the wind, or the bolometric luminosity. We derive this mass self-consistently using the observed gas surface density, the measured region separation length, and the inferred SFE, by writing
\be
\label{eq:mhii}
M_\text{H{\sc ii}}=\pi\left(\frac{\lambda}{2}\right)^2\esf\Sigma_{\rm gas} .
\ee

Finally, these predicted feedback timescales do not consider energy losses due to turbulent mixing or the porosity of the surrounding interstellar medium. These effects decrease the efficiency of feedback\cite{rosen14,chevance16} and thus may increase the timescale over which feedback disrupts the parent GMC.

\vspace{1.5mm} \noindent \emph{Region separation length.}
The measured region separation lengths are compared to the Toomre length and the gas disc scale height. These are defined as follows.

The Toomre length\cite{toomre64} describes the largest scale in a differentially rotating disc that can undergo gravitational collapse against centrifugal forces and is given by
\be
\label{eq:ltoomre}
l_{\rm T}=\frac{4\pi^2 G\Sigma_{\rm gas}}{\kappa^2} ,
\ee
where $G=6.674~{\rm m}^3~{\rm kg}^{-1}~{\rm s}^{-2}$ is the gravitational constant, $\Sigma_{\rm gas}$ is the gas surface density, and $\kappa$ is the epicyclic frequency, which is defined as
\be
\label{eq:kappa}
\kappa=\frac{v_{\rm c}}{R}\left(1+\frac{{\rm d}\ln{v_{\rm c}}}{{\rm d}\ln{R}}\right)^{1/2} ,
\ee
with $v_{\rm c}$ the circular velocity. \autoref{fig:profiles} shows that the Toomre length does not match the observed region separation length. This is not surprising, because the disc is stable against global gravitational collapse ($Q>1$, see \ref{fig:galaxy}) and thus the Toomre length is unlikely to set the separation length of independent star-forming regions.

The gas disc scale height is obtained from the observed surface densities and velocity dispersions by assuming hydrostatic equilibrium. It is given by
\be
\label{eq:hdisc}
h_{\rm gas}=\frac{\sigma_{\rm gas}^2}{\pi G \phi_P \Sigma_{\rm gas}} ,
\ee
where $\sigma_{\rm gas}$ is the gas velocity dispersion (assumed to correspond to {\sc Hi}, because the interstellar medium of NGC300 is dominated by atomic gas) and $\phi_P$ is a quantity reflecting the influence of the stellar gravitational potential. It is given by
\be
\label{eq:phip}
\phi_P=1+\frac{\Sigma_\star}{\Sigma_{\rm gas}}\frac{\sigma_{\rm gas}}{\sigma_\star} ,
\ee
where $\Sigma_\star$ is the stellar surface density and $\sigma_\star$ is the stellar velocity dispersion. Because the latter has not been measured directly for NGC300, we derive it from the stellar scale height $h_\star$ (see {\bf Structural parameters}) and stellar surface density as
\be
\label{eq:sigmastar}
\sigma_\star=2^{1/4}\left(\pi G h_\star\Sigma_\star\right)^{1/2} .
\ee

\vspace{1.5mm} \noindent \emph{Feedback velocity.}
The measured feedback outflow velocity is compared to the sound speed in ionised gas, the ionisation front velocity, the stellar wind shock velocity, and the radiation pressure-driven shell velocity. The sound speed in ionised gas follows directly from the temperature $T$ and the mean molecular weight $\mu$ as in equation~\ref{eq:cs}. The ionisation front velocity, stellar wind shock velocity, and radiation pressure-driven shell velocity are all derived from their corresponding timescales in equations~\ref{eq:tphot}, \ref{eq:twind}, or \ref{eq:trad} as
\be
\label{eq:vshell}
v_{\rm shell}=\frac{r_{\rm stdev}}{t_{\rm shell}} ,
\ee
where $v_{\rm shell}$ represents the predicted feedback outflow velocity due to any of these three mechanisms and $t_{\rm shell}$ refers to the corresponding timescale from equations~\ref{eq:tphot}, \ref{eq:twind}, or \ref{eq:trad}.

\vspace{1.5mm} \noindent \emph{GMC dispersal by stellar feedback.}
The short measured feedback timescale ($\sim1.5~\myr$ or $10{-}15\%$ of the total GMC lifetime) strongly suggests that stellar feedback drives GMC dispersal in NGC300. Otherwise, another physical mechanism would need to conspire such that GMCs are always dispersed briefly after the appearance of massive stars. In principle, GMCs may also disperse dynamically, with or without forming (massive) stars, and then re-form. Within our framework, GMC dispersal without any associated (\halpha-emitting) star formation and the subsequent (re-)formation of a GMC would take place in a single `lifecycle', because \halpha emission is used to calibrate the evolutionary timeline. This means that our interpretation does not assume that every CO peak will evolve into an \halpha peak, but it does assume that each \halpha peak emerged from a CO peak. As a result, we measure the cumulative time spent as a CO-bright region for each \halpha-bright region that forms. Any time before GMC dispersal without an associated \hii region would thus be added onto the next GMC's lifetime. Our measurements show that this rarely happens, because GMCs are found to live for approximately one turbulent crossing time prior to forming massive stars, whereas the combination of GMC dispersal by dynamics, their subsequent (re-)formation, and their final evolution towards (massive) star formation would require multiple crossing times.

At face value, GMC dispersal by feedback seems at odds with the high virial ratios of GMCs in NGC300 (see \ref{fig:gmcs}). However, recent work has shown that GMCs often appear supervirial due to confinement by external pressure\cite{hughes13}. The low molecular gas fraction of NGC300 implies that the hydrostatic balance of the GMCs is governed by external forces\cite{schruba18}. This means that the total force balance is closer to equilibrium than suggested by the direct measurement of the virial ratio. Even if the GMCs are not self-gravitating globally, their substructured nature enables dense clumps within them to undergo local gravitational collapse\cite{clark05,padoan12,dale13,rathborne15}, which in turn induces star formation and stellar feedback. This implies that the GMCs do not need to be bound, nor do they need to be globally collapsing, in order to go through an evolutionary cycle of star formation and feedback-driven dispersal. The measured feedback timescale of 1.5 Myr is much shorter than a GMC crossing time; it would require a considerable coincidence to synchronise the time needed for dense clump and massive star formation to match the time needed for dynamical GMC dispersal. This is supported further by the match between the region separation length and the gas disc scale height, which is indicative of feedback-driven blowout. Finally, we note that our results do not require the complete destruction of GMCs by feedback, but merely their dispersal relative to the resulting \hii regions. In summary, our results suggest an interpretation in which the GMCs in NGC300 are dynamical, transient features embedded in a lower-density gas flow, which are dispersed by feedback from massive stars that formed in gravitationally-collapsing, dense clumps within the GMC.

Without additional information, our analysis does not distinguish between feedback-driven GMC destruction due to a phase transition or due to kinetic dispersal. In the former case, the feedback timescale refers to the time needed to photodissociate or ionise the region. In the latter case, the mass that constituted the GMC remains molecular, but is separated from the young stellar population by feedback-driven motion. In this case, the GMC is likely to break up, such that it no longer corresponds to a single entity. In the context of our analysis method, the descendant(s) of the GMC then represent new GMCs at the beginning of the evolutionary timeline.

\vspace{1.5mm} \noindent {\bf Relation to other work.} \emph{Analysis method.}
During the past four decades, attempts to characterise the cloud lifecycle have followed a wide variety of approaches\cite{scoville79,sanders85,leisawitz89,elmegreen00,hartmann01,engargiola03,blitz07,kawamura09,koda09,murray11,dobbs14,meidt15,kruijssen15,corbelli17,barnes17,kreckel18}. These pioneering works have led to contrasting results, finding GMC lifetimes from $1$ to $100~\myr$. Each of these studies presented important advances in our understanding of the cloud lifecycle, but collectively they faced a number of key shortcomings. Firstly, previous works exhibit major differences in experiment design. Some studies count objects of different classes to statistically infer their lifetimes\cite{kawamura09,corbelli17}, whereas others follow progressions along evolutionary streamlines with known streaming velocities and look for changes in properties along these streamlines\cite{engargiola03,koda09,meidt15,kruijssen15}. Some studies equate the GMC lifetime to the age (spread) of associated young stellar populations and thus exclude any `inert' phase of the GMC preceding massive star formation\cite{leisawitz89,elmegreen00,hartmann01,grasha18}, whereas others consider the lifetimes of H$_2$ molecules rather than those of GMCs\cite{scoville79,koda09}. Secondly, several of these studies were hampered by differing, subjective classifications of objects (e.g.\ the definitions of GMCs and \hii regions). Especially for the studies relying on statistical inference, the definition of these classes of objects directly sets the derived evolutionary timescales. Due to the hierarchical structure of the interstellar medium and young stellar populations\cite{efremov98}, it is not clear how the necessary classes of objects are best defined. Perhaps most crucially, none of these studies offered a unifying framework for translating these timescales to the underlying physics of cloud-scale star formation and feedback. As a result, these measurements have not enabled a predictive theory for explaining galaxy growth from a bottom-up perspective, in terms of the lifecycle of its building blocks.

The method adopted here\cite{kruijssen18} provides a self-consistent physical and statistical framework for characterising the lifecycle of molecular clouds and star-forming regions. It adopts a Lagrangian perspective of following mass elements that reside on an evolutionary timeline independently of their neighbours. The method itself is agnostic about the definition of these `independent regions'; their separation is defined by the separation length $\lambda$, below which the CO-to-\halpha ratio significantly deviates from the galactic average. This approach improves on previous methods by not relying on subjective classifications of `molecular clouds' or `\hii regions' in various evolutionary phases or requiring the idealised situation of evolutionary streamlines with a common zero point at which star formation is initiated. While classification-based methods have needed to resolve GMCs (${\sim}20~\pc$) to distinguish their evolutionary phase, our method only requires their separation ($100{-}150~\pc$ in NGC300) to be resolved, greatly increasing the distance scale over which it can be applied. Even within the single galaxy NGC300, we report the environmental variation of GMC lifetimes, indicating that at least part of the $1{-}100~\myr$ range found in previous work might be physical in nature. Future applications of the methodology adopted here will systematically test this prediction.

\vspace{1.5mm} \noindent \emph{The physical meaning of the region separation length.}
The de-correlation of CO and \halpha emission on spatial scales smaller than the region separation length $\lambda$ implies the existence of a characteristic size scale that defines the structure of the interstellar medium and possibly galaxies at large. It is a key question how this characteristic size scale is consistent with the traditional concept of a `scale-free' interstellar medium\cite{elmegreen96,hopkins12c,hopkins13,krumholz14}. \autoref{fig:profiles} shows that $\lambda$ closely matches the gas disc scale height. In the context of excursion-set models of the scale-free interstellar medium\cite{hopkins13}, this corresponds to the `first crossing' scale, i.e.\ the largest scale at which structures become self-gravitating. As such, the region separation length likely defines the upper bound on the range of scales across which the interstellar medium is scale-free. It may plausibly also correspond to the turbulence driving scale, which is thought to be similar to the disc scale height\cite{maclow04}. Future kinematic studies of the CO emission in NGC300 will be able to test this prediction, by identifying characteristic length scales in the velocity power spectrum\cite{henshaw16b} and comparing these to the measured region separation lengths.

\vspace{1.5mm} \noindent \emph{Spatial scale dependence of the star formation relation.}
This work appears at a time when the theoretical star formation community is increasingly tackling the dependence of the relation between gas and star formation on time\cite{zamoraaviles12,kruijssen14,elmegreen18,burkhart18,vazquez18} and spatial scales\cite{feldmann11,krumholz12,semenov16,semenov17}. The discussion is moving away from the static picture imposed by the galaxy-scale relation between molecular gas and star formation, and instead aims to express the underlying processes in terms of mass flows\cite{kruijssen18,semenov18}. The vigorous evolutionary cycling observed in NGC300 demonstrates that these theoretical efforts are moving towards a formalism that more closely matches the star formation process seen in nature. Our results imply that the cloud-scale star formation process sets in at a gas-dominated stage (with high gas surface densities and low SFR surface densities), cycling through a phase during which the star formation activity increases, after which the parent GMC is disrupted and the region ends up being dominated by young stars (at a high SFR surface density and a low gas surface density). The SFR naturally increases with time during this process\cite{ochsendorf17}, until the gas is expelled. This means that individual GMCs or star-forming regions are moving in the plane spanned by the gas mass (surface density) and SFR (surface density) (see \ref{fig:cycling}). The tight correlation between these quantities observed on galaxy-wide scales\cite{silk97,elmegreen97,kennicutt98} only arises when averaging over a large population of independent regions undergoing this evolutionary lifecycle.\cite{schruba10,kruijssen14} As demonstrated in this work, it is therefore not necessarily the slope of the gas mass-SFR relation, but more prominently its scatter on sub-kpc scales that encodes the underlying physics driving the lifecycle of GMCs and star-forming regions.

\vspace{1.5mm} \noindent \emph{Environmental dependence of the GMC lifecycle.}
These results also reinforce the theoretical expectation that the GMC lifecycle is environmentally dependent. This has been predicted by a variety of numerical simulations\cite{dobbs13,fujimoto14,dobbs15,tasker15} and analytical work\cite{meidt13,meidt18,jeffreson18}, which all take the view that GMC evolution is coupled to the large-scale galactic dynamics. While we find no evidence for the major influence of galactic dynamics in NGC300, this could itself be driven by a dependence of GMC properties on the galactic environment. NGC300 has a low molecular gas surface density (see~\ref{fig:galaxy}), implying that GMCs represent isolated `islands' that are likely embedded within an envelope of lower-density atomic gas. The coupling between GMC evolution and galactic dynamics can be tested better in galaxies with high gas surface densities that are rich in molecular gas, such that CO emission also traces the GMC envelopes. We do find strong evidence for the regulation of the GMC lifecycle by stellar feedback, which carries environmental dependences on the GMC volume density, the GMC radius, and the metallicity (see {\bf Theoretical predictions}). In view of these predictions, it is a high priority for future work in this area to explore the environmental variation of the physical quantities shown in \autoref{fig:profiles} across the galaxy population.

\vspace{1.5mm} \noindent \emph{Regulation by stellar feedback.}
Finally, numerical simulations of galactic star formation have recently emphasised the importance of stellar feedback in regulating the growth and evolution of galaxies\cite{ostriker10,ostriker11,kim11,hopkins14}. Our observational results present a key empirical validation of this prediction. Specifically, we provide important evidence for the prediction that supernova feedback by itself is insufficient for regulating galaxy evolution, and that `early' feedback in the form of photoionisation, stellar winds, or radiation pressure is required\cite{stinson13,agertz13,hopkins14,dale15,gatto15,gatto17,hu17}. Our conclusion that photoionisation plays a major role in the dispersal of low-to-intermediate-mass clouds is also consistent with a variety of theoretical predictions\cite{matzner02,krumholz09d,dale14,matzner15,rahner17,kim18} and observational work within the Local Group\cite{lopez14,chevance16}. The critical step forward made by this work is that it provides a self-consistent framework in which all of the above theoretical predictions can be tested simultaneously using a single set of observational data products and a single analysis method. Now that observational studies are capable of quantitatively testing these predictions, it is to be expected that upcoming systematic empirical studies of the GMC lifecycle and its dependence on the galactic environment will generate a step change by motivating future theoretical models for star formation and feedback in the context of galaxy formation and evolution.

\medskip \noindent
{\small \bf {Code Availability}} {\small 
    {A dedicated publication of the analysis software used in the current study is in preparation. The code is available from the corresponding author (J.M.D.K.) upon reasonable request.}
    }

\medskip \noindent
{\small \bf {Data Availability}} {\small 
    {The ALMA CO(1-0) data used in this work are from projects \mbox{2013.1.00351.S} and \mbox{2015.1.00258.S} (PI A.~Schruba) and are publicly available through the ALMA archive (\url{https://almascience.eso.org/alma-data/archive}). The MPG/ESO \mbox{2.2-m} \halpha data are publicly available as raw data from the ESO archive (\url{http://archive.eso.org/}) under programme ID \mbox{065.N-0076} (PI F.~Bresolin). All other (e.g.\ derived) data that support the findings of this study are available from the corresponding author (J.M.D.K.) upon reasonable request.}
    }

\clearpage


\setcounter{page}{1}
\setcounter{figure}{0}
\setcounter{table}{0}
\captionsetup[figure]{labelformat=empty}
\captionsetup[table]{labelformat=empty}

\renewcommand{\thefigure}{Extended Data Figure~\arabic{figure}}
\renewcommand{\thetable}{Extended Data Table~\arabic{table}}


\begin{figure*}
\centerline{
\includegraphics[width=\textwidth]{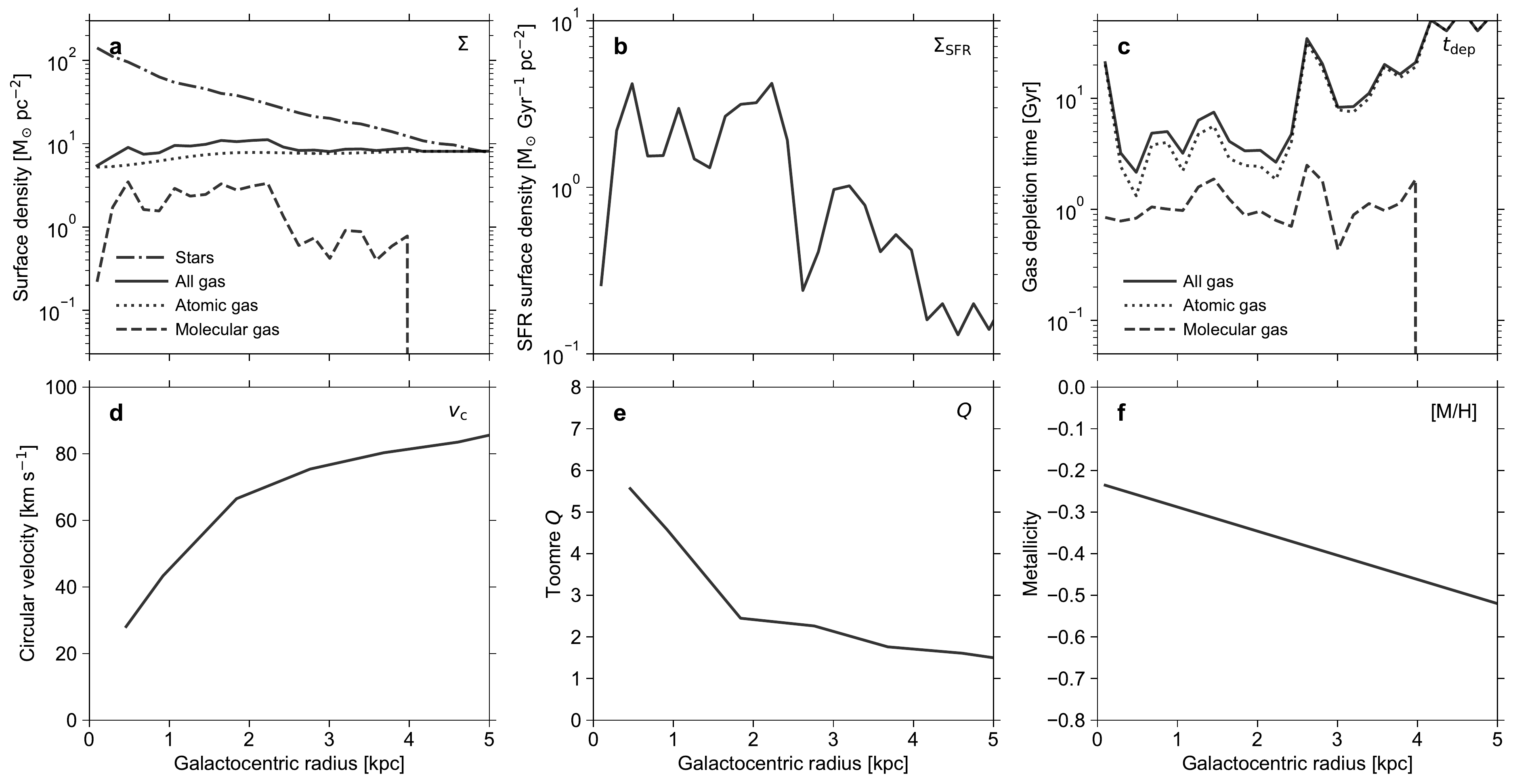}}
\vspace{-2mm}
\caption{\label{fig:galaxy}\textbf{Extended Data Figure 1 $|$ Radial profiles of quantities describing the galactic structure of NGC300.} The panels show the surface densities of molecular gas, atomic gas, total gas, and stars ({\bf a}), the SFR surface density ({\bf b}), the depletion times of molecular, atomic, and total gas ({\bf c}), the rotation curve ({\bf d}), the Toomre $Q$ stability parameter ({\bf e}, where $Q=1$ corresponds to equilibrium), and the metallicity ({\bf f}).}
\vspace{-4mm}
\end{figure*}

\clearpage
\begin{figure*}
\centerline{
\includegraphics[width=\textwidth]{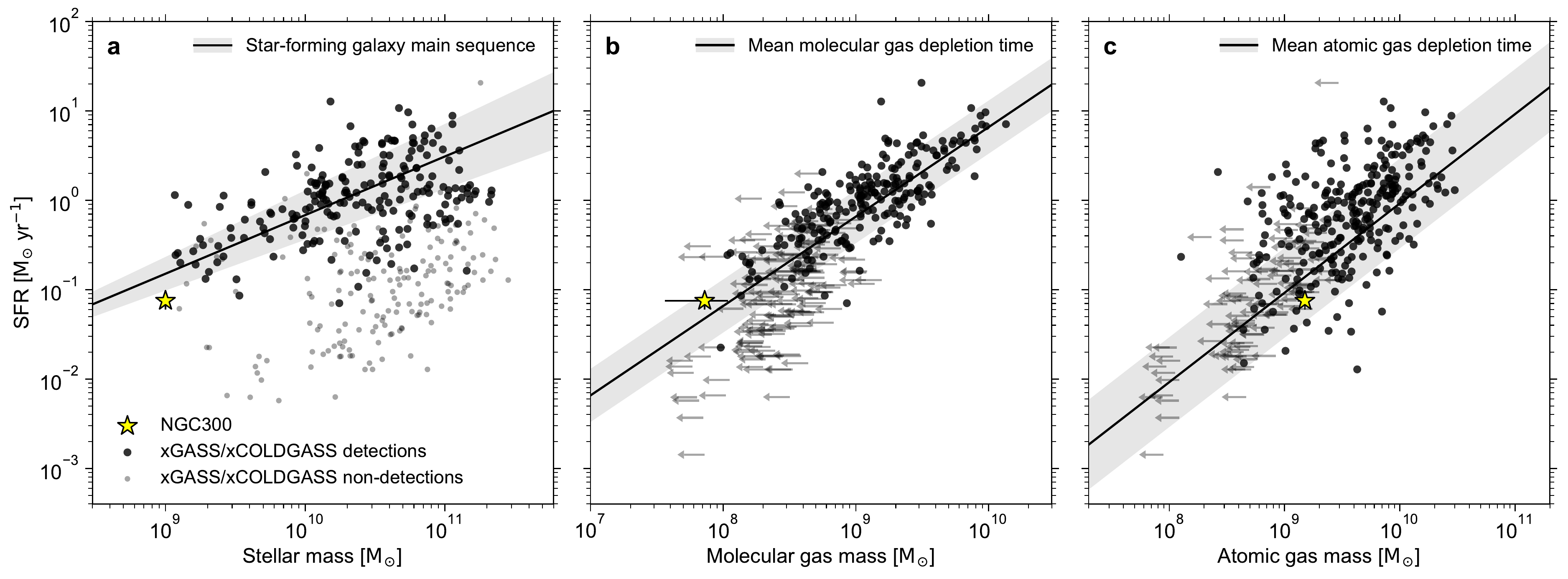}
}
\vspace{-2mm}
\caption{\label{fig:galpop}\textbf{Extended Data Figure 2 $|$ Integrated properties of NGC300 relative to other star-forming galaxies in the nearby Universe.} The panels show the SFR as a function of stellar mass ({\bf a}), molecular gas mass ({\bf b}), and atomic gas mass ({\bf c}), both for NGC300 and nearby galaxies from the xCOLDGASS\cite{saintonge17} and xGASS\cite{catinella18} surveys. In panels {\bf b} and {\bf c}, the arrows indicate $3\sigma$ and $5\sigma$ upper limits of non-detections in xCOLDGASS and xGASS, respectively. The solid lines represent the star-forming galaxy main sequence\cite{catinella18} ({\bf a}), the mean molecular gas depletion time of the xCOLDGASS detections ({\bf b}), and the mean atomic gas depletion time of the xGASS detections ({\bf c}), with the $1\sigma$ scatter shown in grey.}
\vspace{-4mm}
\end{figure*}

\clearpage
\begin{figure*}
\centerline{
\includegraphics[width=\textwidth]{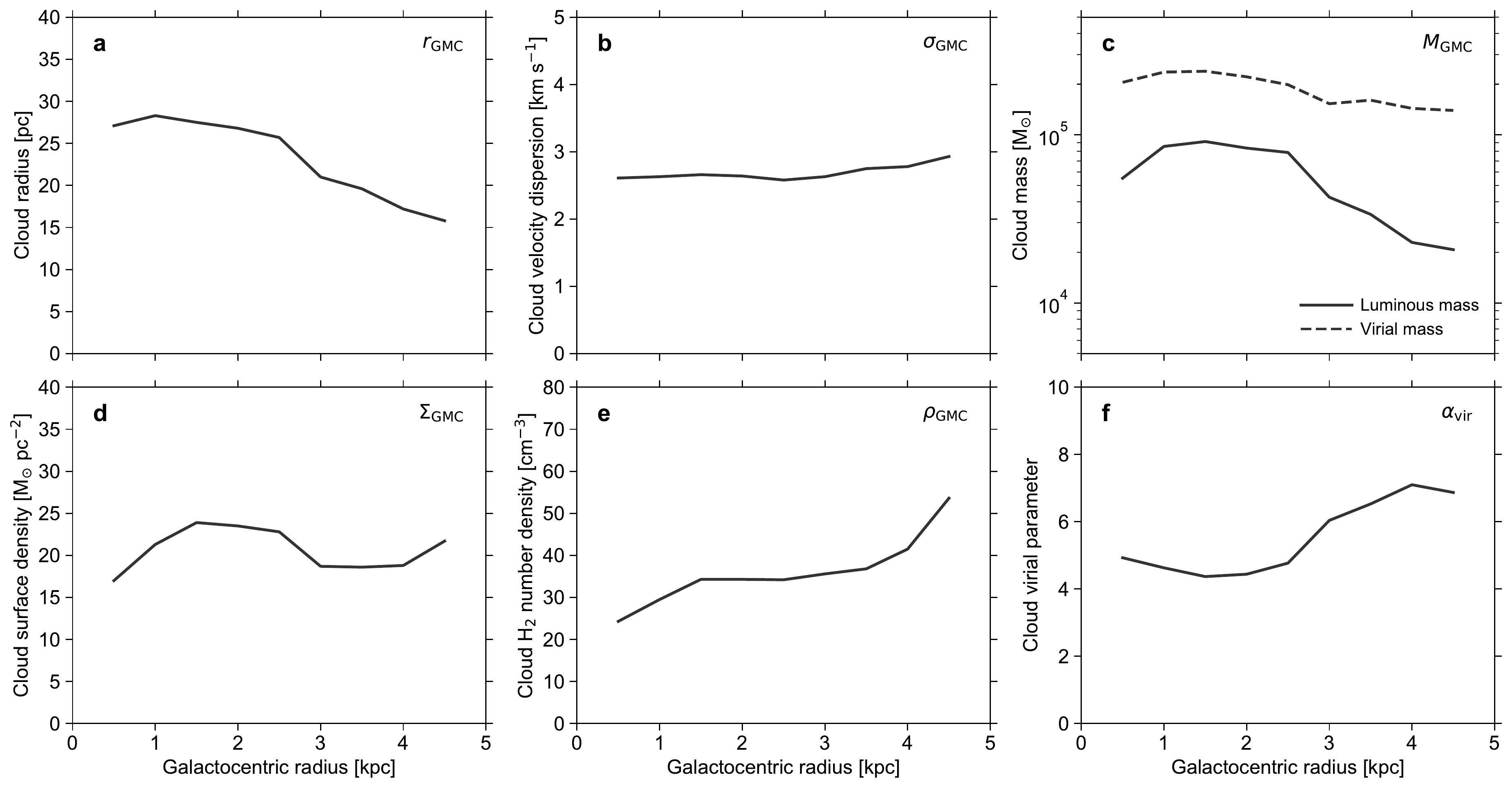}}
\vspace{-2mm}
\caption{\label{fig:gmcs}\textbf{Extended Data Figure 3 $|$ Radial profiles of the average properties of the GMC population in NGC300.} The panels show the GMC radius ({\bf a}), velocity dispersion ({\bf b}), luminous and virial masses ({\bf c}), surface density ({\bf d}), molecular hydrogen number density ({\bf e}), and virial parameter ({\bf f}, where $\alpha_{\rm vir}=1$ corresponds to virial equilibrium).}
\vspace{-4mm}
\end{figure*}

\clearpage
\begin{figure}
\centerline{
\includegraphics[width=0.4\textwidth]{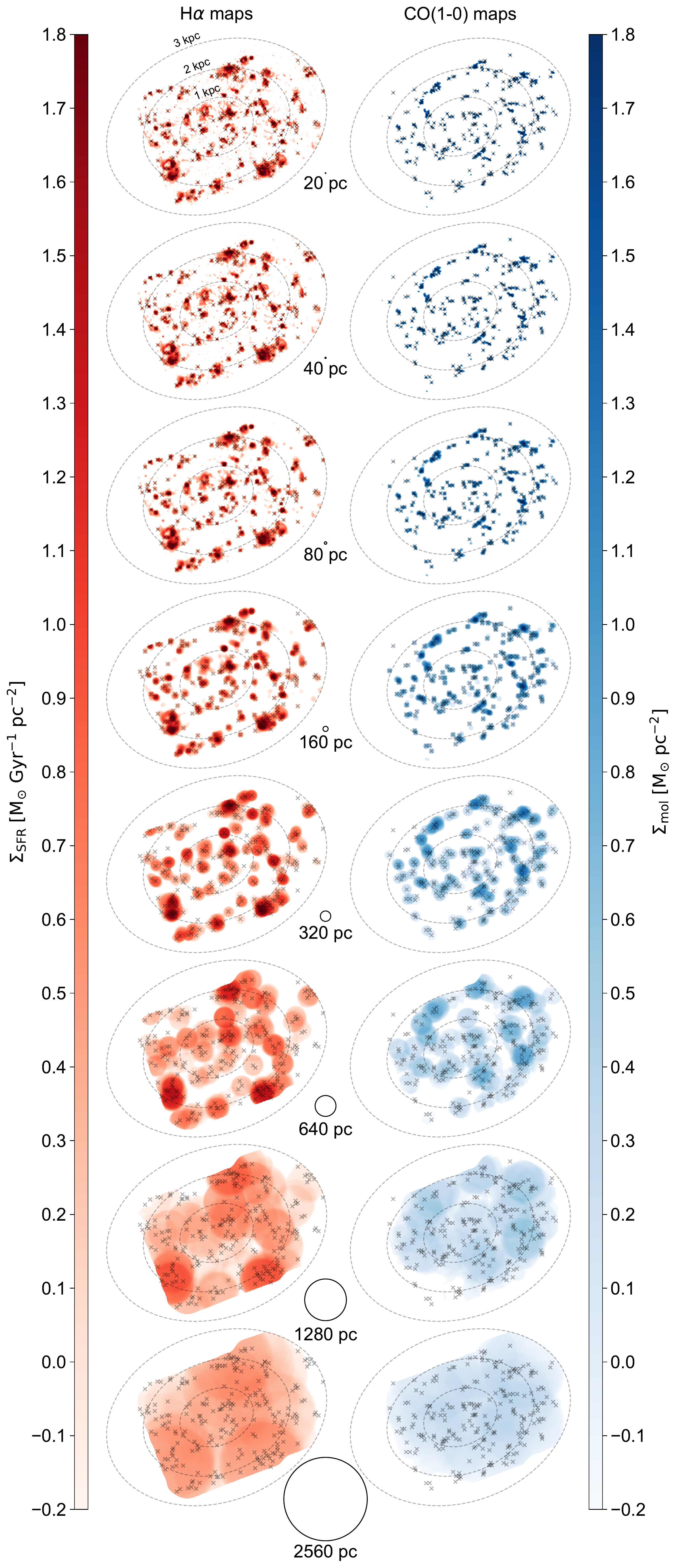}}
\vspace{-2mm}
\caption{\label{fig:maps}\textbf{Extended Data Figure 4 $|$ NGC300 seen at different aperture sizes.} This figure illustrates the image processing of this work. The panels show the H$\alpha$ emission (left) and \co emission (right) from NGC300 convolved with top-hat apertures of diameters increasing from top to bottom from $20~\pc$ to $2560~\pc$ (see the annotated circles). Each panel also shows the locations of the emission peaks identified in the $20~\pc$-resolution images (crosses), at which the flux density measurements are made when deriving the CO-to-\halpha flux ratio as a function of size scale as in \autoref{fig:tuningfork}.}
\vspace{-4mm}
\end{figure}

\clearpage
\begin{figure*}
\centerline{
\includegraphics[width=\textwidth]{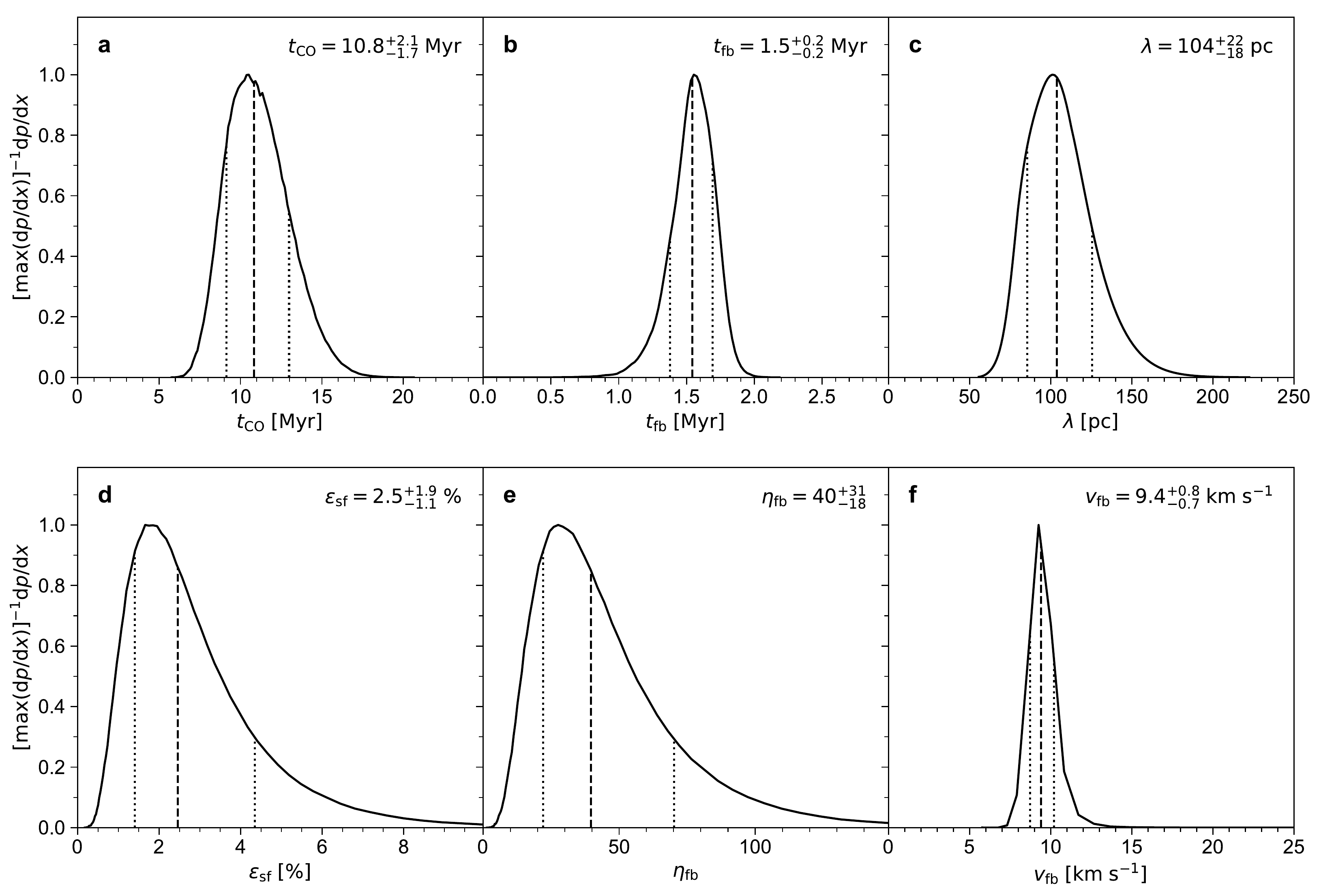}}
\vspace{-2mm}
\caption{\label{fig:pdfs}\textbf{Extended Data Figure 5 $|$ Probability distribution functions of constrained quantities.}  Normalised probability distributions of the six constrained quantities (solid lines), with best-fitting values (dashed lines) and $1\sigma$ uncertainties (dotted lines) indicated in the top-right corner of each panel.}
\vspace{-4mm}
\end{figure*}

\clearpage
\begin{figure}
\centerline{
\includegraphics[width=0.5\textwidth]{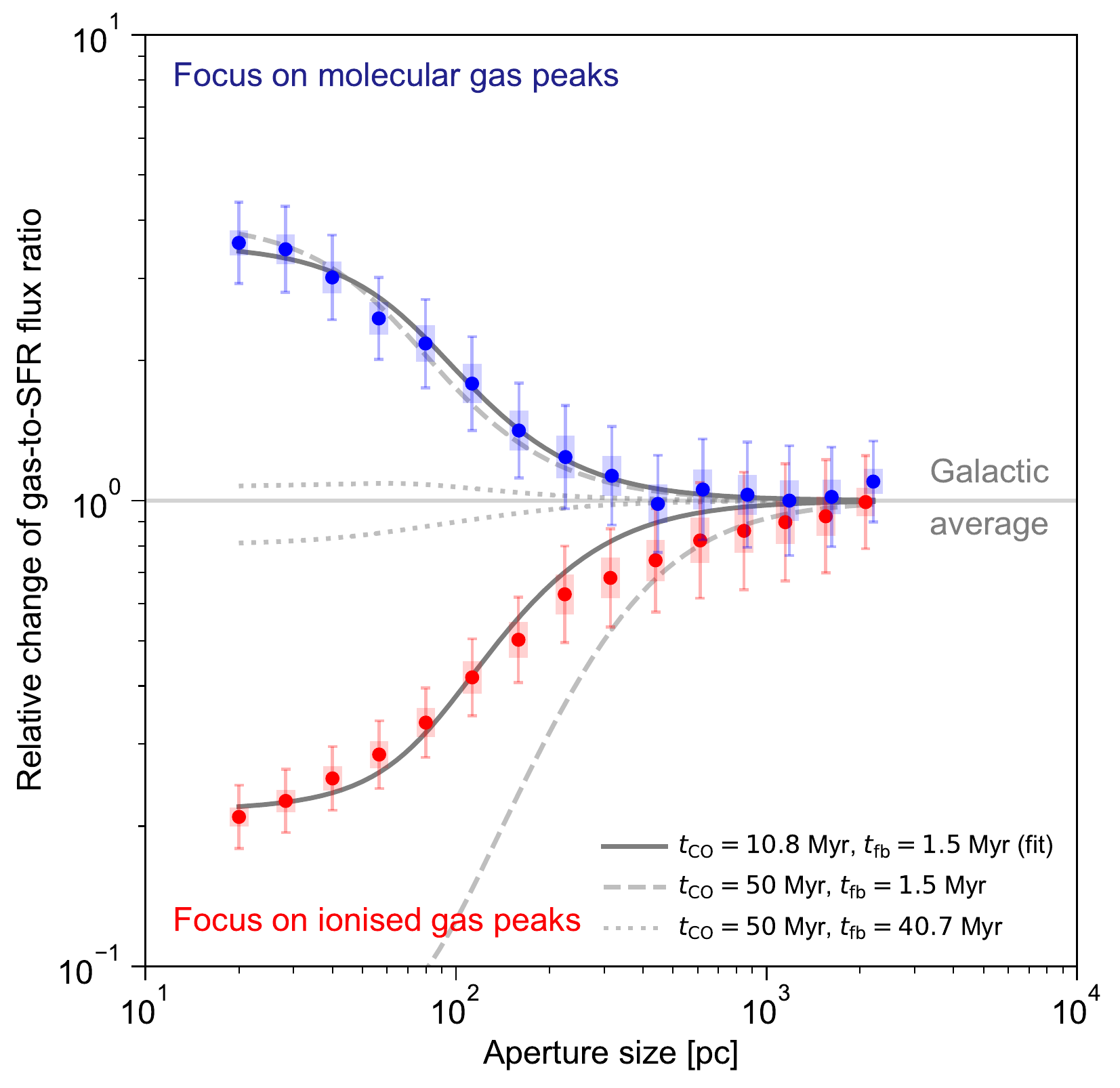}
}
\vspace{-2mm}
\caption{\label{fig:tuningfork_lines}\textbf{Extended Data Figure 6 $|$ Influence of the GMC lifecycle on the de-correlation of molecular gas and young stellar emission.} Shown is the change of the CO-to-\halpha flux ratio relative to the galactic average as a function of spatial scale, for apertures placed on CO emission peaks (top branch) and \halpha emission peaks (bottom branch). The symbols and $1\sigma$ error bars show the CO-to-\halpha flux ratios observed across the entire field-of-view of NGC300 as in \autoref{fig:tuningfork}. The evolutionary timeline of the GMC lifecycle is constrained by fitting the model indicated by the solid lines. Alternative models with long GMC lifetimes are shown by the dashed and dotted lines (see the legend). These alternatives are ruled out by the observations.}
\vspace{-4mm}
\end{figure}

\clearpage
\begin{figure}
\centerline{
\includegraphics[width=0.5\textwidth]{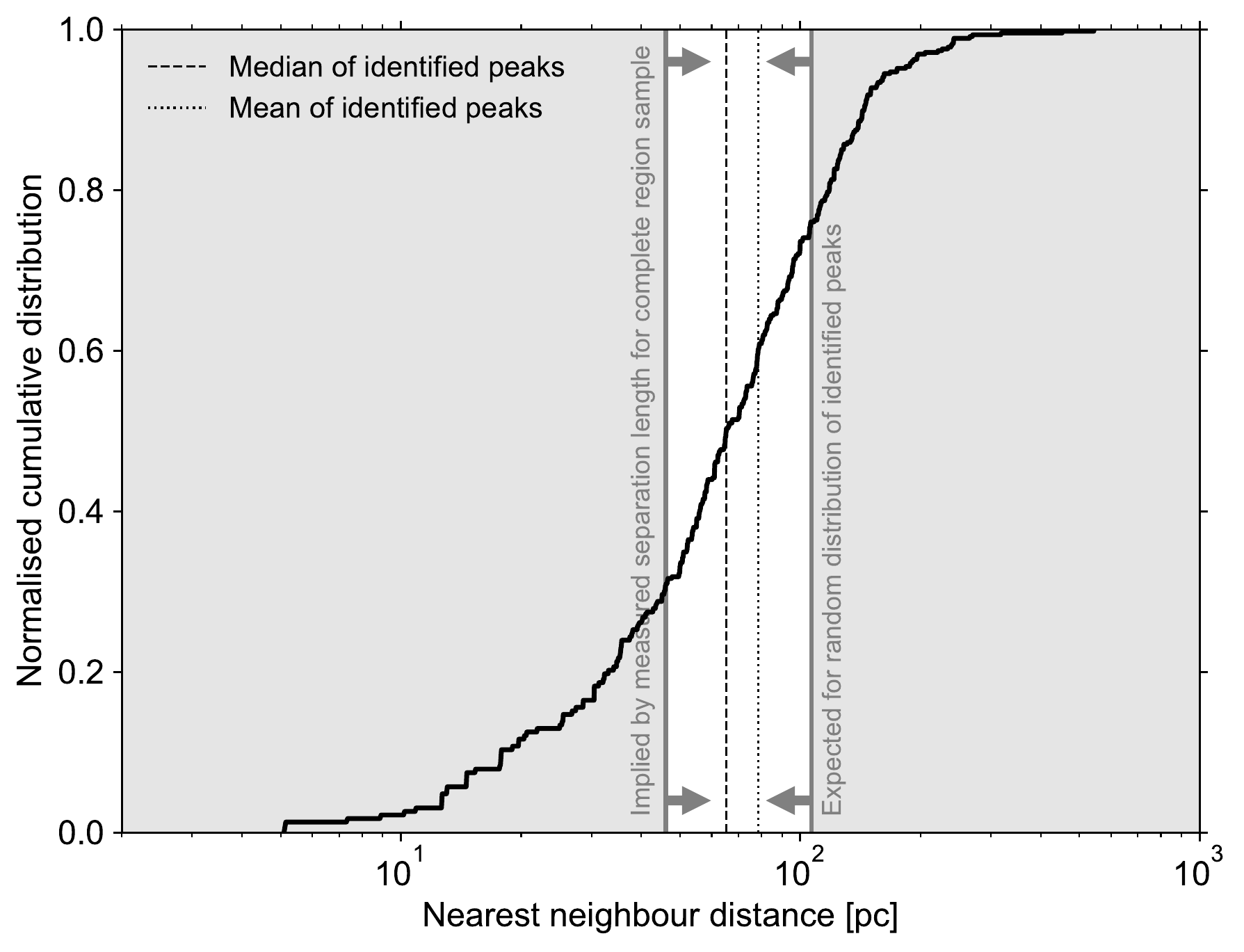}
}
\vspace{-2mm}
\caption{\label{fig:neighbours}\textbf{Extended Data Figure 7 $|$ Distribution of nearest neighbour distances of identified emission peaks.} The black solid line shows the cumulative distribution of the distances to the nearest neighbours across the combined sample of emission peaks identified in the \halpha and CO maps. The median and mean distance are indicated by the vertical dashed and dotted lines, respectively. The vertical grey lines indicate lower and upper limits derived in the Methods section, the first of which is implied by the measured region separation length. The location of the median and mean nearest neighbour distances in between these limits is consistent with the measured separation lengths.}
\vspace{-4mm}
\end{figure}

\clearpage
\begin{figure}
\centerline{
\includegraphics[width=0.5\textwidth]{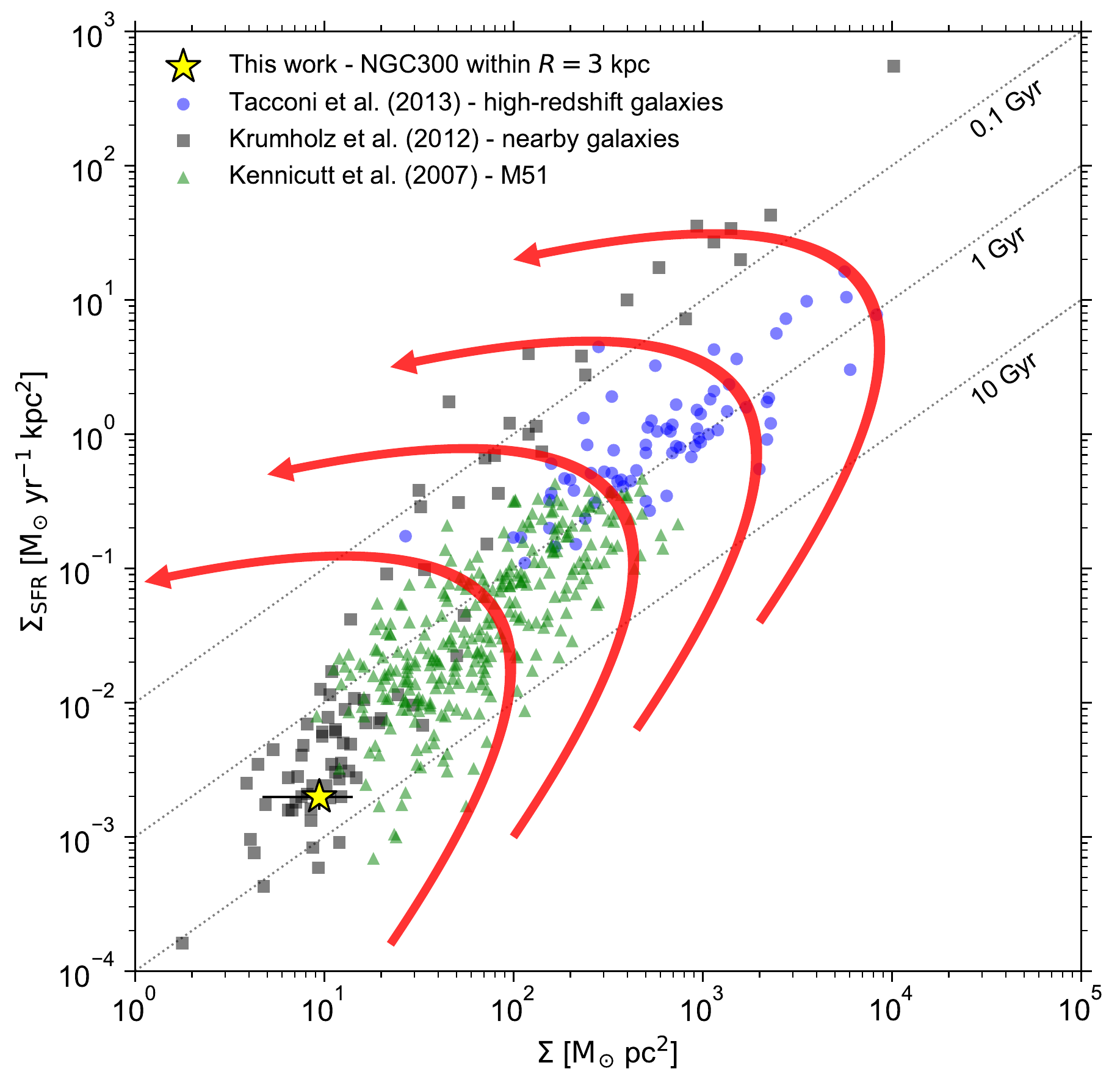}
}
\vspace{-2mm}
\caption{\label{fig:cycling}\textbf{Extended Data Figure 8 $|$ Schematic illustration of cloud-scale evolutionary cycling in the $\Sigma{-}\Sigma_{\rm SFR}$ plane.} The symbols show the observed relation between the total gas surface density ($\Sigma$, where the high-redshift sample is assumed to be fully molecular) and the SFR surface density ($\Sigma_{\rm SFR}$) for galaxies in the local Universe and at high-redshift (see the legend)\cite{kennicutt07,krumholz12,tacconi13}. For NGC300, the error bars show the $1\sigma$ uncertainties. Dotted lines represent constant gas depletion times as indicated by the labels. The results of this work show that GMCs and star-forming regions move through this diagram. As a function of time, they increase their gas density, increase their SFR, expel gas through feedback, and eventually fade by stellar evolution, which is schematically illustrated by the red arrows.}
\vspace{-4mm}
\end{figure}

\clearpage


\setcounter{page}{1}
\setcounter{figure}{0}
\setcounter{table}{0}
\renewcommand{\thefigure}{S\arabic{figure}}
\renewcommand{\thetable}{S\arabic{table}}

\begin{center}
{\bf \Large \uppercase{Supplementary information} }
\end{center}

\noindent {\bf Supplementary Video 1 $|$ Spatial de-correlation between molecular gas and ionised emission from young stars towards small spatial scales in the nearby galaxy NGC300.}
The top panels show the ionised emission (\halpha, left) and molecular gas (CO, right) maps of NGC300, with crosses indicating the emission peaks in each of the maps. The bottom-left panel shows the gas depletion time (the ratio between the top maps). The colour of this map changes from white on large spatial scales (strong CO-\halpha correlation) to bright red and blue on small spatial scales (strong CO-\halpha anti-correlation). The bottom-right panel quantifies this behaviour by showing how the change of the CO-to-\halpha flux ratio relative to the galactic average increases towards small ($<150~\pc$) aperture sizes. The circle and vertical line indicate the spatial scale (`aperture size') at which the galaxy is observed.

\vspace{1.5mm} \noindent {\bf Supplementary Video 2 $|$ Relation between the change of the CO-to-\halpha flux ratio relative to the galactic average and the physical quantities defining the cloud lifecycle.}
The young stellar lifetime ($\tstar$), the cloud lifetime ($\tgas$), the feedback timescale ($\tover$), and the region separation length ($\lambda$) are initially set equal to values measured for NGC300, but are systematically varied to demonstrate their effect on the CO-to-\halpha ratio. The top panels show mock `CO' and `\halpha' maps from a numerical simulation of a disc galaxy, with an inclination and position angle mimicking NGC300. The images are generated using stellar particles in specific age intervals, yielding emission peak lifetimes as indicated in the timeline and annotation at the top of the video. The bottom-left panel shows the `gas depletion time', i.e.\ the ratio between the top maps. The bottom-right panel shows the change of this ratio relative to the galactic average, with $\lambda$ indicated along the bottom axis. This diagram provides a non-degenerate measurement of the three measured quantities ($\tgas$, $\tover$, and $\lambda$), because $\tstar$ is known (see Methods).


\begin{thebibliography}{10}
\expandafter\ifx\csname url\endcsname\relax
  \def\url#1{\texttt{#1}}\fi
\expandafter\ifx\csname urlprefix\endcsname\relax\def\urlprefix{URL }\fi
\providecommand{\bibinfo}[2]{#2}
\providecommand{\eprint}[2][]{\url{#2}}

\bibitem{somerville15}
\bibinfo{author}{{Somerville}, R.~S.} \& \bibinfo{author}{{Dav{\'e}}, R.}
\newblock \bibinfo{title}{{Physical Models of Galaxy Formation in a
  Cosmological Framework}}.
\newblock \emph{\bibinfo{journal}{\araa}} \textbf{\bibinfo{volume}{53}},
  \bibinfo{pages}{51--113} (\bibinfo{year}{2015}).

\bibitem{naab17}
\bibinfo{author}{{Naab}, T.} \& \bibinfo{author}{{Ostriker}, J.~P.}
\newblock \bibinfo{title}{{Theoretical Challenges in Galaxy Formation}}.
\newblock \emph{\bibinfo{journal}{\araa}} \textbf{\bibinfo{volume}{55}},
  \bibinfo{pages}{59--109} (\bibinfo{year}{2017}).

\bibitem{scannapieco12}
\bibinfo{author}{{Scannapieco}, C.} \emph{et~al.}
\newblock \bibinfo{title}{{The Aquila comparison project: the effects of
  feedback and numerical methods on simulations of galaxy formation}}.
\newblock \emph{\bibinfo{journal}{\mnras}} \textbf{\bibinfo{volume}{423}},
  \bibinfo{pages}{1726--1749} (\bibinfo{year}{2012}).

\bibitem{hopkins13b}
\bibinfo{author}{{Hopkins}, P.~F.}, \bibinfo{author}{{Narayanan}, D.} \&
  \bibinfo{author}{{Murray}, N.}
\newblock \bibinfo{title}{{The meaning and consequences of star formation
  criteria in galaxy models with resolved stellar feedback}}.
\newblock \emph{\bibinfo{journal}{\mnras}} \textbf{\bibinfo{volume}{432}},
  \bibinfo{pages}{2647--2653} (\bibinfo{year}{2013}).

\bibitem{kennicutt12}
\bibinfo{author}{{Kennicutt}, R.~C.} \& \bibinfo{author}{{Evans}, N.~J.}
\newblock \bibinfo{title}{{Star Formation in the Milky Way and Nearby
  Galaxies}}.
\newblock \emph{\bibinfo{journal}{\araa}} \textbf{\bibinfo{volume}{50}},
  \bibinfo{pages}{531--608} (\bibinfo{year}{2012}).

\bibitem{krumholz14}
\bibinfo{author}{{Krumholz}, M.~R.}
\newblock \bibinfo{title}{{The big problems in star formation: The star
  formation rate, stellar clustering, and the initial mass function}}.
\newblock \emph{\bibinfo{journal}{Phys.~Rep.}} \textbf{\bibinfo{volume}{539}},
  \bibinfo{pages}{49--134} (\bibinfo{year}{2014}).

\bibitem{grand17}
\bibinfo{author}{{Grand}, R.~J.~J.} \emph{et~al.}
\newblock \bibinfo{title}{{The Auriga Project: the properties and formation
  mechanisms of disc galaxies across cosmic time}}.
\newblock \emph{\bibinfo{journal}{\mnras}} \textbf{\bibinfo{volume}{467}},
  \bibinfo{pages}{179--207} (\bibinfo{year}{2017}).

\bibitem{hopkins18}
\bibinfo{author}{{Hopkins}, P.~F.} \emph{et~al.}
\newblock \bibinfo{title}{{FIRE-2 simulations: physics versus numerics in
  galaxy formation}}.
\newblock \emph{\bibinfo{journal}{\mnras}} \textbf{\bibinfo{volume}{480}},
  \bibinfo{pages}{800--863} (\bibinfo{year}{2018}).

\bibitem{bigiel08}
\bibinfo{author}{{Bigiel}, F.} \emph{et~al.}
\newblock \bibinfo{title}{{The Star Formation Law in Nearby Galaxies on Sub-Kpc
  Scales}}.
\newblock \emph{\bibinfo{journal}{\aj}} \textbf{\bibinfo{volume}{136}},
  \bibinfo{pages}{2846--2871} (\bibinfo{year}{2008}).

\bibitem{calzetti12}
\bibinfo{author}{{Calzetti}, D.}, \bibinfo{author}{{Liu}, G.} \&
  \bibinfo{author}{{Koda}, J.}
\newblock \bibinfo{title}{{Star Formation Laws: The Effects of Gas Cloud
  Sampling}}.
\newblock \emph{\bibinfo{journal}{\apj}} \textbf{\bibinfo{volume}{752}},
  \bibinfo{pages}{98} (\bibinfo{year}{2012}).

\bibitem{zuckerman74b}
\bibinfo{author}{{Zuckerman}, B.} \& \bibinfo{author}{{Palmer}, P.}
\newblock \bibinfo{title}{{Radio radiation from interstellar molecules}}.
\newblock \emph{\bibinfo{journal}{\araa}} \textbf{\bibinfo{volume}{12}},
  \bibinfo{pages}{279--313} (\bibinfo{year}{1974}).

\bibitem{koda09}
\bibinfo{author}{{Koda}, J.} \emph{et~al.}
\newblock \bibinfo{title}{{Dynamically Driven Evolution of the Interstellar
  Medium in M51}}.
\newblock \emph{\bibinfo{journal}{\apjl}} \textbf{\bibinfo{volume}{700}},
  \bibinfo{pages}{L132--L136} (\bibinfo{year}{2009}).

\bibitem{leisawitz89}
\bibinfo{author}{{Leisawitz}, D.}, \bibinfo{author}{{Bash}, F.~N.} \&
  \bibinfo{author}{{Thaddeus}, P.}
\newblock \bibinfo{title}{{A CO survey of regions around 34 open clusters}}.
\newblock \emph{\bibinfo{journal}{\apjs}} \textbf{\bibinfo{volume}{70}},
  \bibinfo{pages}{731--812} (\bibinfo{year}{1989}).

\bibitem{elmegreen00}
\bibinfo{author}{{Elmegreen}, B.~G.}
\newblock \bibinfo{title}{{Star Formation in a Crossing Time}}.
\newblock \emph{\bibinfo{journal}{\apj}} \textbf{\bibinfo{volume}{530}},
  \bibinfo{pages}{277--281} (\bibinfo{year}{2000}).

\bibitem{kruijssen14}
\bibinfo{author}{{Kruijssen}, J.~M.~D.} \& \bibinfo{author}{{Longmore}, S.~N.}
\newblock \bibinfo{title}{{An uncertainty principle for star formation - I. Why
  galactic star formation relations break down below a certain spatial scale}}.
\newblock \emph{\bibinfo{journal}{\mnras}} \textbf{\bibinfo{volume}{439}},
  \bibinfo{pages}{3239--3252} (\bibinfo{year}{2014}).

\bibitem{kruijssen18}
\bibinfo{author}{{Kruijssen}, J.~M.~D.} \emph{et~al.}
\newblock \bibinfo{title}{{An uncertainty principle for star formation - II. A
  new method for characterizing the cloud-scale physics of star formation and
  feedback across cosmic history}}.
\newblock \emph{\bibinfo{journal}{\mnras}} \textbf{\bibinfo{volume}{479}},
  \bibinfo{pages}{1866--1952} (\bibinfo{year}{2018}).

\bibitem{schruba10}
\bibinfo{author}{{Schruba}, A.}, \bibinfo{author}{{Leroy}, A.~K.},
  \bibinfo{author}{{Walter}, F.}, \bibinfo{author}{{Sandstrom}, K.} \&
  \bibinfo{author}{{Rosolowsky}, E.}
\newblock \bibinfo{title}{{The Scale Dependence of the Molecular Gas Depletion
  Time in M33}}.
\newblock \emph{\bibinfo{journal}{\apj}} \textbf{\bibinfo{volume}{722}},
  \bibinfo{pages}{1699--1706} (\bibinfo{year}{2010}).

\bibitem{dobbs13}
\bibinfo{author}{{Dobbs}, C.~L.} \& \bibinfo{author}{{Pringle}, J.~E.}
\newblock \bibinfo{title}{{The exciting lives of giant molecular clouds}}.
\newblock \emph{\bibinfo{journal}{\mnras}} \textbf{\bibinfo{volume}{432}},
  \bibinfo{pages}{653--667} (\bibinfo{year}{2013}).

\bibitem{jeffreson18}
\bibinfo{author}{{Jeffreson}, S.~M.~R.} \& \bibinfo{author}{{Kruijssen},
  J.~M.~D.}
\newblock \bibinfo{title}{{A general theory for the lifetimes of giant
  molecular clouds under the influence of galactic dynamics}}.
\newblock \emph{\bibinfo{journal}{\mnras}} \textbf{\bibinfo{volume}{476}},
  \bibinfo{pages}{3688--3715} (\bibinfo{year}{2018}).

\bibitem{tan00}
\bibinfo{author}{{Tan}, J.~C.}
\newblock \bibinfo{title}{{Star Formation Rates in Disk Galaxies and
  Circumnuclear Starbursts from Cloud Collisions}}.
\newblock \emph{\bibinfo{journal}{\apj}} \textbf{\bibinfo{volume}{536}},
  \bibinfo{pages}{173--184} (\bibinfo{year}{2000}).

\bibitem{schaye15}
\bibinfo{author}{{Schaye}, J.} \emph{et~al.}
\newblock \bibinfo{title}{{The EAGLE project: simulating the evolution and
  assembly of galaxies and their environments}}.
\newblock \emph{\bibinfo{journal}{\mnras}} \textbf{\bibinfo{volume}{446}},
  \bibinfo{pages}{521--554} (\bibinfo{year}{2015}).

\bibitem{pillepich18}
\bibinfo{author}{{Pillepich}, A.} \emph{et~al.}
\newblock \bibinfo{title}{{Simulating galaxy formation with the IllustrisTNG
  model}}.
\newblock \emph{\bibinfo{journal}{\mnras}} \textbf{\bibinfo{volume}{473}},
  \bibinfo{pages}{4077--4106} (\bibinfo{year}{2018}).

\bibitem{mckee77}
\bibinfo{author}{{McKee}, C.~F.} \& \bibinfo{author}{{Ostriker}, J.~P.}
\newblock \bibinfo{title}{{A theory of the interstellar medium - Three
  components regulated by supernova explosions in an inhomogeneous substrate}}.
\newblock \emph{\bibinfo{journal}{\apj}} \textbf{\bibinfo{volume}{218}},
  \bibinfo{pages}{148--169} (\bibinfo{year}{1977}).

\bibitem{hopkins12}
\bibinfo{author}{{Hopkins}, P.~F.}, \bibinfo{author}{{Quataert}, E.} \&
  \bibinfo{author}{{Murray}, N.}
\newblock \bibinfo{title}{{The structure of the interstellar medium of
  star-forming galaxies}}.
\newblock \emph{\bibinfo{journal}{\mnras}} \textbf{\bibinfo{volume}{421}},
  \bibinfo{pages}{3488--3521} (\bibinfo{year}{2012}).

\bibitem{vutisalchavakul14}
\bibinfo{author}{{Vutisalchavakul}, N.}, \bibinfo{author}{{Evans}, N.~J., II}
  \& \bibinfo{author}{{Battersby}, C.}
\newblock \bibinfo{title}{{The Star-formation Relation for Regions in the
  Galactic Plane: The Effect of Spatial Resolution}}.
\newblock \emph{\bibinfo{journal}{\apj}} \textbf{\bibinfo{volume}{797}},
  \bibinfo{pages}{77} (\bibinfo{year}{2014}).

\bibitem{bolatto13}
\bibinfo{author}{{Bolatto}, A.~D.} \emph{et~al.}
\newblock \bibinfo{title}{{Suppression of star formation in the galaxy NGC 253
  by a starburst-driven molecular wind}}.
\newblock \emph{\bibinfo{journal}{\nat}} \textbf{\bibinfo{volume}{499}},
  \bibinfo{pages}{450--453} (\bibinfo{year}{2013}).

\bibitem{muratov15}
\bibinfo{author}{{Muratov}, A.~L.} \emph{et~al.}
\newblock \bibinfo{title}{{Gusty, gaseous flows of FIRE: galactic winds in
  cosmological simulations with explicit stellar feedback}}.
\newblock \emph{\bibinfo{journal}{\mnras}} \textbf{\bibinfo{volume}{454}},
  \bibinfo{pages}{2691--2713} (\bibinfo{year}{2015}).

\bibitem{westmeier11}
\bibinfo{author}{{Westmeier}, T.}, \bibinfo{author}{{Braun}, R.} \&
  \bibinfo{author}{{Koribalski}, B.~S.}
\newblock \bibinfo{title}{{Gas and dark matter in the Sculptor group: NGC
  300}}.
\newblock \emph{\bibinfo{journal}{\mnras}} \textbf{\bibinfo{volume}{410}},
  \bibinfo{pages}{2217--2236} (\bibinfo{year}{2011}).

\bibitem{agertz13}
\bibinfo{author}{{Agertz}, O.}, \bibinfo{author}{{Kravtsov}, A.~V.},
  \bibinfo{author}{{Leitner}, S.~N.} \& \bibinfo{author}{{Gnedin}, N.~Y.}
\newblock \bibinfo{title}{{Toward a Complete Accounting of Energy and Momentum
  from Stellar Feedback in Galaxy Formation Simulations}}.
\newblock \emph{\bibinfo{journal}{\apj}} \textbf{\bibinfo{volume}{770}},
  \bibinfo{pages}{25} (\bibinfo{year}{2013}).

\bibitem{genzel11}
\bibinfo{author}{{Genzel}, R.} \emph{et~al.}
\newblock \bibinfo{title}{{The Sins Survey of z$\sim$2 Galaxy Kinematics:
  Properties of the Giant Star-forming Clumps}}.
\newblock \emph{\bibinfo{journal}{\apj}} \textbf{\bibinfo{volume}{733}},
  \bibinfo{pages}{101} (\bibinfo{year}{2011}).

\end{thebibliography}

\begin{thebibliography}{10}
\setcounter{enumiv}{30}

\expandafter\ifx\csname url\endcsname\relax
  \def\url#1{\texttt{#1}}\fi
\expandafter\ifx\csname urlprefix\endcsname\relax\def\urlprefix{URL }\fi
\providecommand{\bibinfo}[2]{#2}
\providecommand{\eprint}[2][]{\url{#2}}

\bibitem{bolatto13b}
\bibinfo{author}{{Bolatto}, A.~D.}, \bibinfo{author}{{Wolfire}, M.} \&
  \bibinfo{author}{{Leroy}, A.~K.}
\newblock \bibinfo{title}{{The CO-to-H$_{2}$ Conversion Factor}}.
\newblock \emph{\bibinfo{journal}{\araa}} \textbf{\bibinfo{volume}{51}},
  \bibinfo{pages}{207--268} (\bibinfo{year}{2013}).

\bibitem{wong11}
\bibinfo{author}{{Wong}, T.} \emph{et~al.}
\newblock \bibinfo{title}{{The Magellanic Mopra Assessment (MAGMA). I. The
  Molecular Cloud Population of the Large Magellanic Cloud}}.
\newblock \emph{\bibinfo{journal}{\apjs}} \textbf{\bibinfo{volume}{197}},
  \bibinfo{pages}{16} (\bibinfo{year}{2011}).

\bibitem{druard14}
\bibinfo{author}{{Druard}, C.} \emph{et~al.}
\newblock \bibinfo{title}{{The IRAM M 33 CO(2-1) survey. A complete census of
  molecular gas out to 7 kpc}}.
\newblock \emph{\bibinfo{journal}{\aap}} \textbf{\bibinfo{volume}{567}},
  \bibinfo{pages}{A118} (\bibinfo{year}{2014}).

\bibitem{faesi14}
\bibinfo{author}{{Faesi}, C.~M.}, \bibinfo{author}{{Lada}, C.~J.},
  \bibinfo{author}{{Forbrich}, J.}, \bibinfo{author}{{Menten}, K.~M.} \&
  \bibinfo{author}{{Bouy}, H.}
\newblock \bibinfo{title}{{Molecular Cloud-scale Star Formation in NGC 300}}.
\newblock \emph{\bibinfo{journal}{\apj}} \textbf{\bibinfo{volume}{789}},
  \bibinfo{pages}{81} (\bibinfo{year}{2014}).

\bibitem{roussel05}
\bibinfo{author}{{Roussel}, H.} \emph{et~al.}
\newblock \bibinfo{title}{{Extinction Law Variations and Dust Excitation in the
  Spiral Galaxy NGC 300}}.
\newblock \emph{\bibinfo{journal}{\apj}} \textbf{\bibinfo{volume}{632}},
  \bibinfo{pages}{227--252} (\bibinfo{year}{2005}).

\bibitem{luridiana15}
\bibinfo{author}{{Luridiana}, V.}, \bibinfo{author}{{Morisset}, C.} \&
  \bibinfo{author}{{Shaw}, R.~A.}
\newblock \bibinfo{title}{{PyNeb: a new tool for analyzing emission lines. I.
  Code description and validation of results}}.
\newblock \emph{\bibinfo{journal}{\aap}} \textbf{\bibinfo{volume}{573}},
  \bibinfo{pages}{A42} (\bibinfo{year}{2015}).

\bibitem{leitherer99}
\bibinfo{author}{{Leitherer}, C.} \emph{et~al.}
\newblock \bibinfo{title}{{Starburst99: Synthesis Models for Galaxies with
  Active Star Formation}}.
\newblock \emph{\bibinfo{journal}{\apjs}} \textbf{\bibinfo{volume}{123}},
  \bibinfo{pages}{3--40} (\bibinfo{year}{1999}).

\bibitem{haydon18}
\bibinfo{author}{{Haydon}, D.~T.} \emph{et~al.}
\newblock \bibinfo{title}{{An uncertainty principle for star formation -- III.
  The characteristic time-scales of star formation rate tracers}}.
\newblock \emph{\bibinfo{journal}{\mnras~submitted, arXiv:1810.10897}}
  (\bibinfo{year}{2018}).

\bibitem{murphy11}
\bibinfo{author}{{Murphy}, E.~J.} \emph{et~al.}
\newblock \bibinfo{title}{{Calibrating Extinction-free Star Formation Rate
  Diagnostics with 33 GHz Free-free Emission in NGC 6946}}.
\newblock \emph{\bibinfo{journal}{\apj}} \textbf{\bibinfo{volume}{737}},
  \bibinfo{pages}{67} (\bibinfo{year}{2011}).

\bibitem{kroupa01}
\bibinfo{author}{{Kroupa}, P.}
\newblock \bibinfo{title}{{On the variation of the initial mass function}}.
\newblock \emph{\bibinfo{journal}{\mnras}} \textbf{\bibinfo{volume}{322}},
  \bibinfo{pages}{231--246} (\bibinfo{year}{2001}).

\bibitem{kang16}
\bibinfo{author}{{Kang}, X.}, \bibinfo{author}{{Zhang}, F.}, \bibinfo{author}{{Chang}, R.}, \bibinfo{author}{{Wang}, L.} \& \bibinfo{author}{{Cheng}, L.}
\newblock \bibinfo{title}{{The star formation history of low-mass disk galaxies: A case study of NGC 300}}.
\newblock \emph{\bibinfo{journal}{\aap}} \textbf{\bibinfo{volume}{585}},
  \bibinfo{pages}{A20} (\bibinfo{year}{2016}).

\bibitem{dalcanton09}
\bibinfo{author}{{Dalcanton}, J.~J.} \emph{et~al.}
\newblock \bibinfo{title}{{The ACS Nearby Galaxy Survey Treasury}}.
\newblock \emph{\bibinfo{journal}{\apjs}} \textbf{\bibinfo{volume}{183}},
  \bibinfo{pages}{67--108} (\bibinfo{year}{2009}).

\bibitem{koch18}
\bibinfo{author}{{Koch}, E.~W.} \emph{et~al.}
\newblock \bibinfo{title}{{Kinematics of the atomic ISM ifigun M33 on 80 pc
  scales}}.
\newblock \emph{\bibinfo{journal}{\mnras}} \textbf{\bibinfo{volume}{479}},
  \bibinfo{pages}{2505--2533} (\bibinfo{year}{2018}).

\bibitem{dale09}
\bibinfo{author}{{Dale}, D.~A.} \emph{et~al.}
\newblock \bibinfo{title}{{The Spitzer Local Volume Legacy: Survey Description
  and Infrared Photometry}}.
\newblock \emph{\bibinfo{journal}{\apj}} \textbf{\bibinfo{volume}{703}},
  \bibinfo{pages}{517--556} (\bibinfo{year}{2009}).

\bibitem{querejeta15}
\bibinfo{author}{{Querejeta}, M.} \emph{et~al.}
\newblock \bibinfo{title}{{The Spitzer Survey of Stellar Structure in Galaxies
  (S$^{4}$G): Precise Stellar Mass Distributions from Automated Dust Correction
  at 3.6 {$\mu$}m}}.
\newblock \emph{\bibinfo{journal}{\apjs}} \textbf{\bibinfo{volume}{219}},
  \bibinfo{pages}{5} (\bibinfo{year}{2015}).

\bibitem{meidt12}
\bibinfo{author}{{Meidt}, S.~E.} \emph{et~al.}
\newblock \bibinfo{title}{{Reconstructing the Stellar Mass Distributions of
  Galaxies Using S$^{4}$G IRAC 3.6 and 4.5 {$\mu$}m Images. I. Correcting for
  Contamination by Polycyclic Aromatic Hydrocarbons, Hot Dust, and
  Intermediate-age Stars}}.
\newblock \emph{\bibinfo{journal}{\apj}} \textbf{\bibinfo{volume}{744}},
  \bibinfo{pages}{17} (\bibinfo{year}{2012}).

\bibitem{leroy08}
\bibinfo{author}{{Leroy}, A.~K.} \emph{et~al.}
\newblock \bibinfo{title}{{The Star Formation Efficiency in Nearby Galaxies:
  Measuring Where Gas Forms Stars Effectively}}.
\newblock \emph{\bibinfo{journal}{\aj}} \textbf{\bibinfo{volume}{136}},
  \bibinfo{pages}{2782--2845} (\bibinfo{year}{2008}).

\bibitem{meidt14}
\bibinfo{author}{{Meidt}, S.~E.} \emph{et~al.}
\newblock \bibinfo{title}{{Reconstructing the Stellar Mass Distributions of
  Galaxies Using S$^{4}$G IRAC 3.6 and 4.5 {$\mu$}m Images. II. The Conversion
  from Light to Mass}}.
\newblock \emph{\bibinfo{journal}{\apj}} \textbf{\bibinfo{volume}{788}},
  \bibinfo{pages}{144} (\bibinfo{year}{2014}).

\bibitem{faesi18}
\bibinfo{author}{{Faesi}, C.~M.}, \bibinfo{author}{{Lada}, C.~J.} \&
  \bibinfo{author}{{Forbrich}, J.}
\newblock \bibinfo{title}{{The ALMA View of GMCs in NGC 300: Physical
  Properties and Scaling Relations at 10 pc Resolution}}.
\newblock \emph{\bibinfo{journal}{\apj}} \textbf{\bibinfo{volume}{857}},
  \bibinfo{pages}{19} (\bibinfo{year}{2018}).

\bibitem{vanderkruit11}
\bibinfo{author}{{van der Kruit}, P.~C.} \& \bibinfo{author}{{Freeman}, K.~C.}
\newblock \bibinfo{title}{{Galaxy Disks}}.
\newblock \emph{\bibinfo{journal}{\araa}} \textbf{\bibinfo{volume}{49}},
  \bibinfo{pages}{301--371} (\bibinfo{year}{2011}).

\bibitem{gogarten10}
\bibinfo{author}{{Gogarten}, S.~M.} \emph{et~al.}
\newblock \bibinfo{title}{{The Advanced Camera for Surveys Nearby Galaxy Survey
  Treasury. V. Radial Star Formation History of NGC 300}}.
\newblock \emph{\bibinfo{journal}{\apj}} \textbf{\bibinfo{volume}{712}},
  \bibinfo{pages}{858--874} (\bibinfo{year}{2010}).

\bibitem{bresolin09}
\bibinfo{author}{{Bresolin}, F.} \emph{et~al.}
\newblock \bibinfo{title}{{Extragalactic Chemical Abundances: Do H II Regions
  and Young Stars Tell the Same Story? The Case of the Spiral Galaxy NGC 300}}.
\newblock \emph{\bibinfo{journal}{\apj}} \textbf{\bibinfo{volume}{700}},
  \bibinfo{pages}{309--330} (\bibinfo{year}{2009}).

\bibitem{saintonge17}
\bibinfo{author}{{Saintonge}, A.} \emph{et~al.}
\newblock \bibinfo{title}{{xCOLD GASS: The Complete IRAM 30 m Legacy Survey of
  Molecular Gas for Galaxy Evolution Studies}}.
\newblock \emph{\bibinfo{journal}{\apjs}} \textbf{\bibinfo{volume}{233}},
  \bibinfo{pages}{22} (\bibinfo{year}{2017}).

\bibitem{catinella18}
\bibinfo{author}{{Catinella}, B.} \emph{et~al.}
\newblock \bibinfo{title}{{xGASS: total cold gas scaling relations and
  molecular-to-atomic gas ratios of galaxies in the local Universe}}.
\newblock \emph{\bibinfo{journal}{\mnras}} \textbf{\bibinfo{volume}{476}},
  \bibinfo{pages}{875--895} (\bibinfo{year}{2018}).

\bibitem{moster13}
\bibinfo{author}{{Moster}, B.~P.}, \bibinfo{author}{{Naab}, T.} \&
  \bibinfo{author}{{White}, S.~D.~M.}
\newblock \bibinfo{title}{{Galactic star formation and accretion histories from
  matching galaxies to dark matter haloes}}.
\newblock \emph{\bibinfo{journal}{\mnras}} \textbf{\bibinfo{volume}{428}},
  \bibinfo{pages}{3121--3138} (\bibinfo{year}{2013}).

\bibitem{blandhawthorn16}
\bibinfo{author}{{Bland-Hawthorn}, J.} \& \bibinfo{author}{{Gerhard}, O.}
\newblock \bibinfo{title}{{The Galaxy in Context: Structural, Kinematic, and
  Integrated Properties}}.
\newblock \emph{\bibinfo{journal}{\araa}} \textbf{\bibinfo{volume}{54}},
  \bibinfo{pages}{529--596} (\bibinfo{year}{2016}).

\bibitem{rosolowsky06}
\bibinfo{author}{{Rosolowsky}, E.} \& \bibinfo{author}{{Leroy}, A.}
\newblock \bibinfo{title}{{Bias-free Measurement of Giant Molecular Cloud
  Properties}}.
\newblock \emph{\bibinfo{journal}{\pasp}} \textbf{\bibinfo{volume}{118}},
  \bibinfo{pages}{590--610} (\bibinfo{year}{2006}).

\bibitem{leroy15}
\bibinfo{author}{{Leroy}, A.~K.} \emph{et~al.}
\newblock \bibinfo{title}{{ALMA Reveals the Molecular Medium Fueling the
  Nearest Nuclear Starburst}}.
\newblock \emph{\bibinfo{journal}{\apj}} \textbf{\bibinfo{volume}{801}},
  \bibinfo{pages}{25} (\bibinfo{year}{2015}).

\bibitem{solomon87}
\bibinfo{author}{{Solomon}, P.~M.}, \bibinfo{author}{{Rivolo}, A.~R.},
  \bibinfo{author}{{Barrett}, J.} \& \bibinfo{author}{{Yahil}, A.}
\newblock \bibinfo{title}{{Mass, luminosity, and line width relations of
  Galactic molecular clouds}}.
\newblock \emph{\bibinfo{journal}{\apj}} \textbf{\bibinfo{volume}{319}},
  \bibinfo{pages}{730--741} (\bibinfo{year}{1987}).

\bibitem{williams94}
\bibinfo{author}{{Williams}, J.~P.}, \bibinfo{author}{{de Geus}, E.~J.} \&
  \bibinfo{author}{{Blitz}, L.}
\newblock \bibinfo{title}{{Determining structure in molecular clouds}}.
\newblock \emph{\bibinfo{journal}{\apj}} \textbf{\bibinfo{volume}{428}},
  \bibinfo{pages}{693--712} (\bibinfo{year}{1994}).

\bibitem{dasilva12}
\bibinfo{author}{{da Silva}, R.~L.}, \bibinfo{author}{{Fumagalli}, M.} \&
  \bibinfo{author}{{Krumholz}, M.}
\newblock \bibinfo{title}{{SLUG---Stochastically Lighting Up Galaxies. I.
  Methods and Validating Tests}}.
\newblock \emph{\bibinfo{journal}{\apj}} \textbf{\bibinfo{volume}{745}},
  \bibinfo{pages}{145} (\bibinfo{year}{2012}).

\bibitem{krumholz15b}
\bibinfo{author}{{Krumholz}, M.~R.}, \bibinfo{author}{{Fumagalli}, M.},
  \bibinfo{author}{{da Silva}, R.~L.}, \bibinfo{author}{{Rendahl}, T.} \&
  \bibinfo{author}{{Parra}, J.}
\newblock \bibinfo{title}{{SLUG - stochastically lighting up galaxies - III. A
  suite of tools for simulated photometry, spectroscopy, and Bayesian inference
  with stochastic stellar populations}}.
\newblock \emph{\bibinfo{journal}{\mnras}} \textbf{\bibinfo{volume}{452}},
  \bibinfo{pages}{1447--1467} (\bibinfo{year}{2015}).

\bibitem{hygate18}
\bibinfo{author}{{Hygate}, A.~P.~S.} \emph{et~al.}
\newblock \bibinfo{title}{{An uncertainty principle for star formation -- IV.
  On the nature and filtering of diffuse emission}}.
\newblock \emph{\bibinfo{journal}{\mnras~submitted, arXiv:1810.11405}}
  (\bibinfo{year}{2018}).

\bibitem{schruba11}
\bibinfo{author}{{Schruba}, A.} \emph{et~al.}
\newblock \bibinfo{title}{{A Molecular Star Formation Law in the
  Atomic-gas-dominated Regime in Nearby Galaxies}}.
\newblock \emph{\bibinfo{journal}{\aj}} \textbf{\bibinfo{volume}{142}},
  \bibinfo{pages}{37} (\bibinfo{year}{2011}).

\bibitem{leroy13}
\bibinfo{author}{{Leroy}, A.~K.} \emph{et~al.}
\newblock \bibinfo{title}{{Molecular Gas and Star Formation in nearby Disk
  Galaxies}}.
\newblock \emph{\bibinfo{journal}{\aj}} \textbf{\bibinfo{volume}{146}},
  \bibinfo{pages}{19} (\bibinfo{year}{2013}).

\bibitem{pety13}
\bibinfo{author}{{Pety}, J.} \emph{et~al.}
\newblock \bibinfo{title}{{The Plateau de Bure + 30 m Arcsecond Whirlpool
  Survey Reveals a Thick Disk of Diffuse Molecular Gas in the M51 Galaxy}}.
\newblock \emph{\bibinfo{journal}{\apj}} \textbf{\bibinfo{volume}{779}},
  \bibinfo{pages}{43} (\bibinfo{year}{2013}).

\bibitem{krumholz12}
\bibinfo{author}{{Krumholz}, M.~R.}, \bibinfo{author}{{Dekel}, A.} \&
  \bibinfo{author}{{McKee}, C.~F.}
\newblock \bibinfo{title}{{A Universal, Local Star Formation Law in Galactic
  Clouds, nearby Galaxies, High-redshift Disks, and Starbursts}}.
\newblock \emph{\bibinfo{journal}{\apj}} \textbf{\bibinfo{volume}{745}},
  \bibinfo{pages}{69} (\bibinfo{year}{2012}).

\bibitem{elmegreen87}
\bibinfo{author}{{Elmegreen}, B.~G.}
\newblock \bibinfo{title}{{Supercloud formation by nonaxisymmetric
  gravitational instabilities in sheared magnetic galaxy disks}}.
\newblock \emph{\bibinfo{journal}{\apj}} \textbf{\bibinfo{volume}{312}},
  \bibinfo{pages}{626--639} (\bibinfo{year}{1987}).

\bibitem{leitherer14}
\bibinfo{author}{{Leitherer}, C.} \emph{et~al.}
\newblock \bibinfo{title}{{The Effects of Stellar Rotation. II. A Comprehensive
  Set of Starburst99 Models}}.
\newblock \emph{\bibinfo{journal}{\apjs}} \textbf{\bibinfo{volume}{212}},
  \bibinfo{pages}{14} (\bibinfo{year}{2014}).

\bibitem{spitzer78}
\bibinfo{author}{{Spitzer}, L.}
\newblock \emph{\bibinfo{title}{{Physical processes in the interstellar
  medium}}} (\bibinfo{publisher}{Wiley-Interscience}, \bibinfo{year}{1978}).

\bibitem{hosokawa06}
\bibinfo{author}{{Hosokawa}, T.} \& \bibinfo{author}{{Inutsuka}, S.-i.}
\newblock \bibinfo{title}{{Dynamical Expansion of Ionization and Dissociation
  Front around a Massive Star. II. On the Generality of Triggered Star
  Formation}}.
\newblock \emph{\bibinfo{journal}{\apj}} \textbf{\bibinfo{volume}{646}},
  \bibinfo{pages}{240--257} (\bibinfo{year}{2006}).

\bibitem{tielens05}
\bibinfo{author}{{Tielens}, A.~G.~G.~M.}
\newblock \emph{\bibinfo{title}{{The Physics and Chemistry of the Interstellar
  Medium}}} (\bibinfo{publisher}{Cambridge University Press},
  \bibinfo{year}{2005}).

\bibitem{weaver77}
\bibinfo{author}{{Weaver}, R.}, \bibinfo{author}{{McCray}, R.},
  \bibinfo{author}{{Castor}, J.}, \bibinfo{author}{{Shapiro}, P.} \&
  \bibinfo{author}{{Moore}, R.}
\newblock \bibinfo{title}{{Interstellar bubbles. II - Structure and
  evolution}}.
\newblock \emph{\bibinfo{journal}{\apj}} \textbf{\bibinfo{volume}{218}},
  \bibinfo{pages}{377--395} (\bibinfo{year}{1977}).

\bibitem{maclow88}
\bibinfo{author}{{Mac Low}, M.-M.} \& \bibinfo{author}{{McCray}, R.}
\newblock \bibinfo{title}{{Superbubbles in disk galaxies}}.
\newblock \emph{\bibinfo{journal}{\apj}} \textbf{\bibinfo{volume}{324}},
  \bibinfo{pages}{776--785} (\bibinfo{year}{1988}).

\bibitem{thompson05}
\bibinfo{author}{{Thompson}, T.~A.}, \bibinfo{author}{{Quataert}, E.} \&
  \bibinfo{author}{{Murray}, N.}
\newblock \bibinfo{title}{{Radiation Pressure-supported Starburst Disks and
  Active Galactic Nucleus Fueling}}.
\newblock \emph{\bibinfo{journal}{\apj}} \textbf{\bibinfo{volume}{630}},
  \bibinfo{pages}{167--185} (\bibinfo{year}{2005}).

\bibitem{krumholz12b}
\bibinfo{author}{{Krumholz}, M.~R.} \& \bibinfo{author}{{Thompson}, T.~A.}
\newblock \bibinfo{title}{{Direct Numerical Simulation of Radiation
  Pressure-driven Turbulence and Winds in Star Clusters and Galactic Disks}}.
\newblock \emph{\bibinfo{journal}{\apj}} \textbf{\bibinfo{volume}{760}},
  \bibinfo{pages}{155} (\bibinfo{year}{2012}).

\bibitem{rosen14}
\bibinfo{author}{{Rosen}, A.~L.}, \bibinfo{author}{{Lopez}, L.~A.},
  \bibinfo{author}{{Krumholz}, M.~R.} \& \bibinfo{author}{{Ramirez-Ruiz}, E.}
\newblock \bibinfo{title}{{Gone with the wind: Where is the missing stellar
  wind energy from massive star clusters?}}
\newblock \emph{\bibinfo{journal}{\mnras}} \textbf{\bibinfo{volume}{442}},
  \bibinfo{pages}{2701--2716} (\bibinfo{year}{2014}).

\bibitem{chevance16}
\bibinfo{author}{{Chevance}, M.} \emph{et~al.}
\newblock \bibinfo{title}{{A milestone toward understanding PDR properties in
  the extreme environment of LMC-30 Doradus}}.
\newblock \emph{\bibinfo{journal}{\aap}} \textbf{\bibinfo{volume}{590}},
  \bibinfo{pages}{A36} (\bibinfo{year}{2016}).

\bibitem{toomre64}
\bibinfo{author}{{Toomre}, A.}
\newblock \bibinfo{title}{{On the gravitational stability of a disk of stars}}.
\newblock \emph{\bibinfo{journal}{\apj}} \textbf{\bibinfo{volume}{139}},
  \bibinfo{pages}{1217--1238} (\bibinfo{year}{1964}).

\bibitem{schruba17}
\bibinfo{author}{{Schruba}, A.} \emph{et~al.}
\newblock \bibinfo{title}{{Physical Properties of Molecular Clouds at 2 pc
  Resolution in the Low-metallicity Dwarf Galaxy NGC 6822 and the Milky Way}}.
\newblock \emph{\bibinfo{journal}{\apj}} \textbf{\bibinfo{volume}{835}},
  \bibinfo{pages}{278} (\bibinfo{year}{2017}).

\bibitem{hughes13}
\bibinfo{author}{{Hughes}, A.} \emph{et~al.}
\newblock \bibinfo{title}{{A Comparative Study of Giant Molecular Clouds in
  M51, M33, and the Large Magellanic Cloud}}.
\newblock \emph{\bibinfo{journal}{\apj}} \textbf{\bibinfo{volume}{779}},
  \bibinfo{pages}{46} (\bibinfo{year}{2013}).

\bibitem{schruba18}
\bibinfo{author}{{Schruba}, A.}, \bibinfo{author}{{Kruijssen}, J.~M.~D.} \&
  \bibinfo{author}{{Leroy}, A.~K.}
\newblock \bibinfo{title}{{How Galactic Environment affects the Dynamical State
  of Molecular Clouds and their Star Formation Efficiency}}.
\newblock \emph{\bibinfo{journal}{\apj~submitted}}  (\bibinfo{year}{2018}).

\bibitem{clark05}
\bibinfo{author}{{Clark}, P.~C.}, \bibinfo{author}{{Bonnell}, I.~A.},
  \bibinfo{author}{{Zinnecker}, H.} \& \bibinfo{author}{{Bate}, M.~R.}
\newblock \bibinfo{title}{{Star formation in unbound giant molecular clouds:
  the origin of OB associations?}}
\newblock \emph{\bibinfo{journal}{\mnras}} \textbf{\bibinfo{volume}{359}},
  \bibinfo{pages}{809--818} (\bibinfo{year}{2005}).

\bibitem{padoan12}
\bibinfo{author}{{Padoan}, P.}, \bibinfo{author}{{Haugb{\o}lle}, T.} \&
  \bibinfo{author}{{Nordlund}, {\AA}.}
\newblock \bibinfo{title}{{A Simple Law of Star Formation}}.
\newblock \emph{\bibinfo{journal}{\apjl}} \textbf{\bibinfo{volume}{759}},
  \bibinfo{pages}{L27} (\bibinfo{year}{2012}).

\bibitem{dale13}
\bibinfo{author}{{Dale}, J.~E.}, \bibinfo{author}{{Ercolano}, B.} \&
  \bibinfo{author}{{Bonnell}, I.~A.}
\newblock \bibinfo{title}{{Ionizing feedback from massive stars in massive
  clusters - III. Disruption of partially unbound clouds}}.
\newblock \emph{\bibinfo{journal}{\mnras}} \textbf{\bibinfo{volume}{430}},
  \bibinfo{pages}{234--246} (\bibinfo{year}{2013}).

\bibitem{rathborne15}
\bibinfo{author}{{Rathborne}, J.~M.} \emph{et~al.}
\newblock \bibinfo{title}{{A Cluster in the Making: ALMA Reveals the Initial
  Conditions for High-mass Cluster Formation}}.
\newblock \emph{\bibinfo{journal}{\apj}} \textbf{\bibinfo{volume}{802}},
  \bibinfo{pages}{125} (\bibinfo{year}{2015}).

\bibitem{scoville79}
\bibinfo{author}{{Scoville}, N.~Z.}, \bibinfo{author}{{Solomon}, P.~M.} \&
  \bibinfo{author}{{Sanders}, D.~B.}
\newblock \bibinfo{title}{{CO observations of spiral structure and the lifetime
  of giant molecular clouds}}.
\newblock In \bibinfo{editor}{{Burton}, W.~B.} (ed.)
  \emph{\bibinfo{booktitle}{The Large-Scale Characteristics of the Galaxy}},
  vol.~\bibinfo{volume}{84} of \emph{\bibinfo{series}{IAU Symposium}},
  \bibinfo{pages}{277--282} (\bibinfo{year}{1979}).

\bibitem{sanders85}
\bibinfo{author}{{Sanders}, D.~B.}, \bibinfo{author}{{Scoville}, N.~Z.} \&
  \bibinfo{author}{{Solomon}, P.~M.}
\newblock \bibinfo{title}{{Giant molecular clouds in the Galaxy. II -
  Characteristics of discrete features}}.
\newblock \emph{\bibinfo{journal}{\apj}} \textbf{\bibinfo{volume}{289}},
  \bibinfo{pages}{373--387} (\bibinfo{year}{1985}).

\bibitem{hartmann01}
\bibinfo{author}{{Hartmann}, L.}, \bibinfo{author}{{Ballesteros-Paredes}, J.}
  \& \bibinfo{author}{{Bergin}, E.~A.}
\newblock \bibinfo{title}{{Rapid Formation of Molecular Clouds and Stars in the
  Solar Neighborhood}}.
\newblock \emph{\bibinfo{journal}{\apj}} \textbf{\bibinfo{volume}{562}},
  \bibinfo{pages}{852--868} (\bibinfo{year}{2001}).

\bibitem{engargiola03}
\bibinfo{author}{{Engargiola}, G.}, \bibinfo{author}{{Plambeck}, R.~L.},
  \bibinfo{author}{{Rosolowsky}, E.} \& \bibinfo{author}{{Blitz}, L.}
\newblock \bibinfo{title}{{Giant Molecular Clouds in M33. I. BIMA All-Disk
  Survey}}.
\newblock \emph{\bibinfo{journal}{\apjs}} \textbf{\bibinfo{volume}{149}},
  \bibinfo{pages}{343--363} (\bibinfo{year}{2003}).

\bibitem{blitz07}
\bibinfo{author}{{Blitz}, L.} \emph{et~al.}
\newblock \bibinfo{title}{{Giant Molecular Clouds in Local Group Galaxies}}.
\newblock \emph{\bibinfo{journal}{Protostars and Planets V}}
  \bibinfo{pages}{81--96} (\bibinfo{year}{2007}).

\bibitem{kawamura09}
\bibinfo{author}{{Kawamura}, A.} \emph{et~al.}
\newblock \bibinfo{title}{{The Second Survey of the Molecular Clouds in the
  Large Magellanic Cloud by NANTEN. II. Star Formation}}.
\newblock \emph{\bibinfo{journal}{\apjs}} \textbf{\bibinfo{volume}{184}},
  \bibinfo{pages}{1--17} (\bibinfo{year}{2009}).

\bibitem{murray11}
\bibinfo{author}{{Murray}, N.}
\newblock \bibinfo{title}{{Star Formation Efficiencies and Lifetimes of Giant
  Molecular Clouds in the Milky Way}}.
\newblock \emph{\bibinfo{journal}{\apj}} \textbf{\bibinfo{volume}{729}},
  \bibinfo{pages}{133} (\bibinfo{year}{2011}).

\bibitem{dobbs14}
\bibinfo{author}{{Dobbs}, C.~L.} \emph{et~al.}
\newblock \bibinfo{title}{{Formation of Molecular Clouds and Global Conditions
  for Star Formation}}.
\newblock \emph{\bibinfo{journal}{Protostars and Planets VI}}
  \bibinfo{pages}{3--26} (\bibinfo{year}{2014}).

\bibitem{meidt15}
\bibinfo{author}{{Meidt}, S.~E.} \emph{et~al.}
\newblock \bibinfo{title}{{Short GMC Lifetimes: An Observational Estimate with
  the PdBI Arcsecond Whirlpool Survey (PAWS)}}.
\newblock \emph{\bibinfo{journal}{\apj}} \textbf{\bibinfo{volume}{806}},
  \bibinfo{pages}{72} (\bibinfo{year}{2015}).

\bibitem{kruijssen15}
\bibinfo{author}{{Kruijssen}, J.~M.~D.}, \bibinfo{author}{{Dale}, J.~E.} \&
  \bibinfo{author}{{Longmore}, S.~N.}
\newblock \bibinfo{title}{{The dynamical evolution of molecular clouds near the
  Galactic Centre - I. Orbital structure and evolutionary timeline}}.
\newblock \emph{\bibinfo{journal}{\mnras}} \textbf{\bibinfo{volume}{447}},
  \bibinfo{pages}{1059--1079} (\bibinfo{year}{2015}).

\bibitem{corbelli17}
\bibinfo{author}{{Corbelli}, E.} \emph{et~al.}
\newblock \bibinfo{title}{{From molecules to young stellar clusters: the star
  formation cycle across the disk of M 33}}.
\newblock \emph{\bibinfo{journal}{\aap}} \textbf{\bibinfo{volume}{601}},
  \bibinfo{pages}{A146} (\bibinfo{year}{2017}).

\bibitem{barnes17}
\bibinfo{author}{{Barnes}, A.~T.} \emph{et~al.}
\newblock \bibinfo{title}{{Star formation rates and efficiencies in the
  Galactic Centre}}.
\newblock \emph{\bibinfo{journal}{\mnras}} \textbf{\bibinfo{volume}{469}},
  \bibinfo{pages}{2263--2285} (\bibinfo{year}{2017}).

\bibitem{kreckel18}
\bibinfo{author}{{Kreckel}, K.} \emph{et~al.}
\newblock \bibinfo{title}{{A 50 pc Scale View of Star Formation Efficiency
  across NGC 628}}.
\newblock \emph{\bibinfo{journal}{\apjl}} \textbf{\bibinfo{volume}{863}},
  \bibinfo{pages}{L21} (\bibinfo{year}{2018}).

\bibitem{grasha18}
\bibinfo{author}{{Grasha}, K.} \emph{et~al.}
\newblock \bibinfo{title}{{Connecting young star clusters to CO molecular gas
  in NGC 7793 with ALMA-LEGUS}}.
\newblock \emph{\bibinfo{journal}{\mnras}} \textbf{\bibinfo{volume}{481}},
  \bibinfo{pages}{1016--1027} (\bibinfo{year}{2018}).

\bibitem{efremov98}
\bibinfo{author}{{Efremov}, Y.~N.} \& \bibinfo{author}{{Elmegreen}, B.~G.}
\newblock \bibinfo{title}{{Hierarchical star formation from the time-space
  distribution of star clusters in the Large Magellanic Cloud}}.
\newblock \emph{\bibinfo{journal}{\mnras}} \textbf{\bibinfo{volume}{299}},
  \bibinfo{pages}{588--594} (\bibinfo{year}{1998}).

\bibitem{elmegreen96}
\bibinfo{author}{{Elmegreen}, B.~G.} \& \bibinfo{author}{{Falgarone}, E.}
\newblock \bibinfo{title}{{A Fractal Origin for the Mass Spectrum of
  Interstellar Clouds}}.
\newblock \emph{\bibinfo{journal}{\apj}} \textbf{\bibinfo{volume}{471}},
  \bibinfo{pages}{816} (\bibinfo{year}{1996}).

\bibitem{hopkins12c}
\bibinfo{author}{{Hopkins}, P.~F.}
\newblock \bibinfo{title}{{An excursion-set model for the structure of giant
  molecular clouds and the interstellar medium}}.
\newblock \emph{\bibinfo{journal}{\mnras}} \textbf{\bibinfo{volume}{423}},
  \bibinfo{pages}{2016--2036} (\bibinfo{year}{2012}).

\bibitem{hopkins13}
\bibinfo{author}{{Hopkins}, P.~F.}
\newblock \bibinfo{title}{{A general theory of turbulent fragmentation}}.
\newblock \emph{\bibinfo{journal}{\mnras}} \textbf{\bibinfo{volume}{430}},
  \bibinfo{pages}{1653--1693} (\bibinfo{year}{2013}).

\bibitem{maclow04}
\bibinfo{author}{{Mac Low}, M.-M.} \& \bibinfo{author}{{Klessen}, R.~S.}
\newblock \bibinfo{title}{{Control of star formation by supersonic
  turbulence}}.
\newblock \emph{\bibinfo{journal}{Reviews of Modern Physics}}
  \textbf{\bibinfo{volume}{76}}, \bibinfo{pages}{125--194}
  (\bibinfo{year}{2004}).

\bibitem{henshaw16b}
\bibinfo{author}{{Henshaw}, J.~D.}, \bibinfo{author}{{Longmore}, S.~N.} \&
  \bibinfo{author}{{Kruijssen}, J.~M.~D.}
\newblock \bibinfo{title}{{Seeding the Galactic Centre gas stream:
  gravitational instabilities set the initial conditions for the formation of
  protocluster clouds}}.
\newblock \emph{\bibinfo{journal}{\mnras}} \textbf{\bibinfo{volume}{463}},
  \bibinfo{pages}{L122--L126} (\bibinfo{year}{2016}).

\bibitem{zamoraaviles12}
\bibinfo{author}{{Zamora-Avil{\'e}s}, M.},
  \bibinfo{author}{{V{\'a}zquez-Semadeni}, E.} \&
  \bibinfo{author}{{Col{\'{\i}}n}, P.}
\newblock \bibinfo{title}{{An Evolutionary Model for Collapsing Molecular
  Clouds and Their Star Formation Activity}}.
\newblock \emph{\bibinfo{journal}{\apj}} \textbf{\bibinfo{volume}{751}},
  \bibinfo{pages}{77} (\bibinfo{year}{2012}).

\bibitem{elmegreen18}
\bibinfo{author}{{Elmegreen}, B.~G.}
\newblock \bibinfo{title}{{On the Appearance of Thresholds in the Dynamical
  Model of Star Formation}}.
\newblock \emph{\bibinfo{journal}{\apj}} \textbf{\bibinfo{volume}{854}},
  \bibinfo{pages}{16} (\bibinfo{year}{2018}).

\bibitem{burkhart18}
\bibinfo{author}{{Burkhart}, B.}
\newblock \bibinfo{title}{{The Star Formation Rate in the Gravoturbulent
  Interstellar Medium}}.
\newblock \emph{\bibinfo{journal}{\apj}} \textbf{\bibinfo{volume}{863}},
  \bibinfo{pages}{118} (\bibinfo{year}{2018}).

\bibitem{vazquez18}
\bibinfo{author}{{V{\'a}zquez-Semadeni}, E.},
  \bibinfo{author}{{Zamora-Avil{\'e}s}, M.},
  \bibinfo{author}{{Galv{\'a}n-Madrid}, R.} \& \bibinfo{author}{{Forbrich}, J.}
\newblock \bibinfo{title}{{Molecular cloud evolution - VI. Measuring cloud
  ages}}.
\newblock \emph{\bibinfo{journal}{\mnras}} \textbf{\bibinfo{volume}{479}},
  \bibinfo{pages}{3254--3263} (\bibinfo{year}{2018}).

\bibitem{feldmann11}
\bibinfo{author}{{Feldmann}, R.}, \bibinfo{author}{{Gnedin}, N.~Y.} \&
  \bibinfo{author}{{Kravtsov}, A.~V.}
\newblock \bibinfo{title}{{How Universal is the $\Sigma_{SFR} - \Sigma _{H_2}$
  Relation?}}
\newblock \emph{\bibinfo{journal}{\apj}} \textbf{\bibinfo{volume}{732}},
  \bibinfo{pages}{115} (\bibinfo{year}{2011}).

\bibitem{semenov16}
\bibinfo{author}{{Semenov}, V.~A.}, \bibinfo{author}{{Kravtsov}, A.~V.} \&
  \bibinfo{author}{{Gnedin}, N.~Y.}
\newblock \bibinfo{title}{{Nonuniversal Star Formation Efficiency in Turbulent
  ISM}}.
\newblock \emph{\bibinfo{journal}{\apj}} \textbf{\bibinfo{volume}{826}},
  \bibinfo{pages}{200} (\bibinfo{year}{2016}).

\bibitem{semenov17}
\bibinfo{author}{{Semenov}, V.~A.}, \bibinfo{author}{{Kravtsov}, A.~V.} \&
  \bibinfo{author}{{Gnedin}, N.~Y.}
\newblock \bibinfo{title}{{The Physical Origin of Long Gas Depletion Times in
  Galaxies}}.
\newblock \emph{\bibinfo{journal}{\apj}} \textbf{\bibinfo{volume}{845}},
  \bibinfo{pages}{133} (\bibinfo{year}{2017}).

\bibitem{semenov18}
\bibinfo{author}{{Semenov}, V.~A.}, \bibinfo{author}{{Kravtsov}, A.~V.} \&
  \bibinfo{author}{{Gnedin}, N.~Y.}
\newblock \bibinfo{title}{{How Galaxies Form Stars: The Connection between
  Local and Global Star Formation in Galaxy Simulations}}.
\newblock \emph{\bibinfo{journal}{\apj}} \textbf{\bibinfo{volume}{861}},
  \bibinfo{pages}{4} (\bibinfo{year}{2018}).

\bibitem{ochsendorf17}
\bibinfo{author}{{Ochsendorf}, B.~B.}, \bibinfo{author}{{Meixner}, M.},
  \bibinfo{author}{{Roman-Duval}, J.}, \bibinfo{author}{{Rahman}, M.} \&
  \bibinfo{author}{{Evans}, N.~J., II}.
\newblock \bibinfo{title}{{What Sets the Massive Star Formation Rates and
  Efficiencies of Giant Molecular Clouds?}}
\newblock \emph{\bibinfo{journal}{\apj}} \textbf{\bibinfo{volume}{841}},
  \bibinfo{pages}{109} (\bibinfo{year}{2017}).

\bibitem{silk97}
\bibinfo{author}{{Silk}, J.}
\newblock \bibinfo{title}{{Feedback, Disk Self-Regulation, and Galaxy
  Formation}}.
\newblock \emph{\bibinfo{journal}{\apj}} \textbf{\bibinfo{volume}{481}},
  \bibinfo{pages}{703} (\bibinfo{year}{1997}).

\bibitem{elmegreen97}
\bibinfo{author}{{Elmegreen}, B.~G.} \& \bibinfo{author}{{Efremov}, Y.~N.}
\newblock \bibinfo{title}{{A Universal Formation Mechanism for Open and
  Globular Clusters in Turbulent Gas}}.
\newblock \emph{\bibinfo{journal}{\apj}} \textbf{\bibinfo{volume}{480}},
  \bibinfo{pages}{235--245} (\bibinfo{year}{1997}).

\bibitem{kennicutt98}
\bibinfo{author}{{Kennicutt}, R.~C., Jr.}
\newblock \bibinfo{title}{{Star Formation in Galaxies Along the Hubble
  Sequence}}.
\newblock \emph{\bibinfo{journal}{\araa}} \textbf{\bibinfo{volume}{36}},
  \bibinfo{pages}{189--232} (\bibinfo{year}{1998}).

\bibitem{fujimoto14}
\bibinfo{author}{{Fujimoto}, Y.}, \bibinfo{author}{{Tasker}, E.~J.},
  \bibinfo{author}{{Wakayama}, M.} \& \bibinfo{author}{{Habe}, A.}
\newblock \bibinfo{title}{{Do giant molecular clouds care about the galactic
  structure?}}
\newblock \emph{\bibinfo{journal}{\mnras}} \textbf{\bibinfo{volume}{439}},
  \bibinfo{pages}{936--953} (\bibinfo{year}{2014}).

\bibitem{dobbs15}
\bibinfo{author}{{Dobbs}, C.~L.}, \bibinfo{author}{{Pringle}, J.~E.} \&
  \bibinfo{author}{{Duarte-Cabral}, A.}
\newblock \bibinfo{title}{{The frequency and nature of `cloud-cloud collisions'
  in galaxies}}.
\newblock \emph{\bibinfo{journal}{\mnras}} \textbf{\bibinfo{volume}{446}},
  \bibinfo{pages}{3608--3620} (\bibinfo{year}{2015}).

\bibitem{tasker15}
\bibinfo{author}{{Tasker}, E.~J.}, \bibinfo{author}{{Wadsley}, J.} \&
  \bibinfo{author}{{Pudritz}, R.}
\newblock \bibinfo{title}{{Star Formation in Disk Galaxies. III. Does Stellar
  Feedback Result in Cloud Death?}}
\newblock \emph{\bibinfo{journal}{\apj}} \textbf{\bibinfo{volume}{801}},
  \bibinfo{pages}{33} (\bibinfo{year}{2015}).

\bibitem{meidt13}
\bibinfo{author}{{Meidt}, S.~E.} \emph{et~al.}
\newblock \bibinfo{title}{{Gas Kinematics on Giant Molecular Cloud Scales in
  M51 with PAWS: Cloud Stabilization through Dynamical Pressure}}.
\newblock \emph{\bibinfo{journal}{\apj}} \textbf{\bibinfo{volume}{779}},
  \bibinfo{pages}{45} (\bibinfo{year}{2013}).

\bibitem{meidt18}
\bibinfo{author}{{Meidt}, S.~E.} \emph{et~al.}
\newblock \bibinfo{title}{{A Model for the Onset of Self-gravitation and Star
  Formation in Molecular Gas Governed by Galactic Forces. I. Cloud-scale Gas
  Motions}}.
\newblock \emph{\bibinfo{journal}{\apj}} \textbf{\bibinfo{volume}{854}},
  \bibinfo{pages}{100} (\bibinfo{year}{2018}).

\bibitem{ostriker10}
\bibinfo{author}{{Ostriker}, E.~C.}, \bibinfo{author}{{McKee}, C.~F.} \&
  \bibinfo{author}{{Leroy}, A.~K.}
\newblock \bibinfo{title}{{Regulation of Star Formation Rates in Multiphase
  Galactic Disks: A Thermal/Dynamical Equilibrium Model}}.
\newblock \emph{\bibinfo{journal}{\apj}} \textbf{\bibinfo{volume}{721}},
  \bibinfo{pages}{975--994} (\bibinfo{year}{2010}).

\bibitem{ostriker11}
\bibinfo{author}{{Ostriker}, E.~C.} \& \bibinfo{author}{{Shetty}, R.}
\newblock \bibinfo{title}{{Maximally Star-forming Galactic Disks. I. Starburst
  Regulation Via Feedback-driven Turbulence}}.
\newblock \emph{\bibinfo{journal}{\apj}} \textbf{\bibinfo{volume}{731}},
  \bibinfo{pages}{41} (\bibinfo{year}{2011}).

\bibitem{kim11}
\bibinfo{author}{{Kim}, C.-G.}, \bibinfo{author}{{Kim}, W.-T.} \&
  \bibinfo{author}{{Ostriker}, E.~C.}
\newblock \bibinfo{title}{{Regulation of Star Formation Rates in Multiphase
  Galactic Disks: Numerical Tests of the Thermal/Dynamical Equilibrium Model}}.
\newblock \emph{\bibinfo{journal}{\apj}} \textbf{\bibinfo{volume}{743}},
  \bibinfo{pages}{25} (\bibinfo{year}{2011}).

\bibitem{hopkins14}
\bibinfo{author}{{Hopkins}, P.~F.} \emph{et~al.}
\newblock \bibinfo{title}{{Galaxies on FIRE (Feedback In Realistic
  Environments): stellar feedback explains cosmologically inefficient star
  formation}}.
\newblock \emph{\bibinfo{journal}{\mnras}} \textbf{\bibinfo{volume}{445}},
  \bibinfo{pages}{581--603} (\bibinfo{year}{2014}).

\bibitem{stinson13}
\bibinfo{author}{{Stinson}, G.~S.} \emph{et~al.}
\newblock \bibinfo{title}{{Making Galaxies In a Cosmological Context: the need
  for early stellar feedback}}.
\newblock \emph{\bibinfo{journal}{\mnras}} \textbf{\bibinfo{volume}{428}},
  \bibinfo{pages}{129--140} (\bibinfo{year}{2013}).

\bibitem{dale15}
\bibinfo{author}{{Dale}, J.~E.}
\newblock \bibinfo{title}{{The modelling of feedback in star formation
  simulations}}.
\newblock \emph{\bibinfo{journal}{New.\ Astron.\ Rev.}}
  \textbf{\bibinfo{volume}{68}}, \bibinfo{pages}{1--33} (\bibinfo{year}{2015}).

\bibitem{gatto15}
\bibinfo{author}{{Gatto}, A.} \emph{et~al.}
\newblock \bibinfo{title}{{Modelling the supernova-driven ISM in different
  environments}}.
\newblock \emph{\bibinfo{journal}{\mnras}} \textbf{\bibinfo{volume}{449}},
  \bibinfo{pages}{1057--1075} (\bibinfo{year}{2015}).

\bibitem{gatto17}
\bibinfo{author}{{Gatto}, A.} \emph{et~al.}
\newblock \bibinfo{title}{{The SILCC project - III. Regulation of star
  formation and outflows by stellar winds and supernovae}}.
\newblock \emph{\bibinfo{journal}{\mnras}} \textbf{\bibinfo{volume}{466}},
  \bibinfo{pages}{1903--1924} (\bibinfo{year}{2017}).

\bibitem{hu17}
\bibinfo{author}{{Hu}, C.-Y.}, \bibinfo{author}{{Naab}, T.},
  \bibinfo{author}{{Glover}, S.~C.~O.}, \bibinfo{author}{{Walch}, S.} \&
  \bibinfo{author}{{Clark}, P.~C.}
\newblock \bibinfo{title}{{Variable interstellar radiation fields in simulated
  dwarf galaxies: supernovae versus photoelectric heating}}.
\newblock \emph{\bibinfo{journal}{\mnras}} \textbf{\bibinfo{volume}{471}},
  \bibinfo{pages}{2151--2173} (\bibinfo{year}{2017}).

\bibitem{matzner02}
\bibinfo{author}{{Matzner}, C.~D.}
\newblock \bibinfo{title}{{On the Role of Massive Stars in the Support and
  Destruction of Giant Molecular Clouds}}.
\newblock \emph{\bibinfo{journal}{\apj}} \textbf{\bibinfo{volume}{566}},
  \bibinfo{pages}{302--314} (\bibinfo{year}{2002}).

\bibitem{krumholz09d}
\bibinfo{author}{{Krumholz}, M.~R.} \& \bibinfo{author}{{Matzner}, C.~D.}
\newblock \bibinfo{title}{{The Dynamics of Radiation-pressure-dominated H II
  Regions}}.
\newblock \emph{\bibinfo{journal}{\apj}} \textbf{\bibinfo{volume}{703}},
  \bibinfo{pages}{1352--1362} (\bibinfo{year}{2009}).

\bibitem{dale14}
\bibinfo{author}{{Dale}, J.~E.}, \bibinfo{author}{{Ngoumou}, J.},
  \bibinfo{author}{{Ercolano}, B.} \& \bibinfo{author}{{Bonnell}, I.~A.}
\newblock \bibinfo{title}{{Before the first supernova: combined effects of H II
  regions and winds on molecular clouds}}.
\newblock \emph{\bibinfo{journal}{\mnras}} \textbf{\bibinfo{volume}{442}},
  \bibinfo{pages}{694--712} (\bibinfo{year}{2014}).

\bibitem{matzner15}
\bibinfo{author}{{Matzner}, C.~D.} \& \bibinfo{author}{{Jumper}, P.~H.}
\newblock \bibinfo{title}{{Star Cluster Formation with Stellar Feedback and
  Large-scale Inflow}}.
\newblock \emph{\bibinfo{journal}{\apj}} \textbf{\bibinfo{volume}{815}},
  \bibinfo{pages}{68} (\bibinfo{year}{2015}).

\bibitem{rahner17}
\bibinfo{author}{{Rahner}, D.}, \bibinfo{author}{{Pellegrini}, E.~W.},
  \bibinfo{author}{{Glover}, S.~C.~O.} \& \bibinfo{author}{{Klessen}, R.~S.}
\newblock \bibinfo{title}{{Winds and radiation in unison: a new semi-analytic
  feedback model for cloud dissolution}}.
\newblock \emph{\bibinfo{journal}{\mnras}} \textbf{\bibinfo{volume}{470}},
  \bibinfo{pages}{4453--4472} (\bibinfo{year}{2017}).

\bibitem{kim18}
\bibinfo{author}{{Kim}, J.-G.}, \bibinfo{author}{{Kim}, W.-T.} \&
  \bibinfo{author}{{Ostriker}, E.~C.}
\newblock \bibinfo{title}{{Modeling UV Radiation Feedback from Massive Stars.
  II. Dispersal of Star-forming Giant Molecular Clouds by Photoionization and
  Radiation Pressure}}.
\newblock \emph{\bibinfo{journal}{\apj}} \textbf{\bibinfo{volume}{859}},
  \bibinfo{pages}{68} (\bibinfo{year}{2018}).

\bibitem{lopez14}
\bibinfo{author}{{Lopez}, L.~A.} \emph{et~al.}
\newblock \bibinfo{title}{{The Role of Stellar Feedback in the Dynamics of H II
  Regions}}.
\newblock \emph{\bibinfo{journal}{\apj}} \textbf{\bibinfo{volume}{795}},
  \bibinfo{pages}{121} (\bibinfo{year}{2014}).

\bibitem{kennicutt07}
\bibinfo{author}{{Kennicutt}, R.~C., Jr.} \emph{et~al.}
\newblock \bibinfo{title}{{Star Formation in NGC 5194 (M51a). II. The Spatially
  Resolved Star Formation Law}}.
\newblock \emph{\bibinfo{journal}{\apj}} \textbf{\bibinfo{volume}{671}},
  \bibinfo{pages}{333--348} (\bibinfo{year}{2007}).

\bibitem{tacconi13}
\bibinfo{author}{{Tacconi}, L.~J.} \emph{et~al.}
\newblock \bibinfo{title}{{Phibss: Molecular Gas Content and Scaling Relations
  in z \~{} 1-3 Massive, Main-sequence Star-forming Galaxies}}.
\newblock \emph{\bibinfo{journal}{\apj}} \textbf{\bibinfo{volume}{768}},
  \bibinfo{pages}{74} (\bibinfo{year}{2013}).

\end{thebibliography}
\end{document}